\pdfoutput=1
\documentclass[aps,prd,amsmath,floats,floatfix, twocolumn,
superscriptaddress,nofootinbib,showpacs]{revtex4-1}

\usepackage[T1]{fontenc}
\usepackage[utf8]{inputenc}
\usepackage{lmodern}
\usepackage[utf8]{inputenc}
\usepackage{gensymb}
\usepackage{verbatim}
\usepackage{tabularx}

\usepackage[dvipsnames, usenames]{xcolor}
\definecolor{linkcolor}{rgb}{0.0,0.3,0.5}
\usepackage[hypertexnames=false, unicode, colorlinks=true, linkcolor=linkcolor,
citecolor=linkcolor, filecolor=linkcolor,urlcolor=linkcolor,
pdfusetitle]{hyperref}

\usepackage[all]{hypcap}
\usepackage{graphicx}
\usepackage{xspace}
\usepackage{amssymb}
\usepackage[normalem]{ulem} 
\usepackage{bm} 
\usepackage{multirow}
\usepackage{url}
\usepackage{microtype}

\graphicspath{%
  {figs/}%
}

\DeclareMathAlphabet{\mathpzc}{OT1}{pzc}{m}{it}

\newcommand{\mc}{\mathcal{M}}
\newcommand{\omd}{\dot{\Omega}_r}

\newcommand{\vect}[1]{\boldsymbol{#1}}
\newcommand{\uvect}[1]{\vect{\hat{#1}}}

\newcommand{\mnras}{Mon. Not. R. Astron Soc.}
\newcommand{\aap}{Astron. Astrophys.}

\newcommand{\sovast}{{\it Sov. Astronom.}}

\newcommand{\hatt}{\hat{t}}

\newcommand{\etal}{\textit{et al.\ }}

\newcommand{\phidot}{\dot{\phi}}

\begin{document}

\title{Detecting resonant tidal excitations of Rossby modes in coalescing neutron-star binaries with third-generation gravitational-wave detectors}

\newcommand\caltech{\affiliation{TAPIR, Walter Burke Institute for Theoretical Physics, California Institute of Technology, Pasadena, California 91125, USA}}

\author{Sizheng Ma}
\email{sma@caltech.edu}
\author{Hang Yu}
\email{hangyu@caltech.edu}
\author{Yanbei Chen}
\email{yanbei@caltech.edu}
\caltech

\hypersetup{pdfauthor={Ma et al.}}

\date{\today}

\begin{abstract}
Rossby modes ($r$-modes) of rotating neutron stars can be excited by the gravitomagnetic forces in  coalescing binary systems. The previous study by Flanagan and Racine [Phys. Rev. D 75, 044001 (2007)] showed that this kind of dynamical tide (DT) can induce phase shifts of $\sim 0.1$\,rad on gravitational waveforms, which is detectable by third-generation (3G) detectors. In this paper, we study the impact of this DT on measuring neutron-star parameters in the era of 3G detectors. 
We incorporate two universal relations among neutron star properties predicted by different equations of state: (i) the well-known I-Love relation between momentum of inertia and ($f$-mode) tidal Love number, and (ii) a relation between the $r$-mode overlap and tidal Love number, which is newly explored in this paper. We find that $r$-mode DT will provide rich information about slowly rotating neutron stars with frequency $10-100$ Hz and spin inclination angle $18^\circ-110^\circ$. For a binary neutron star system (with a signal-to-noise ratio $\sim1500$ in the Cosmic Explorer), the spin frequency of each individual neutron star can be constrained to 6\% (fractional error) in the best-case scenario. The degeneracy between the Love numbers of individual neutron stars is dramatically reduced:  each individual Love number can be constrained to around 20\% in the best case, while the fractional error for both symmetric and anti-symmetric Love numbers are reduced by factors of around 300. Furthermore, DT also allows us to  measure the spin inclination angles of the neutron stars, to 0.09\,rad in the best case, and thus place constraints on NS natal kicks and supernova explosion models. Besides parameter estimation, we have also developed a semi-analytic method that accurately describes detailed features of the binary evolution that arise due to the DT.

\end{abstract}

\maketitle

\section{Introduction} 
\label{sec:introduction}
Gravitational-wave (GW) astronomy recently provided us a new way to study neutron star (NS) physics, with events GW170817 \cite{TheLIGOScientific:2017qsa,Monitor:2017mdv} and GW190425 \cite{Abbott:2020uma}  already imposing new constraints on NS properties~\cite{Abbott:2018exr,LIGOScientific:2019eut,Annala:2017llu,Most:2018hfd,Margalit:2017dij,Rezzolla:2017aly,Ruiz:2017due,Shibata:2017xdx,Pratten:2019sed}.  With future upgrades for Advanced LIGO and Virgo \cite{TheLIGOScientific:2014jea,Tse:2019wcy,Harry:2010zz,Aasi:2013wya,ligo4,ligo5,LIGOScientific:2019vkc,voyager,virgo1,TheVirgo:2014hva,Acernese:2019sbr,virgo2}, LIGO-India \cite{ligo3}, KAGRA \cite{Aso:2013eba,Somiya:2011np}, as well as third-generation detectors like the Einstein Telescope \cite{Punturo:2010zza,Hild:2010id,et2,ET} and the Cosmic Explorer (CE) \cite{Evans:2016mbw,Reitze:2019iox}, we expect to detect more events, as well as achieving much higher signal-to-noise ratios \cite{Baibhav:2019gxm}. These new opportunities motivate the more accurate modeling of NSs in coalescing binaries, and the GWs they emit \cite{Hinderer:2016eia,Dietrich:2017aum,Akcay:2018yyh,Nagar:2018zoe,Barkett:2019tus,Messina:2019uby}.

During the last few minutes of a binary neutron star inspiral, the orbital frequency sweeps through tens of Hertz to hundreds of Hertz; internal fluid motions of NSs may get resonantly excited by tidal forces exerted by their companions.  Such fluid motion will also act back onto the orbital motion. This phenomenon is called dynamical tide (DT), which was first investigated by Cowling \cite{cowling1941non}. For non-spinning NSs in binaries with circular orbits, only $g$-modes can be resonantly excited, but their effect on gravitational waveforms is negligible~\cite{reisenegger1994excitation,Lai+94, Yu:17a, Yu:2017cxe}. However, DT can be enhanced by orbital eccentricity \cite{Chirenti:2016xys,Yang:2018bzx,Yang:2019kmf,Vick:2019cun} and rotation \cite{Lai:1996pn,Lai_1997,Ho+Lai+99,Ma:2020rak}. 

The effect of rotation on DTs was first studied by Lai \etal \cite{Lai:1996pn,Lai_1997}, who aimed to explain the orbital decay of the PSR J0045-7319/B binary system. They pointed out that stellar rotation can  excite $f$-modes and lower-order $g$-modes. Since these modes couple more strongly to the tidal field, DTs should be more pronounced. The same formalism was then applied to coalescing binaries by Ho and Lai \cite{Ho+Lai+99} to investigate the impact of DTs on GW.  It was found that  $f$-mode resonances require NSs to rotate at very high frequencies --- and only in this case it induces significant GW phase shift.  On the other hand, $g$-mode resonances are still too weak to be detected. 
Several authors later extended the above Newtonians studies, and discussed the excitation of $f$-modes and $g$-modes in the context of general relativity \cite{Gualtieri:2001cm,Pons:2001xs,Miniutti:2002bh,Steinhoff:2016rfi,Hinderer:2016eia,Schmidt:2019wrl}. 

Although $f$-mode resonances can 
significantly influence GW, the high rotation rate required here is disfavored by the astrophysics of formation scenarios. 
It is generally believed that NSs in binaries have already spun down to low frequencies ($\lesssim 40$ Hz) when their GW frequencies enter the band of ground-based detectors \cite{Lyne1153,Andersson+Ho+18}. For example, recent events GW170817 \cite{TheLIGOScientific:2017qsa} and GW190425 \cite{Abbott:2020uma} were all consistent with low spins. 
In this way, $f$-mode resonance might not be very promising for GW observations. 

For spinning neutron stars, besides $f$-, $p$- and $g$-modes, there also exist inertial modes (hybrid modes) that are mainly supported by the Coriolis force \cite{Lindblom:1998ka,Lockitch:1998nq,Yoshida:1999jz}. As pointed out by Lockitch and Friedman \cite{Lockitch:1998nq}, for isentropic stars\footnote{Isentropic stars have no buoyancy; both $g$-modes and $r$-modes (purely axial modes) have vanishing mode frequency in the non-spinning limit, and they are mixed even without rotation.}, rotation mixes purely axial modes and purely polar modes, and leads to a class of modes which have hybrid parity, where each of them can be classified into axial-led or polar-led 
For Newtonian stars, there is a special subclass of modes that has a purely axial parity, which is usually referred to as the Rossby modes (or $r$-modes) \cite{Papaloizou:1978zz,1982ApJ...256..717S}, although this particular subclass also obtains a polar part for relativistic stars \cite{Lockitch:2000aa,Lockitch:2002sy,Idrisy:2014qca}. On the other hand, for non-isentropic NSs, the mixing of $g$-modes and $r$-modes takes place at relatively high rotation rate (where the Coriolis force dominates over the buoyancy), leading to the so-called inertial-gravity modes.

Previous studies have shown that inertial modes (including $r$-modes)  and inertial-gravity modes can be excited by the (gravito-electric) Newtonian tidal field \cite{Ho+Lai+99,Lai:2006pr,Xu:2017hqo}. The resonance takes place as the orbital angular velocity becomes comparable to the spin frequency, which requires a minimum NS spin of only tens of Hertz.  Although this spin requirement is more likely fulfilled, the effect of such resonances are too weak to be detectable.  However, inertial modes (including $r$-modes) can also be excited by the gravitomagnetic force \cite{Schenk+01,Flanagan:2006sb,Poisson:2020eki,Poisson:2020mdi,Poisson:2020ify}. As pointed out by Flanagan \etal (hereafter FR07) \cite{Flanagan:2006sb}, this kind of DT can induce $\sim0.1$\,rad of GW phase shift ---   detectable by  third-generation detectors, like ET and CE. Their studies, though, did not provide detailed discussions on how those DTs impact parameter estimation. Later on, Yu \etal \cite{Yu:2017zgi} proposed that the $r$-mode DTs can improve the accuracy in measuring tidal Love numbers. The aim of this paper is to build upon FR07 and Ref.\ \cite{Yu:2017zgi}, and study the $r$-mode DTs with more details, including a Fisher-matrix analysis. Especially, we will investigate the impact of universal relations among NS  properties \cite{Yagi:2013awa,2013Sci...341..365Y}.


This paper is organized as follows. We first give a summary on the implication of $r$-mode DT on GW parameter estimation in Sec.\ \ref{sec:summary}, and the rest of this paper presents the details. In Sec.\ \ref{sec:EOM}, we briefly review the coupling between the $r$-mode and the gravitomagnetic force. In Sec.\ \ref{sec:dyna-r-mode}, we discuss features of $r$-mode excitation and the $r$-mode's impact on the orbital evolution. We first focus on the orbital part in Sec.\ \ref{sec:EOM-orbit}, where we compare the model presented in FR07 [Eq.\ (\ref{Ddot-pp-intro})] with numerical integration. Next we study the tidal excitation in Sec.\ \ref{sec:EOM-tide}, where we give an analytic formula for $r$-mode tidal evolution that is valid in the entire evolution regime.  In Sec.\ \ref{sec:universal}, we explore the universal relation between the normalized $r$-mode overlap and the normalized tidal Love number. Sec.\ \ref{sec:GW} focuses on the construction of gravitational waveforms.  We first propose a hybrid GW waveform in Sec.\ \ref{sec:GW-hybrid}, which incorporates both $r$-mode  and other PN effects. Next in Sec.\ \ref{sec:toy-model} we provide a model for the GW phase of $r$-mode, following FR07. In Sec.\ \ref{sec:res}, we use the Fisher information matrix formalism to discuss
the influence of $r$-mode excitation on parameter estimation. Finally in Sec.\ \ref{sec:conclusion} we summarize our results.

Throughout this paper we use the geometric units with
$G = c = 1$.

\section{Summary of parameter dependencies}
\label{sec:summary}

In this section, we shall give a brief summary of $r$-mode resonance's impact on the gravitational waveform, and on how physical parameters of neutron-star and neutron-star--black-hole binaries influence the gravitational waveform, via point-particle, spin-orbit, $f$-mode tide, and $r$-mode resonance effects.  We shall divide binaries into four categories, and discuss in each category how some, or all, of the physical parameters of the neutron stars can be measured. 

\subsection{The role of $r$-mode dynamical tide}
\label{sec:intro-dt}
As discussed in FR07 \cite{Flanagan:2006sb}, $r$-mode excitation is mainly controlled by the spin frequency $\Omega_s$ and $r$-mode coupling coefficient $\mathcal{I}$ [Eq.\ (\ref{mathI})], which play different roles: (i) spin frequency determines the location (both in time and frequency domains) of resonance during orbital evolution, because resonance takes place as the tidal driving frequency (related to the orbital frequency) coincides with the pattern frequency of the mode (determined by the spin frequency) in the inertial frame; (ii) the $r$-mode coupling coefficient $\mathcal{I}$ characterizes the strength of the $r$-mode DT, therefore determines the effect of this DT on orbital evolution [See Sec.~\ref{sec:dyna-r-mode} for details]. As proposed in FR07, the orbital evolution before and after the $r$-mode resonance can described as two different point-particle (PP) orbits, with orbital-frequency evolution given by 
\begin{align}
\phidot(t)=\begin{cases}
    \phidot^{(\text{pre})}_\text{PP}(t),& \text{if } t< t_r\\
    \phidot^{(\text{post})}_\text{PP}(t),              & \text{if } t> t_r
\end{cases},\label{Ddot-pp-intro}
\end{align}
with $t_r$ the time of resonance\footnote{The resonance condition is given by Eq.~(\ref{reson-condi}). Hereafter we use the subscript $r$ to refer to the value at the resonance time.}. These two frequency evolutions satisfy the same evolution equation, with the same set of  parameters (including mass, spin, inclination angles, etc) which, at Newtonian order, reads
\begin{align}
\frac{d}{dt}\phidot_{(\text{PP})}=\frac{96}{5}\mc^{5/3}\phidot_{(\text{PP})}^{11/3}, 
\end{align}
where $\mc$ is the chirp mass.  The two orbits are related to each other by  a ``jump'' at $t_r$
\begin{align}
\phidot^{(\text{post})}_\text{PP}(t_r)-\phidot^{(\text{pre})}_\text{PP}(t_r)=\Delta \phidot_{\rm tid}, \label{jump-intro}
\end{align}
where $\Delta \phidot_{\rm tid}$ depends on $\mathcal{I}$ (see Sec.\ \ref{sec:EOM-orbit} for more details). As a result, by studying the GW emitted from the orbit in Eq.\ (\ref{Ddot-pp-intro}), we can obtain the constraints on $\Omega_s$ and $\mathcal{I}$.

\subsection{The impact of $r$-mode DT on parameter estimation}

In order to explore how the additional information from the $r$-mode DT can improve parameter estimation, we consider both BNS and BHNS binaries, and distinguish whether {\it universal relations} between neutron-star properties are used as input. This will divide these binaries into four categories. 


\subsubsection{BNS systems without $r$-mode DT and universal relations}

\begin{figure}[htb]
        \includegraphics[width=\columnwidth,height=6.9cm,clip=true]{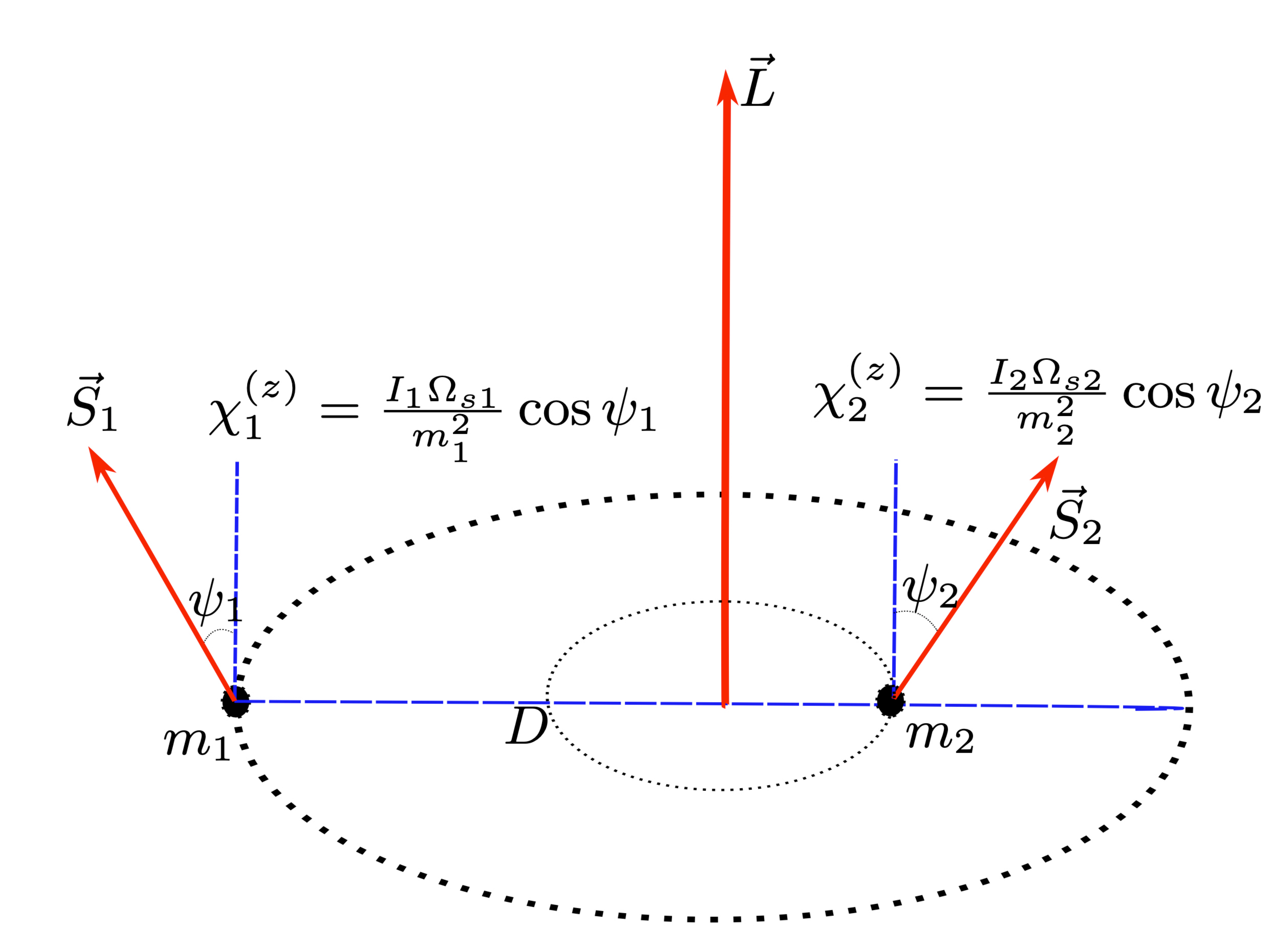} 
  \caption{A BNS system $m_1-m_2$ with two spin vectors $\vec{S}_1$ and $\vec{S}_2$. The neutron stars' spin axis are tilted by angles $\psi_{1,2}$ with respect to the direction of the orbital angular momentum $\vec{L}$. Here the azimuthal angle of the spins are unimportant, because the effect of precession is negligible.}
 \label{fig:orbit-illu}
\end{figure}

A spinning BNS system is sketched in Fig.~\ref{fig:orbit-illu}. 
Without $r$-mode DT (but with $f$-mode adiabatic tide), the system has 6 intrinsic parameters: chirp mass $\mc$; mass ratio $\Xi=m_1/(m_1+m_2)$; two individual tidal Love numbers $\lambda_f^{(i)}$; and two dimensionless spin along the direction of orbital angular momentum $\chi_i^{(z)}$ (The index $i=1,2$ labels the two NSs). In absence of DT, the system evolves under four main effects: the Newtonian gravity, post-Newtonian (PN) corrections from mass ratio,  PN corrections from spin-orbit (SO) coupling, and the adiabatic ($f$-mode) tidal effect, as summarized in Table \ref{table:pp-intro}. Here we do not include the PN spin-spin coupling \cite{Nagar:2018plt} and PN spin precession \cite{Schmidt:2014iyl}, because their effects are negligible even for 3G detectors for the range of spins we consider ($10\sim80$ Hz, see Appendix \ref{app:precession} for more details). For a BNS system, the waveform only depends on a combination of $\lambda_f^{(i)}$ and a combination of $\chi_i^{z}$, not individually; 
as a result, we can only measure four parameters to meaningful accuracy: $\mc$, $\Xi$, $\chi^\text{eff}$, and $\lambda_f^{\text{eff}}$, with $\lambda_f^{\text{eff}}$ and $\chi^\text{eff}$ defined by \cite{Flanagan:2007ix,Racine:2008qv}
 \begin{align}
&\chi^\text{eff}=\frac{m_1\chi_1^{(z)}+m_2\chi_2^{(z)}}{M}, \notag \\
&\lambda_f^{\rm eff}=\left(11\frac{m_2}{m_1}+\frac{M}{m_1}\right)\lambda_f^{(1)}+\left(11\frac{m_1}{m_2}+\frac{M}{m_2}\right)\lambda_f^{(2)}, \notag 
\end{align}
where $M=m_1+m_2$. The two individual Love numbers $\lambda_f^{(i)}$, as well as the two $\chi_i^{(z)}$, are degenerate. Even among the constrainable parameters, errors in $\Xi$, $\chi^\text{eff}$, and $\lambda_f^{\text{eff}}$ are highly correlated with each other. In order to obtain a good constraint on $\lambda_f^{\text{eff}}$, a low-spin prior based on the observed Galactic binary NS population has to be assumed~\cite{TheLIGOScientific:2017qsa,Abbott:2018exr}. As we shall discuss later, such assumption is not necessary if  $r$-mode resonance can be incorporated.

\begin{table}
    \centering
    \caption{Parameters and constraints for a BNS system without  $r$-mode DT and not applying the universal relations between NS properties. We have 6 parameters and 4 independent constraints; as a result, the two individual Love numbers are degenerate, so are the two individual dimensionless spins.}
    \begin{tabular}{c c c c} \hline\hline
   All  variables  & \multirow{2}{*}{Effect} &   Variables in   & Constrainable    \\ 
  in GW    &  & the effect &variable \\ \hline
    \multirow{5}{*}{\shortstack{$\mc,\Xi$ \\$\lambda_f^{(1)},\lambda_f^{(2)}$ \\ $\chi^{(z)}_1,\chi^{(z)}_2$} }    &  PN (PP part)  & $\mc$ & $\mc$ \\ \cline{2-4}
   &   PN (PP part) & $\Xi$ & $\Xi$ \\ \cline{2-4}
  &  SO coupling & $\chi^{(z)}_1,\chi^{(z)}_2$ & $\chi^\text{eff}$ \\ \cline{2-4}
     &   Adiabatic tide& \multirow{2}{*}{$\lambda_f^{(1)},\lambda_f^{(2)}$} &  \multirow{2}{*}{$\lambda_f^{\text{eff}}$} \\ 
      &  ($f$-mode) & & \\ \hline\hline
     \end{tabular}
     \label{table:pp-intro}
\end{table}

\subsubsection{BNS systems with the I-Love universal relation but without DT}
\label{sec:intro-BNS-no-r-uni}
For the same BNS system (without DT), the {\it I-Love} universal relation of NSs \cite{Yagi:2013awa,2013Sci...341..365Y} can be used to improve  parameter estimation. This is a relation between momentum of inertia and tidal Love number, that is  insensitive to the EoS. To use this relation, we express  $\chi^{(z)}$ of each NS as
\begin{align}
\chi^{(z)}_i=\frac{I_i\Omega_{si}}{m_i^2}\cos\psi_i.
\end{align}
In this case, we have a total  of 10 parameters: chirp mass $\mc$; mass ratio $\Xi$; two Love numbers $\lambda_f^{(i)}$; two moments of inertia $I_i$; two inclination angles $\psi_{i}$; and two spin frequencies $\Omega_{si}$. As listed in Table \ref{table:rmode-uni-intro} (``Non-DT sector''), there are 6 constraints on these parameters: four are from GW $(\mc,\Xi,\chi^{\rm eff},\lambda_f^{\rm eff})$, as discussed above; and two are from universal relations (each star contributes one constraint). This is still not enough to independently constrain all 10 parameters.



Introducing the universal relations did not reduce the number of degeneracies because more parameters, namely spin frequency and inclination angle for each NS, are needed to be introduced in order to use these relations.  The situation will change as we consider $r$-mode DT. 


\subsubsection{Resonant BNS systems with universal relations}
Now take the $r$-mode resonance into account.
As shown in Table \ref{table:rmode-uni-intro} , we have in total 12 parameters: chirp mass $\mc$; mass ratio $\Xi$; two Love numbers $\lambda_f^{(i)}$; two momentum of inertia $I_i$; two spin frequencies $\Omega_{si}$; two inclination angles $\psi_{i}$; and two $r$-mode coupling coefficients $\mathcal{I}_i$. Meanwhile, we can obtain 12 constraints from GW and the universal relations. Six of them $(\mc,\Xi,\chi^{\rm eff},\lambda_f^{\rm eff},I_{1,2})$ have already been discussed in Sec.\ \ref{sec:intro-BNS-no-r-uni}. The rest of 6 constraints involve the $r$-mode. Four are constraints on $\Omega_{si}$ and $\mathcal{I}_i$, as we discussed in Sec.\ \ref{sec:intro-dt}. The other two constraints come from a new universal relation found in this paper, namely  the relation between the normalized $r$-mode overlap and the normalized Love number (cf. Sec.\ \ref{sec:universal}). As a result, the number of constraints are the same as the number of parameters. In principle, this breaks degeneracy, and allows us to estimate all parameters independently, given high enough signal-to-noise ratio.  However, as it later turns out, even in the era of 3G detectors, for each individual binary, we still may not fully constrain all parameters with relative error less than 100\%, since the DT sector is not strong enough.


\begin{table}
    \centering
    \caption{Parameters and constraints for a BNS system with universal relations (for both NSs). In the ``Non-DT sector'', when $r$-mode DTs does not take place, we have 10 parameters and 6 constraints, with 4 degeneracies. In prescence of $r$-mode DTs, we have 2 more parameters, but 6 more constraints. As a result, we have in total 12 parameters and 12 constraints; the system can in principle be decoded without degeneracy.}
    \begin{tabular}{c c c c c} \hline\hline
  Variables &  \multirow{2}{*}{Sector}  &   \multirow{2}{*}{Effect} &   Variables  & Constrainable  \\ 
in GW & &       & in  effect &variable \\ \hline
   \multirow{17}{*}{\shortstack{$\mc,\Xi$\\$\lambda_f^{(1)},\lambda_f^{(2)}$\\$I_1,I_2$ \\$\psi_1,\psi_2$\\$\Omega_{s1},\Omega_{s2}$ }} &\multirow{12}{*}{\shortstack{Non-\\DT\\sector}}  &   PN & \multirow{2}{*}{$\mc$} & \multirow{2}{*}{$\mc$} \\ 
   & & (PP part)& & \\ \cline{3-5}
   &    &  PN& \multirow{2}{*}{$\Xi$} & \multirow{2}{*}{$\Xi$} \\ 
     & & (PP part)& & \\ \cline{3-5}
  & &    SO- & $I_1,\Omega_{s1},\psi_1,\mc$ & \multirow{2}{*}{$\chi^\text{eff}$} \\
  & &    coupling  & $I_2,\Omega_{s2},\psi_2,\Xi$ & \\ \cline{3-5}
 &   &     Adiabatic & \multirow{3}{*}{$\lambda_f^{(1)},\lambda_f^{(2)}$} &  \multirow{3}{*}{$\lambda_f^{\text{eff}}$} \\ 
  & \multirow{12}{*}{} &  tide & & \\ 
 & & ($f$-mode) & & \\ \cline{3-5}
& &     I-Love  &$\lambda_f^{(1)},\lambda_f^{(2)}$ &\multirow{3}{*}{$I_1,I_2$} \\ 
&  &     universal & $I_1,I_2$ & \\
& & relation& $\Xi$& \\  \cline{2-5}
\multirow{5}{*}{$\mathcal{I}_1,\mathcal{I}_2$} & \multirow{5}{*}{\shortstack{DT\\sector}} &   $r$-mode  & $\Omega_{s1},\Omega_{s2}$ & $\Omega_{s1},\Omega_{s2}$ \\ 
  &  &  resonances & $\mathcal{I}_{1},\mathcal{I}_{2}$ & $\mathcal{I}_{1},\mathcal{I}_{2}$  \\ \cline{3-5}
  &   &  $r$-mode  & $\mathcal{I}_1,\psi_1,\lambda_f^{(1)}$ &\multirow{3}{*}{ $\psi_1,\psi_2$ } \\
 &  &    universal  &  $\mathcal{I}_2,\psi_2,\lambda_f^{(2)}$ &  \\
 & & relation& $\Xi$& \\
       \hline\hline
     \end{tabular}
     \label{table:rmode-uni-intro}
\end{table}

\subsubsection{Resonant BHNS systems with universal relations}
\label{sec:intro-BHNS}
For BHNS systems, the $r$-mode resonance only takes place once before merger. As shown in Table\ \ref{table:rmode-uni-bhns-intro}, the system has 8 parameters: chirp mass $\mc$; mass ratio $\Xi$; NS Love number $\lambda_f^{(1)}$; NS moment of inertia $I_1$; NS spin frequency of NS $\Omega_{s1}$; NS inclination angle $\psi_1$; NS $r$-mode coupling coefficient $\mathcal{I}_1$, and the spin of BH along the direciton of orbital angular momentum $\chi_2^{(z)}$. We can obtain 8 constraints on these parameters, with 6 of them from GW $(\mc,\Xi,\chi^{\rm eff},\lambda_f^{(1)},\Omega_{s1},\mathcal{I}_1)$, and 2 from the two universal relations. In this way, the BHNS system is also expected to be decoded without degeneracy, given enough signal-to-noise ratio.

\begin{table}
    \centering
    \caption{Parameters and constraints for a BHNS system with universal relations and $r$-mode DT. We have 8 parameters and 8 constraints, and the system can be decoded without degeneracy.}
    \begin{tabular}{c c c c} \hline\hline
  All variables    &   \multirow{2}{*}{Effect} &   Variables  & Constrainable  \\ 
in GW &       & in the effect &variable \\ \hline
   \multirow{8}{*}{$\mc,\Xi$} &     PN (PP part) & $\mc$ & $\mc$ \\ \cline{2-4}
  \multirow{8}{*}{$\lambda_f^{(1)},I_1$}     &  PN (PP part) & $\Xi$ & $\Xi$ \\ \cline{2-4}
   \multirow{8}{*}{$\psi_1,\Omega_{s1}$}  &    SO & $I_1,\Omega_{s1},\psi_1$ & \multirow{2}{*}{$\chi^\text{eff}$} \\
\multirow{8}{*}{$\mathcal{I}_1,\chi_2^{(z)}$}   &    coupling  & $\chi_2^{(z)},\Xi,\mc$ & \\ \cline{2-4}
 \multirow{8}{*}{}    &     Adiabatic tide& \multirow{2}{*}{$\lambda_f^{(1)}$} &  \multirow{2}{*}{$\lambda_f^{(1)}$} \\ 
  &    ($f$-mode) & & \\ \cline{2-4}
 &     I-Love  &\multirow{2}{*}{$\lambda_f^{(1)},I_1$}&\multirow{2}{*}{$I_1$} \\ 
  &     universal relation&  & \\ \cline{2-4}
   &   $r$-mode  & \multirow{2}{*}{$\Omega_{s1},\mathcal{I}_{1}$}  & \multirow{2}{*}{$\Omega_{s1},\mathcal{I}_{1}$}  \\ 
    &  resonances & &   \\ \cline{2-4}
     &  $r$-mode  & $\mathcal{I}_1,\psi_1$ &\multirow{2}{*}{ $\psi_1$ } \\
   &    universal relation &  $\Xi,\lambda_f^{(1)}$ &  \\
       \hline\hline
     \end{tabular}
     \label{table:rmode-uni-bhns-intro}
\end{table}


%


\section{Basic equations of dynamical tides} 
\label{sec:EOM}
In this section, we briefly review the coupling between $r$-modes and gravitomagnetic force in coalescing binary systems. We refer the reader to Refs. \cite{Flanagan:2006sb,Racine:2004hs} by Flanagan and Racine for further details. In Sec. \ref{EOM:rmode}, we first provide some basic information about $r$-modes in NSs. All equations are kept to linear order in spin frequency. In Sec. \ref{EOM:stellar}, we show how $r$-modes are driven by the gravitomagnetic force. Finally in Sec. \ref{EOM:orbital}, we discuss the tidal back reaction to the orbit and present the full equation of motion (EOM). In this section, as done in FR07, the slowly rotating NS is treated in Newtonian gravity.  This will lead to the correct form of evolution equations, although parameters may eventually need relativistic corrections. We shall incorporate PN and $f$-mode adiabatic tide corrections using a hybrid approach described in Sec.~\ref{sec:GW}.

\subsection{Rossby modes}
\label{EOM:rmode}

For a rotating NS with mass $m_1$, we introduce a co-rotating frame $(x^\prime,y^\prime,z^\prime)$, in which the spin of the star $\bm{\Omega}_{s1}$ is along the $z^\prime$ direction. 
In this coordinate system, the perturbation equation of the rotating star is given by
\begin{align}
\frac{\partial^2 \bm{\xi}}{\partial t^2}+2\bm{\Omega}_{s1}\times\frac{\partial \bm{\xi}}{\partial t}+\bm{C}\cdot\bm{\xi}=\bm{a}_\text{ext}, \label{NS-eq}
\end{align}
where $\bm{\xi}$ is the Lagrangian displacement of fluid elements, and $\bm{C}$ is a self-adjoint operator, and $\bm{a}_\text{ext}$ the external driving, which will arise from gravitomagnetism in our case. We refer the interested reader to Ref.~\cite{1981A&A....94..126P} for more details.

In the phase-space mode expansion framework~\cite{Schenk+01}, $\bm{\xi}$ and its time derivative $\dot{\bm{\xi}}$ can be expanded as a summation of modes
\begin{align}
\left(\begin{matrix}
\bm{\xi} \\
\bm{\dot{\xi}}
\end{matrix}\right)=\sum_{lm}c_{lm}(t)
\left(\begin{matrix}
\bm{\xi}_{lm} \\
-i\omega_{lm}\bm{\xi}_{lm}
\end{matrix}\right), \label{mode-exp}
\end{align}
where the angular quantum numbers $l$ and $m$ are integers with $m=\pm l,\pm (l-1)\ldots0$. Here $\omega_{lm}$ is the co-rotating frame eigenfrequency of the mode. Each mode $\bm{\xi}_{lm}$ satisfies the following eigenvalue problem, 
\begin{equation}
    -\omega_{lm}^2\bm{\xi}_{lm}  -2i\omega_{lm} \bm{\Omega}_{s1}\times\bm{\xi}_{lm} +\mathbf{C}\cdot\bm{\xi}_{lm} =0\,,
\end{equation}
and modes with $(l,m)\neq (l',m')$ satisfy the following orthogonality condition,
\begin{align}
\left<\bm{\xi}_{lm},2i\bm{\Omega}_{s1}\times\bm{\xi}_{l^\prime m^\prime}\right>+(\omega_{lm}+\omega_{l^\prime m^\prime})\left<\bm{\xi}_{lm},\bm{\xi}_{l^\prime m^\prime}\right>=0,
\end{align}
with the inner product defined by
\begin{align}
\left<\bm{u},\bm{v}\right>=\int d^3x^\prime\rho\bm{u}^*\cdot\bm{v}, \label{inner_product}
\end{align}
where $\rho$ is the density profile of NS.

In presence of driving, the mode amplitudes satisfy 
\begin{align}
\dot{c}_{lm}(t)+i\omega_{lm} c_{lm}(t)=\frac{i}{b_{lm}}\left<\bm{\xi}_{lm},\bm{a}_\text{ext}\right>, \label{c-evo}
\end{align}
 where the coefficient $b_{lm}$ is given by
\begin{align}
b_{lm}=\left<\bm{\xi}_{lm},2\omega_{lm}\bm{\xi}_{lm}+2i\bm{\Omega}_{s1}\times\bm{\xi}_{lm}\right>. \label{blm}
\end{align}
To linear order in spin frequency, the $(l,m)$ $r$-mode has Lagrangian displacement vector field as  \begin{align}
\bm{\xi}_{lm}(r^\prime,\theta^\prime,\phi^\prime)=-\frac{if_{lm}(r^\prime)}{\sqrt{l(l+1)}}\bm{r}^\prime\times\nabla Y_{lm}(\theta^\prime,\phi^\prime), \label{xi}
\end{align}
where $Y_{lm}(\theta^\prime,\phi^\prime)$ are spherical harmonics, and (co-rotating frame) eigenfrequency given by~\cite{1981A&A....94..126P}
\begin{align}
\omega_{lm}=-\frac{2m\Omega_{s1}}{l(l+1)}. \label{mode-fre}
\end{align}
The radial profile function $f_{lm}(r^\prime)$ can only be determined if one includes the next-order correction of spin frequency. For barotropic stars, only the modes $|m|=l$ exist, whose coefficients $f_{lm}(r^\prime)$ are given by \cite{1981A&A....94..126P}
\begin{align}
f_{lm}(r^\prime)\propto r^{\prime l}. \label{f-baro}
\end{align}
The constant of proportionality can be  determined by the normalization condition
\begin{align}
\left<\bm{\xi}_{lm},\bm{\xi}_{lm}\right>=m_1R_1^2, \label{xi-norm}
\end{align}
where $R_1$ the radius of the star. As for other modes with $|m|\neq l$, $f_{lm}=0$. Henceforth, we restrict our discussions to barotropic stars, and only focus on the $(2,2)$ $r$-mode, since it gives the strongest effect \cite{Flanagan:2006sb}.

%

\subsection{Gravitomagnetic coupling: stellar part}
\label{EOM:stellar}

\begin{figure}[htb]
  \includegraphics[width=\columnwidth,height=7.2cm,clip=true]{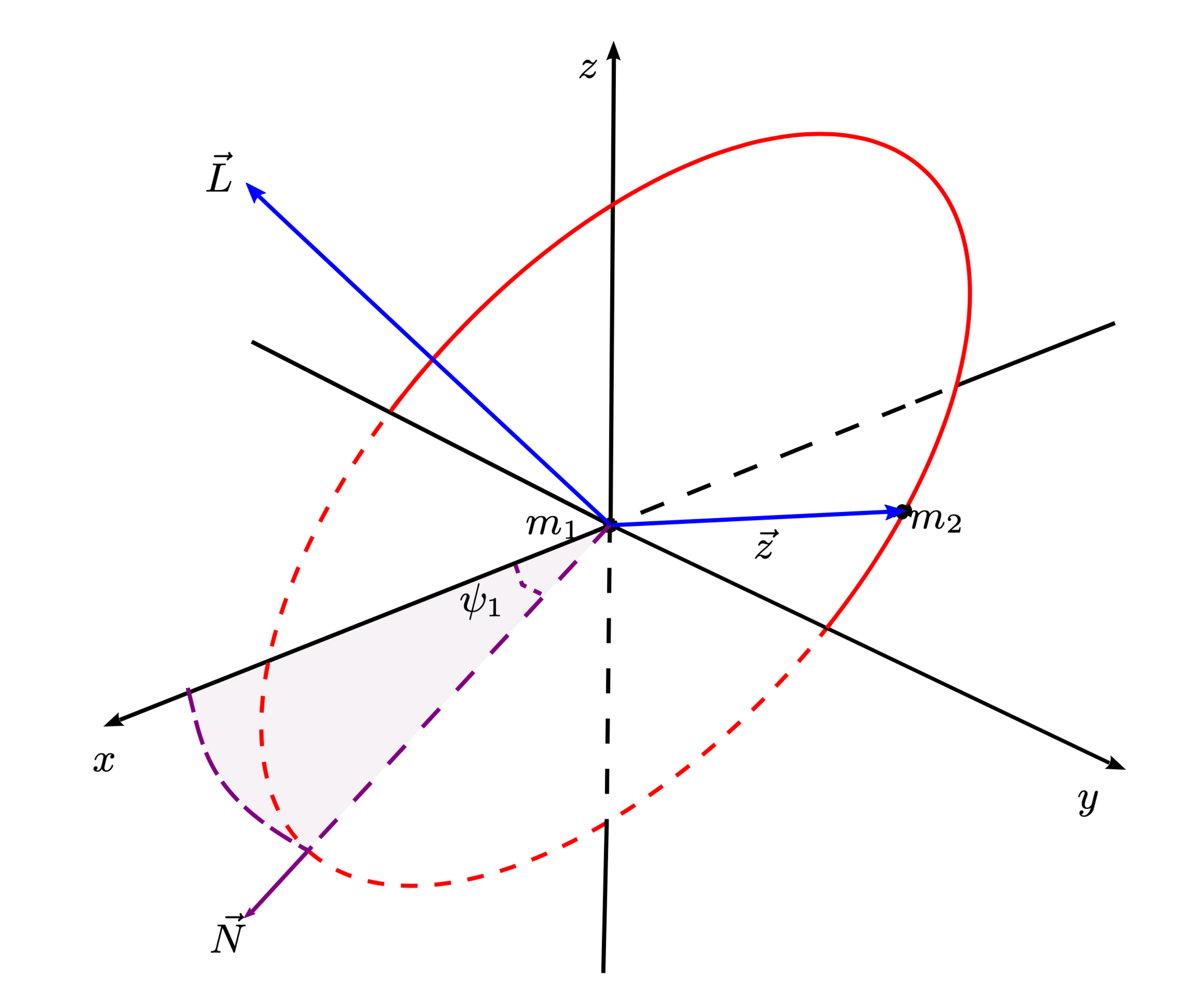} 
  \caption{A co-moving coordinate system $(x,y,z)$ that centers at $m_1$. The companion NS $m_2$ orbits around $m_1$, whose orbital plane intersects with the $x-z$ plane at $\vec{N}$, and intersects with the $x-y$ plane at the $y$-axis. The orbital phase $\phi(t)$ is the angle between $\vec{N}$ and the location of $m_2$, $\vec{z}$. The orbital angular momentum $\vec{L}$ is in the $x-z$ plane, with polar angle $\psi_1$. }
 \label{fig:orbit-illustration}
\end{figure}

To investigate the coupling between the $(2,2)$ $r$-mode and gravitomagnetic force, i.e., the RHS of Eq.\ (\ref{c-evo}), we place the  spinning NS, $m_1$, in a binary system. It is perturbed by the gravitomagnetic tidal field $\mathcal{B}_{ij}(t)$ exerted by the companion $m_2$. For convenience, we  introduce a non-rotating, co-moving coordinate system $(x,y,z)$ centered at $m_1$, as shown in Fig.\ \ref{fig:orbit-illustration}, with $z$ axis placed along the spin axis. The co-moving frame is related to the co-rotating frame (the primed frame) though a rotation about the $z$ ($z^\prime$) axis by $\Omega_{s1}t$.  (Note that the co-moving frame can be non-rotating because in this paper we treat the spin directions of NSs as fixed.) 

Using the same convention as Ref.\ \cite{Flanagan:2006sb}, we parameterize the location of $m_2$, $\bm{z}$, as  
\begin{align}
\bm{z}(t)=D(t)[\cos\psi_1\cos\phi(t),\sin\phi(t),\sin\psi_1\cos\phi(t)],
\end{align}
where $\psi_1$ is the inclination angle between the stellar spin vector and the orbital angular momentum; $D(t)$ is the separation between two NSs. With this parametrization, the orbital angular moment, $\vec{L}$, is in the $x-z$ plane. This plane and the orbital plane intersect at $\vec{N}$. The orbital phase of $m_2$, $\phi(t)$, is the angle between $\vec{z}$ and $\vec{N}$.

The expressions of $\mathcal{B}_{ij}(t)$ and $\bm{a}_\text{ext}$ are given in Sec.~V C of Ref.\ \cite{Flanagan:2006sb}. Plugging them and Eq.\ (\ref{xi}) into Eq.\ (\ref{c-evo}), we obtain the evolution equation for $c_{22}$
\begin{align}
\dot{c}_{22}+i\omega_{22}c_{22}&=\frac{i}{\omega_{22}}\sqrt{\frac{16\pi}{5}}\frac{I_{22}^rm_2R_1}{D^2}\dot{\phi}(\dot{\phi}-2\Omega_{s1}) \notag \\
&\times\sin\psi_1\cos^2\frac{\psi_1}{2}e^{2i\Omega_{s1}t-i\phi}, \label{c-evo-final}
\end{align}
where we have ignored the $\exp[+i\phi]$ part of driving force that does not give rise to resonance, since its effect is negligible. On the RHS of Eq.\ (\ref{c-evo-final}), 
we have used $I^r_{22}$ to denote the $(2,2)$ $r$-mode overlap.  For Newtonian NS, $I^r_{22}$ is given by 
\begin{align}
I^r_{22}=\sqrt{\frac{1}{m_1R_1^4}\int_0^{R_1}\rho r^6dr}. \label{Ir-def}
\end{align}
This is an analogue of tidal Love number for $f$-mode.

In Eq.\ (\ref{c-evo-final}), $\psi_1$ is treated as a constant, because for a typical BNS system, this angle changes on a timescale of $\sim 70$ s, which is much longer than the duration of resonance ($\sim 0.52$ s) and the time to merger ($\sim 4$ s); see Appendix \ref{app:precession} for more details.

From Eq.\ (\ref{c-evo-final}), we can obtain the resonant condition in terms of  orbital frequency $\dot{\phi}_r$
\begin{align}
\dot{\phi}_r=2\Omega_{s1}+\omega_{22}=\frac{4}{3}\Omega_{s1},  \label{reson-condi}
\end{align}
where we have used Eq.\ (\ref{mode-fre}) to replace the mode frequency by spin frequency. Note that the driving force is porportional to $\sim\sin\psi_1\cos^2\psi_1/2$,  therefore no $r$-mode DT takes place for $\psi_1=0,\pi$, when spins are aligned or anti-aligned with the orbital angular momentum.

\subsection{Gravitomagnetic coupling: orbital part}
\label{EOM:orbital}

Let us now use the standard osculating equations to study the orbital evolution. Assume that there is a perturbing force $\bm{F}$ on the Newtonian binary system, which takes the form of
\begin{align}
\frac{\bm{F}}{\mu}=\mathcal{S}\bm{\lambda},
\end{align}
where $\mu$ is the reduced mass, and $\bm{\lambda}$ is the unit azimuthal vector in the orbital plane. The evolution of orbital separation $D(t)$ is given by \cite{Poisson+Will+14}
\begin{align}
\frac{dD}{dt}=2\sqrt{\frac{D^3}{M}}\mathcal{S}. \label{rdot}
\end{align}
In our case, $\mathcal{S}$ contains two parts. The first one is the back-reaction force due to GW radiation, given by the averaged (over orbital timescale) Burke-Thorne dissipation term \cite{Poisson+Will+14}:
\begin{align}
\mathcal{S}_\text{GW}=-\frac{32}{5}\eta\left(\frac{M}{D}\right)^3\sqrt{\frac{M}{D^3}}.\label{sgw}
\end{align}
The other part is the stellar induced force. The current quadrupole moment $J_{ij}$ induced by the $r$-mode (to the linear order in $\bm{\xi}_{22}$) is given by\footnote{This is different from the $f$-mode dynamical tides, where the leading term is mass quadrupole moment.} \cite{Flanagan:2006sb}
\begin{align}
J_{ij}\sim I_{22}^rm_1R_1^3\omega_{22}e^{-2i\Omega_{s1}t}c_{22}+c.c., \label{current-quad-moment}
\end{align}
leading to an azimuthal acceleration $\mathcal{S}_\text{tide}$
\cite{Racine:2004hs}
\begin{align}
\mathcal{S}_\text{tide}=&\frac{8}{3}\sqrt{\frac{\pi}{5}}I_{22}^r\frac{\Omega_{s1}(\omega_{22}+2\Omega_{s1})}{D^3}MR_1^3\sin\psi_1\cos^2\frac{\psi_1}{2} \notag \\
&\times 2\Re(-ie^{i\phi-2i\Omega_st}c_{22}). \label{stide}
\end{align}
The total acceleration is given by  
\begin{equation} 
\mathcal{S}=\mathcal{S}_\text{tide}+\mathcal{S}_\text{GW}.
\end{equation}
Note that the current quadrupole moment also induces forces along the radial axis, and along the axis of orbital angular momentum. These will lead to the evolution in orbital eccentricity and semi-latus rectum. However, as argued in Ref.\ \cite{Flanagan:2006sb}, these effects are negligible,  and the orbit keeps quasicircular within the entire evolution process. We have also confirmed this numerically. Here we do not include those effects.

So far we have only considered the $r$-mode of $m_1$. The resonance for $m_2$ can be treated by a direct replacement $\left(1\leftrightarrow2\right)$. Since we only consider the $(2,2)$ $r$-mode in the two stars, henceforth we will drop the labels of $(l,m)$, and add a new subscript to refer to individual star. For example, $I^r_1$ and $\omega_1$ stand for the $r$-mode overlap and frequency of $m_1$, respectively.

Combining Eqs.\ (\ref{c-evo-final}), (\ref{rdot}), (\ref{sgw}) and (\ref{stide}), we finally arrive at a complete set of EOM for the binary system
\begin{subequations}
\begin{align}
\frac{d\phi}{dt}= &\sqrt{\frac{M}{D^3}}, \label{EOM-phidot}\\
\frac{dD}{dt}= & -\frac{64}{5}\eta\left(\frac{M}{D}\right)^3\nonumber\\
&+\frac{16}{3}\sqrt{\frac{\pi}{5}}I_{1}^r\frac{\Omega_{s1}(\omega_{1}+2\Omega_{s1})}{D^{3/2}}M^{1/2}R_1^3 \notag \\
&\qquad\; \times2 \sin\psi_1\cos^2\frac{\psi_1}{2}\Re(-ie^{i\phi-2i\Omega_{s1}t}c_{1}) \notag \\
&+\frac{16}{3}\sqrt{\frac{\pi}{5}}I_{2}^r\frac{\Omega_{s2}(\omega_{2}+2\Omega_{s2})}{D^{3/2}}M^{1/2}R_2^3 \nonumber \\
&\qquad\; \times 2\sin\psi_2\cos^2\frac{\psi_2}{2} \Re(-ie^{i\phi-2i\Omega_{s2}t}c_{2}), \\
\dot{c}_{1} =& - i\omega_{1}c_{1}+ \frac{i}{\omega_{1}}\sqrt{\frac{16\pi}{5}}\frac{I_{1}^rm_2R_1}{D^2}\dot{\phi}(\dot{\phi}-2\Omega_{s1}) \notag \\
&\qquad\qquad \times\sin\psi_1\cos^2\frac{\psi_1}{2}e^{2i\Omega_{s1}t-i\phi},  \\
\dot{c}_{2}= & -i\omega_{2}c_{2}+\frac{i}{\omega_{2}}\sqrt{\frac{16\pi}{5}}\frac{I_{2}^rm_1R_2}{D^2}\dot{\phi}(\dot{\phi}-2\Omega_{s2}) \notag \\
&\qquad \qquad \times\sin\psi_2\cos^2\frac{\psi_2}{2}e^{2i\Omega_{s2}t-i\phi}. \label{EOM-cdot}
\end{align}
\label{EOM-no-norm}%
\end{subequations}

By simultaneously integrating stellar and orbital parts, we can obtain the evolution of the system, and the gravitational waves it emits.

\section{Dynamics of r-mode excitation} 
\label{sec:dyna-r-mode}
In this section, we discuss features of $r$-mode excitation and its back action onto orbital motion.  In Sec.\ \ref{sec:GW-EOM}, we first cast equations of motion Eqs.~(\ref{EOM-no-norm}) into forms that only depend on independent parameters, and provide the appropriate initial conditions. In Sec.\ \ref{sec:EOM-orbit}, 
we analytically describe orbital motion across tidal resonance, and compare with results in FR07. 
In Sec.\ \ref{sec:EOM-tide}, we 
provide analytical formulas for tidal excitation. 

\subsection{Equation of motion}
\label{sec:GW-EOM}
Although Eqs.\ (\ref{EOM-no-norm}) are complete, and can be integrated straightforwardly, it is difficult to see the actual dependence of solutions on specific parameters. For example, it is well known that the inspiraling process of a binary system is controlled by the chirp mass $\mc=M\eta^{3/5}$ at the leading order, since the direct observable is GW phase as a function of frequency. However both the total mass $M$ and the symmetric mass ratio $\eta$ appears in equations. The degeneracy between them is hidden. This problem can be fixed by rescaling parameters in the following way:
\begin{align}
&c_1=\bar{c}_1\frac{I_{1}^rm_2R_1}{\omega_1M^{2/3}}\sin\psi_1\cos^2\frac{\psi_1}{2}, \\
&c_2=\bar{c}_2\frac{I_{2}^rm_1R_2}{\omega_2M^{2/3}}\sin\psi_2\cos^2\frac{\psi_2}{2},
\end{align}
and replace separation $D$ by orbital frequency $\dot{\phi}$.
In terms of these new parameters, the equations of motion become:
\begin{subequations}
\begin{align}
&\frac{d\dot{\phi}}{dt}=\frac{96}{5}\mathcal{M}^{5/3}\dot{\phi}^{11/3}+32\sqrt{\frac{\pi}{5}}\mathcal{I}_1\Omega_{s1}\dot{\phi}^{8/3} \Re(-ie^{i\phi-2i\Omega_{s1}t}\bar{c}_{1})\notag \\
&+32\sqrt{\frac{\pi}{5}}\mathcal{I}_2\Omega_{s2}\dot{\phi}^{8/3}  \Re(-ie^{i\phi-2i\Omega_{s2}t}\bar{c}_{2}), \\
&\dot{\bar{c}}_{1}-\frac{2}{3}i\Omega_{s1}\bar{c}_{1}=i\sqrt{\frac{16\pi}{5}}(\dot{\phi}-2\Omega_{s1})\dot{\phi}^{7/3}e^{2i\Omega_{s1}t-i\phi},  \\
&\dot{\bar{c}}_{2}-\frac{2}{3}i\Omega_{s2}\bar{c}_{2}=i\sqrt{\frac{16\pi}{5}}(\dot{\phi}-2\Omega_{s2}) \dot{\phi}^{7/3}e^{2i\Omega_{s2}t-i\phi}.
\end{align}
\label{EOM-rescale}%
\end{subequations}
Here we have defined the $r$-mode coupling coefficient for each individual NS as :
\begin{align}
\mathcal{I}_i=\bar{I}_{i}^{r2}m_i^4\sin^2\psi_i\cos^4\frac{\psi_i}{2}\left(1-\frac{m_i}{M}\right), \label{mathI}
\end{align}
with the normalized $r$-mode overlap $\bar{I}^r$ defined by
\begin{align}
\bar{I}^r=\sqrt{\frac{1}{m_{\rm NS}^5}\int_0^{R_{\rm NS}}\rho r^6dr}. \label{def-Irbar}
\end{align}
where $m_{\rm NS}$ and $R_{\rm NS}$ are the mass and the radius of the NS.
In Eqs.~\eqref{EOM-rescale},  $\mathcal{I}_i$ characterizes the tidal backreaction of NS $i$ onto the orbit. Note that $\mathcal{I}$ (the effect of $r$-mode DT) vanishes when $\psi=0,\pi$, and its maximized when $\psi=\pi/3$.

Initial conditions for Eqs.~\eqref{EOM-rescale} are chosen such that the orbit is quasicircular and tides are in equilibrium:
\begin{align}
&\dot{\phi}^{(0)}=2\pi F_0,\quad\phi^{(0)}=0,  \notag \\
&\bar{c}_1^{(0)}=(2\pi F_0)^{4/3}\sqrt{\frac{16\pi}{5}}\frac{(2\pi F_0)^2-4\Omega_{s1}\pi F_0}{\frac{4}{3}\Omega_{s1}-2\pi F_0}, \notag \\
&\bar{c}_2^{(0)}=(2\pi F_0)^{4/3}\sqrt{\frac{16\pi}{5}}\frac{(2\pi F_0)^2-4\Omega_{s2}\pi F_0}{\frac{4}{3}\Omega_{s2}-2\pi F_0}. \notag 
\end{align}
Written in this way, the EOM (as well as initial conditions) depend on five parameters of the binary system: $\mc$, $\mathcal{I}_i$, $\Omega_{si}$, with $i=1,2$.

In the rest of this section, we shall numerically integrate the EOM and discuss features of solutions. For example, we choose a $(1.4,1.4)M_\odot$ BNS system, with one of the stars spinning at frequency $\Omega_{s1}=2\pi\times30$Hz, and the other one non-spinning. We choose an inclination angle $\psi_1$ of $\pi/3$. The EoS is polytopic with $\Gamma=2$, and the radii of both stars are chosen to be $R_{\rm NS}=13$km. This gives the  $r$-mode overlap of $I^r_1=0.185$. There is only one resonance during the inspiral whose resonant orbital frequency is given by 
\begin{align}
\dot{\phi}(t_r)=\frac{4}{3}\Omega_s=2\pi\times40 ~\text{Hz}.
\end{align}
We set an initial orbital frequency $F_0$ as 4.5 Hz, and evolve the system up to the contact separation of $r^\text{stop}=2R_{\rm NS}$.

\subsection{Orbital evolution}
\label{sec:EOM-orbit}

As argued in FR07, the pre- and post-resonance orbit can well approximated  as two point-particle (PP) orbits, with orbital phase given by
\begin{align}
\phi^{(\text{ana})}(t)=\begin{cases}
    \phi^{(\text{pre})}_\text{PP}(t),& \text{if } t< t_r\\
    \phi^{(\text{post})}_\text{PP}(t),              & \text{if } t> t_r
\end{cases},
\end{align}
Both $\phi^{(\text{post},\text{pre})}(t)$ evolve without being affected by tides, following the {\it same equation}:
\begin{align}
 \frac{d}{dt}\phidot^{(\text{post},\text{pre})}=\frac{96}{5}\mc^{5/3}\left[\phidot^{(\text{post},\text{pre})}\right]^{11/3}. \label{Ddot-pp}
\end{align}
The initial condition for pre-resonance orbit is simply (same as the one we used for numerical integration)
\begin{align}
\phi^{\rm (pre)}(t=0)=0\,,\quad \phidot^{(\text{pre})}(t=0)=2\pi F_0,
\end{align}
the resonance orbital frequency is given by 
\begin{equation}
\dot{\phi}^{\rm (pre)}_{\rm PP}(t_r)=\frac{4}{3}\Omega_{s1}\,.   
\end{equation}
At resonance, $\phi_{\rm PP}^{(\text{post})}(t_r)$ is related to $\phi_{\rm PP}^{(\text{pre})}(t_r)$ with continuity in value and a jump in derivative, and one can write
\begin{align}
&\phi_{\rm PP}^{\rm post} (t_r) = \phi_{\rm PP}^{\rm pre}(t_r),\label{phase-continue}\\
      &\Delta\dot{\phi}_{\rm tid} = \dot{\phi}_{\rm PP}^{\rm (post)} (t_r)- \dot{\phi}_{\rm PP}^{\rm (pre)}(t_r) = \ddot{\phi}_{\rm PP}^{\rm (pre)}(t_r) \delta t_r. \label{orbit-jump}
\end{align}
In fact, Eqs.~(\ref{phase-continue}) and (\ref{orbit-jump}) can be grouped into a compact form [cf. Eq.\ (1.6) in FR07]
\begin{align}
    \phi_{\rm PP}^{\rm post} (t) = \phi_{\rm PP}^{\rm pre}(t+\delta t_r) -\delta\phi_r, 
\end{align}
with [cf. Eq.\ (5.37) in FR07]
\begin{subequations}
\begin{align}
&\delta t_r=-\frac{5\pi^2}{192}\left(\frac{4}{3}\right)^{-1/3}\frac{\Omega_{s1}^{-1/3}}{\mc^{10/3}}\mathcal{I}_1 \notag \\
&=-2.64\times10^{-4}~\rm s\left(\frac{\Omega_{s1}}{2\pi\times30 ~\rm Hz}\right)^{-1/3}\left(\frac{\sin^2\psi_1\cos^4\psi_1/2}{0.422}\right) \notag \\
&\times\left(\frac{m_1}{1.4~M_\odot}\right)^3\left(\frac{M}{2.8~M_\odot}\right)^{-7/3}\left(\frac{\eta}{0.25}\right)^{-1}\left(\frac{\bar{I}^r_1}{7.32}\right)^2 . \label{deltat-flanagan}\\
&\delta\phi_r=\dot{\phi}(t_r)\delta t_r=-\frac{5\pi^2}{192}\left(\frac{4}{3}\right)^{2/3}\frac{\Omega_{s1}^{2/3}}{\mc^{10/3}}\mathcal{I}_1 \notag \\
&=-6.65\times10^{-2}~\rm rad \left(\frac{\Omega_{s1}}{2\pi\times30 ~\rm Hz}\right)^{2/3}\left(\frac{\sin^2\psi_1\cos^4\psi_1/2}{0.422}\right) \notag \\
&\times\left(\frac{m_1}{1.4~M_\odot}\right)^3\left(\frac{M}{2.8~M_\odot}\right)^{-7/3}\left(\frac{\eta}{0.25}\right)^{-1}\left(\frac{\bar{I}^r_1}{7.32}\right)^2, \label{deltaphi-flanagan}
\end{align}
\label{delta-flanagan}%
\end{subequations}
Here $\delta t_r$ and $\delta \phi_r$ are effective orbital time and phase shifts between the pre- and post-resonance PP orbits.

\begin{figure}[htb]
        \includegraphics[width=\columnwidth,height=6.5cm,clip=true]{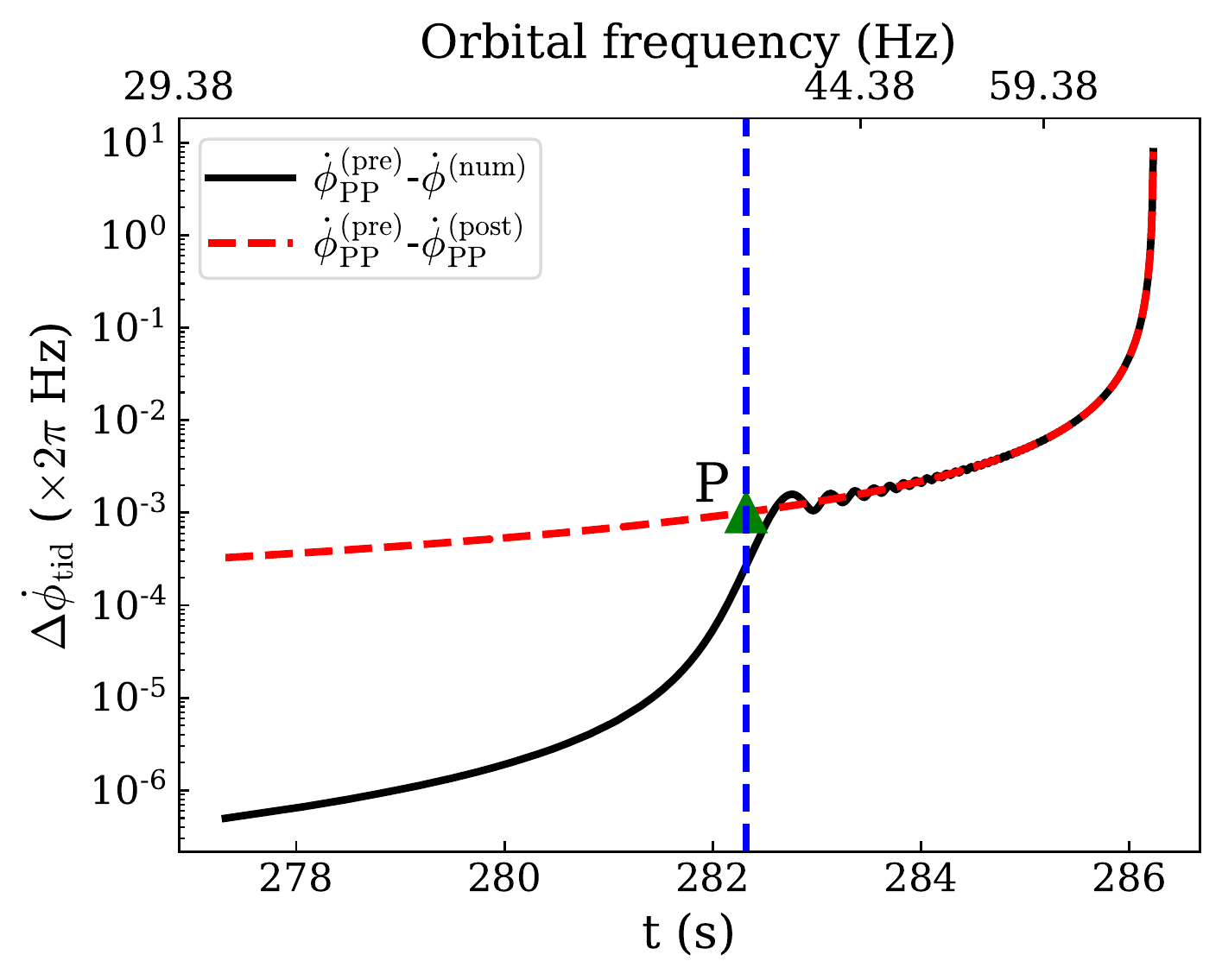} 
  \caption{The orbital dynamics near the $r$-mode resonance. The BNS system has individual masses $(1.4,1.4)M_\odot$. One of them spins at $\Omega_{s1}=30$Hz with the inclination angle $\psi_1=\pi/3$, whereas the other one is non-spinning. The EoS is $\Gamma=2$ polytrope with radius $R_{\rm NS}=13$km. The vertical dashed line represents for the time of resonance $t_r$. The pre-resonance PP orbit $\phidot^{(\text{pre})}_\text{PP}(t)$, which has the same initial condition as the numerical one, is compared with the numerical integration, as shown by black curve. Before the resonance, $|\Delta\phidot_{\rm tid}|$ is below $\sim 2\pi\times 10^{-4}~\rm Hz$, which is mainly caused by the adiabatic $r$-mode. After the resonance, there are some oscillatory features, which are from the $r$-mode oscillation. Eq.\ (\ref{orbit-jump}) gives the new orbital frequency after the $r$-mode is excited, which is labeled by ``P'' in the figure. Using ``P'' as a new initial condition, we obtain the other PP orbit $\phidot^{(\text{post})}_\text{PP}(t)$. The difference between $\phidot^{(\text{pre})}_\text{PP}(t)$ and $\phidot^{(\text{post})}_\text{PP}(t)$ is shown as red dashed line, which tracks the averaged numerical result very well.}
 \label{fig:orbit}
\end{figure}

Note that $\Delta\dot{\phi}_{\rm tid}<0$, and $\phidot^{(\text{post})}_\text{PP}(t_r)<\phidot^{(\text{pre})}_\text{PP}(t_r)$, the resonance transfers the energy from star to orbit, which is different from the case for $f$-mode (cf. Fig. 8 in Ref.\ \cite{Ma:2020rak}). This can be understood by the relation \cite{Schenk+01,Flanagan:2006sb}
\begin{align}
&\Delta E_\text{star}=\frac{E_\text{mode}}{\omega_{lm}}(\omega_{lm}+m\Omega_{s1}) \notag \\
&=-9.9\times10^{46}~{\rm erg}\left(\frac{\Omega_{s1}}{2\pi\times30 ~\rm Hz}\right)^{3}\left(\frac{\bar{I}^r_1}{7.32}\right)^2\left(\frac{m_1}{1.4M_\odot}\right)^4 \notag \\
&\times \left(\frac{\sin^2\psi_1\cos^4\psi_1/2}{0.422}\right) \left(\frac{1-\Xi}{0.5}\right),
\end{align}
where $E_\text{mode}=b_{lm}\omega_{lm}|c_{lm}|^2>0$, and $\Delta E_\text{star}$ is the change of stellar energy. Therefore, for any mode that is subjected to the Chandrasekhar-Friedman-Schutz instability condition \cite{1970PhRvL..24..611C,1978ApJ...222..281F}
\begin{align}
\omega_{lm}(\omega_{lm}+m\Omega_{s1})<0,
\end{align}
the tidal excitation leads to the loss of stellar energy, and the decrease of spin frequency.\footnote{For a given baryon number and total angular momentum, uniform angular velocity is the minimum-energy state \cite{BOYER1966504,1967ApJ...147..317H}.} For completeness, the angular momentum transfer can be written as
\begin{align}
&\Delta L_\text{star}=\frac{E_\text{mode}}{\omega_{lm}} \notag \\
&=-3.9\times10^{44}~{\rm erg\cdot s}\left(\frac{\Omega_{s1}}{2\pi\times30 ~\rm Hz}\right)^{2}\left(\frac{\bar{I}^r_1}{7.32}\right)^2\left(\frac{m_1}{1.4M_\odot}\right)^4 \notag \\
&\times \left(\frac{\sin^2\psi_1\cos^4\psi_1/2}{0.422}\right) \left(\frac{1-\Xi}{0.5}\right),
\end{align}
and the change of spin frequency is given by
\begin{align}
&\Delta\Omega_{s}=-2\pi\times3.6\times10^{-2}~{\rm  Hz}\left(\frac{\Omega_{s1}}{2\pi\times30 ~\rm Hz}\right)^{2}\left(\frac{\bar{I}^r_1}{7.32}\right)^2 \notag \\
&\times\left(\frac{m_1}{1.4M_\odot}\right)^4 \left(\frac{\sin^2\psi_1\cos^4\psi_1/2}{0.422}\right) \left(\frac{1-\Xi}{0.5}\right)\left(\frac{\bar{I}}{14.6}\right)^{-1},
\end{align}
which is negligible in our study.

Using the binary system above, we first investigate how the numerical integration of the EOM deviates from $\phidot^{(\text{pre})}_\text{PP}(t)$. In Fig.\ \ref{fig:orbit}, we plot $\phidot^{(\text{pre})}_\text{PP}(t)-\phidot^{(\text{num})}(t)$ in the black line. The vertical, blue-dashed line stands for the time of resonance. In the pre-resonance regime $t<t_r$, $|\Delta\phidot_{\rm tid}|<2\pi\times10^{-4} ~\rm Hz$, and is mainly contributed by the adiabatic tide. During the resonance, $|\Delta\phidot_{\rm tid}|$ quickly grows to $2\pi\times10^{-3}~\rm Hz$. Then in the post-resonance regime, there is a small oscillation on the top of major deviation, which is caused by the $r$-mode oscillation. We also saw this feature in the case of $f$-mode \cite{Ma:2020rak}. Note that $\phidot^{(\text{num})}\sim2\pi\times10^2-2\pi\times10^3 ~\rm Hz$ for the time interval we present, the deviation caused by the resonance is extremely small.

Eqs.\ (\ref{delta-flanagan}) show that typical values for $\delta\phi_r$ and $\delta t_r$ are $-6.65\times10^{-2}$ rad and $-2.64\times10^{-4}$ s, respectively. The induced change in orbital frequency $|\Delta\phidot_{\rm tid}(t_r)|$ is $2\pi\times 10^{-3}~\rm Hz$, which corresponds to ``P'' in Fig.\ \ref{fig:orbit}. With ``P'' as a new initial condition, we obtain $\phidot^{(\text{post})}_\text{PP}(t)$ by solving Eq.\ (\ref{Ddot-pp}). In Fig.\ \ref{fig:orbit}, we show $\phidot^{(\text{pre})}_\text{PP}(t)-\phidot^{(\text{post})}_\text{PP}(t)$ as red dashed line. We can see $\phidot^{(\text{post})}_\text{PP}(t)$ tracks the averaged $\phidot^{(\text{num})}(t)$ well in the post-resonance regime.

\subsection{Tidal evolution}
\label{sec:EOM-tide}
Let us move on to the stellar part. Following the procedure in Sec. III of Ref.\ \cite{Ma:2020rak}, we define two real-valued quadratures,  $A$ and $B$, from the $r$-mode amplitude,
\begin{align}
A+iB=\bar{c}_1e^{i\phi-im\Omega_{s1}t} \quad {\rm and} \quad m=2,
\end{align}
with $A$ determining the radial force, and $B$ the torque back-reacting onto the orbit, respectively. Here $m=2$ since we are focusing on the $(l=2,m=2)$ mode. Using integration by parts, we obtain analytic expressions of $A$ and $B$ as 
\begin{subequations}
\begin{align}
&A=-\dot{\phi}^{7/3}\sqrt{\frac{16\pi}{5}} \frac{(2\Omega_{s1}-\dot{\phi})}{-\dot{\phi}+\frac{4}{3}\Omega_{s1}}\notag \\
&+\frac{\dot{\phi}_r^{10/3}}{\sqrt{\omd}}\sqrt{\frac{4\pi}{5}}\left[\sqrt{\pi}FC\left(\frac{\hatt}{\sqrt{\pi}}\right)\sin\Theta+\sqrt{\frac{\pi}{2}}\sin\left(\Theta-\frac{\pi}{4}\right)\right.\notag \\
&-\sqrt{\pi}FS\left(\frac{\hatt}{\sqrt{\pi}}\right)\cos\Theta  \left.-\frac{\cos(\Theta-\frac{1}{2}\hatt^2)}{\hatt}\right], \label{A}\\
&B=-\frac{\dot{\phi}_r^{10/3}}{\sqrt{\omd}}\sqrt{\frac{4\pi}{5}}\left[\sqrt{\pi}FC\left(\frac{\hatt}{\sqrt{\pi}}\right)\cos\Theta \right.\notag \\
&+\sqrt{\frac{\pi}{2}}\cos\left(\Theta-\frac{\pi}{4}\right)\left.+\sqrt{\pi}FS\left(\frac{\hatt}{\sqrt{\pi}}\right)\sin\Theta+\frac{\sin(\Theta-\frac{1}{2}\hatt^2)}{\hatt}\right]. \label{B}
\end{align}
\label{AB-ana}%
\end{subequations}
where [Eq.~(\ref{Ddot-pp})]
\begin{align}
&\frac{1}{\omd^{1/2}}=\left[\frac{96}{5}\mc^{5/3}\left(\frac{4}{3}\Omega_{s1}\right)^{11/3}\right]^{-1/2}\notag \\
&=0.2~{\rm s}\left(\frac{\mc}{1.22~M_\odot}\right)^{-5/6}\left(\frac{\Omega_{s1}}{2\pi\times30~\rm Hz}\right)^{-11/6}.
\end{align}
The phases of $A$ and $B$ in the  post-resonance regime is controlled by two quantities,
\begin{subequations}
\begin{align}
& \Theta=\chi_r-\omega_1t+\phi-2\Omega_{s1}t,\\
&\hatt=\sqrt{\omd}(t-t_r), \label{ttilde}
\end{align}
\end{subequations}
with the constant defined by $\chi_r=\omega_1t_r-\phi_r+2\Omega_{s1}t_r$, which results in $\Theta_r=0$ on resonance. Note that $1/\sqrt{\omd}\sim 0.2~{\rm s}$ in Eq.~(\ref{ttilde}) is the duration of tidal excitation, which is much longer than the effective orbital time shift $\delta t_r\sim10^{-4}~{\rm s}$ induced by tide [Eq.~(\ref{deltat-flanagan})].

\begin{figure}[htb]
        \includegraphics[width=\columnwidth,height=6.5cm,clip=true]{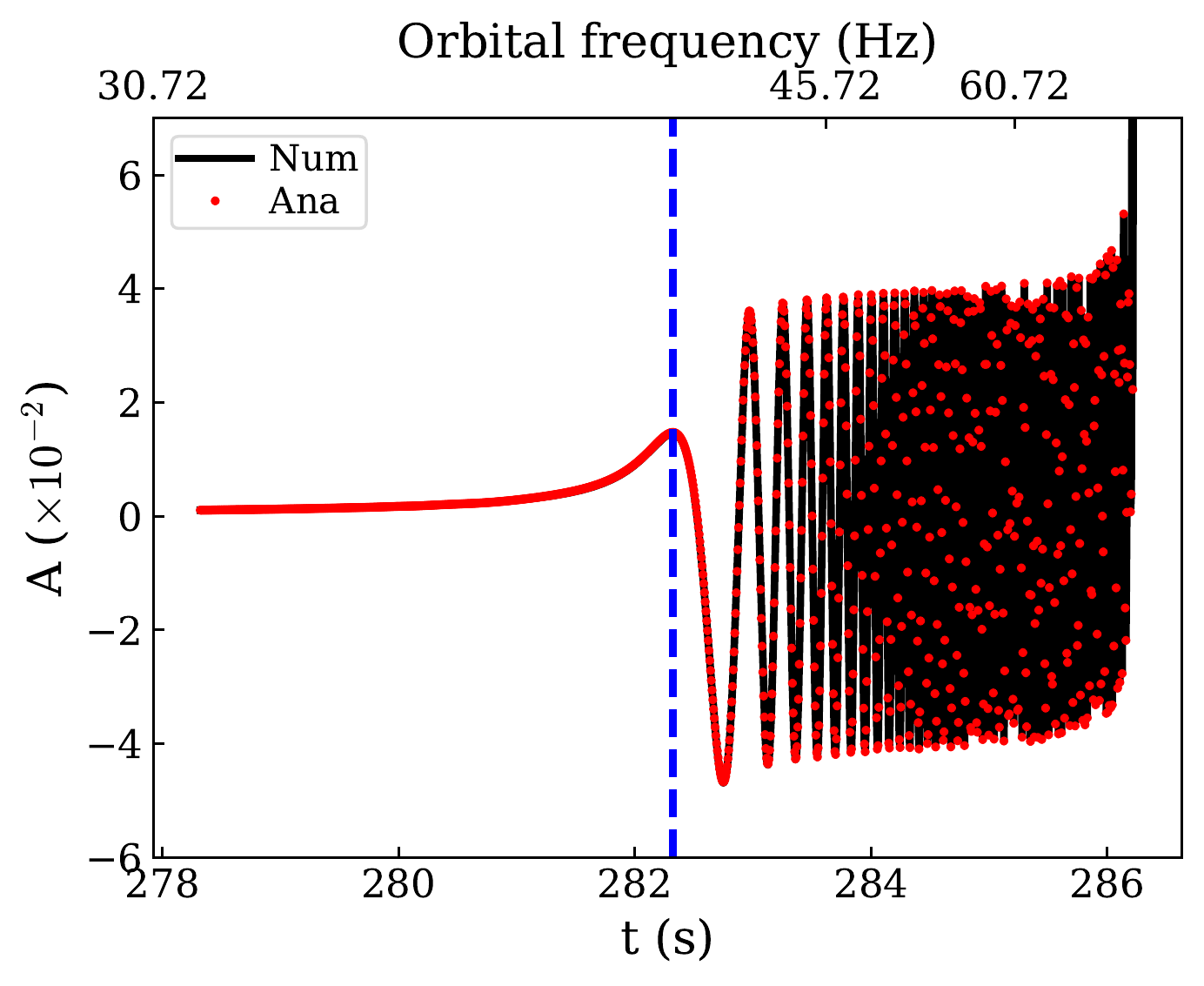} \\
        \includegraphics[width=\columnwidth,height=6.5cm,clip=true]{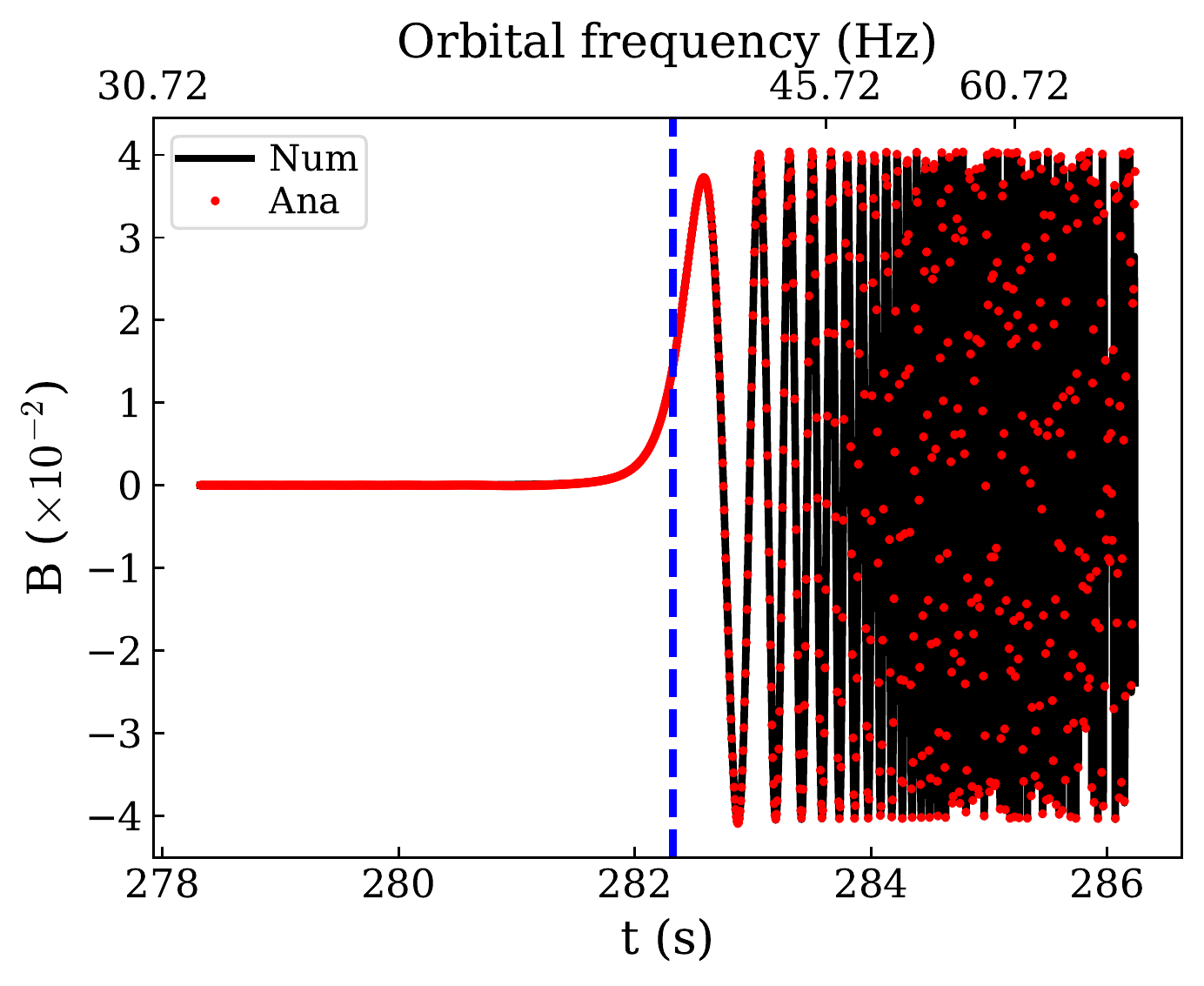}
  \caption{Time evolution of $r$-mode amplitudes, $A$ (upper panel) and $B$ (lower panel). Black curves are from the numerical integration of Eqs.\ (\ref{EOM-rescale}), and red dots are from our analytic approximations in Eqs.\ (\ref{AB-ana}). The vertical dashed line is the time of resonance from numerical simulation. Our analytic results agree with numerical ones to high accuracies. Unlike $f$-mode, the variable $A$ diverges as two NSs become close to each other, this is caused by differences in adiabatic tide. [see first line of Eq.\ (\ref{A})].}
 \label{fig:AB}
\end{figure}

In Fig.\ \ref{fig:AB}, we compare Eqs.\ (\ref{AB-ana}) with numerical integration, for the same binary system mentioned above. We can see that our formulas are accurate in all regimes: adiabatic, resonance and post-resonance. The evolutions of $A$ and $B$ are similar to those of $f$-mode \cite{Ma:2020rak}, except the fact that $A$ increases toward infinity as the two stars become close to each other (recall that the amplitude of $A$ for $f$-mode remains constant after the resonance.) This can be easily understood from Eq.\ (\ref{A}). The first line of Eq.\ (\ref{A}), i.e, the adiabatic part, diverges as $\dot{\phi}^{7/3}$ when $\dot{\phi}\to\infty$. On the other hand, the term remains $\sim1$ for $f$-mode, which stays constant as two stars contact. In spite of the diverging feature of $r$-mode in the late time evolution, our numerical calculation shows that it does not lead to any detectable effect (for 3G detectors).

\section{Rossby-mode Overlaps for Different Equations of State: A New universal relation} 
\label{sec:universal}
\begin{figure}[htb]
        \includegraphics[width=\columnwidth,height=6.5cm,clip=true]{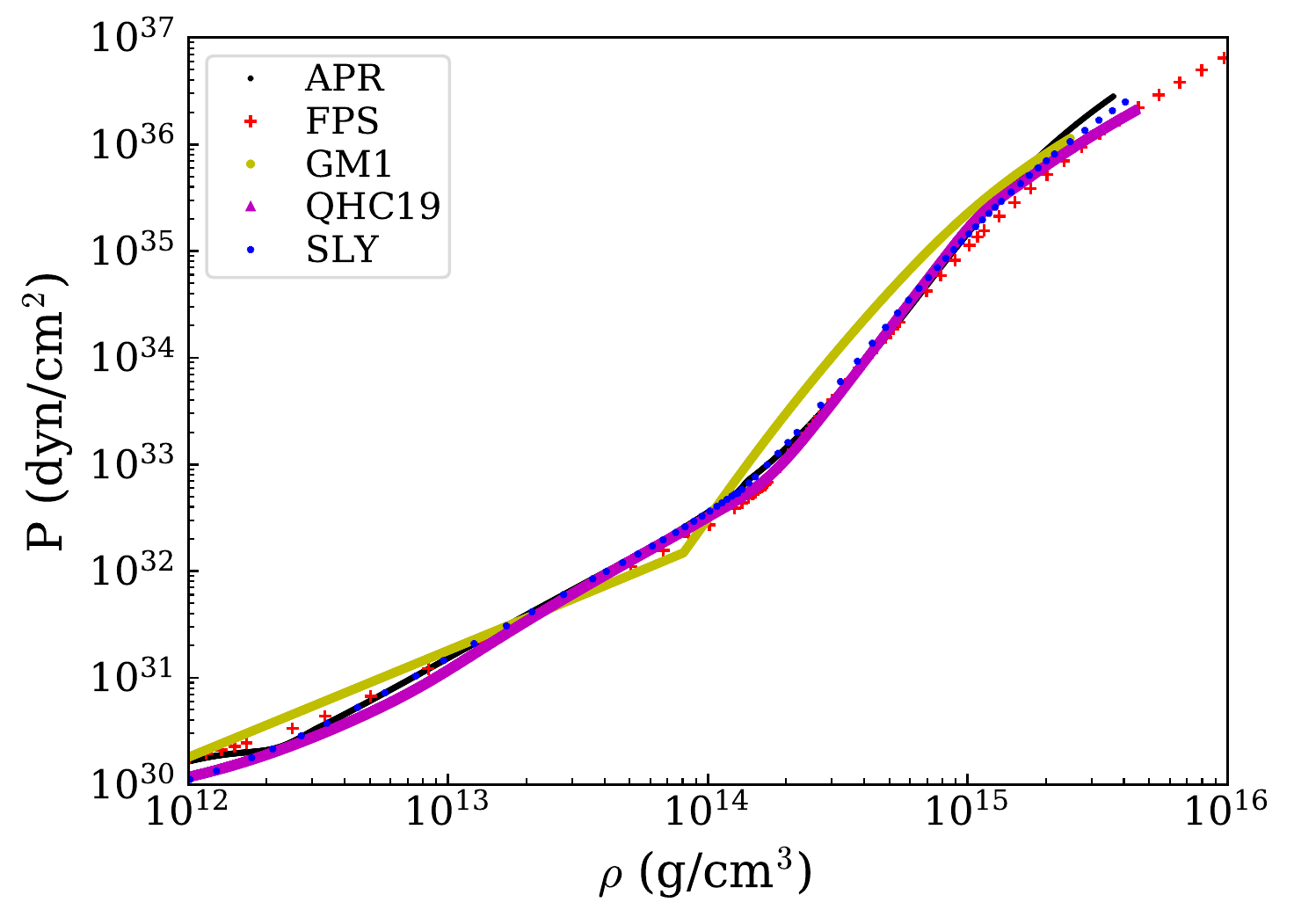}
  \caption{Several EoS for NSs used in this paper.}
 \label{fig:EOS}
\end{figure}

\begin{figure}[htb]
        \includegraphics[width=\columnwidth,height=6.1cm,clip=true]{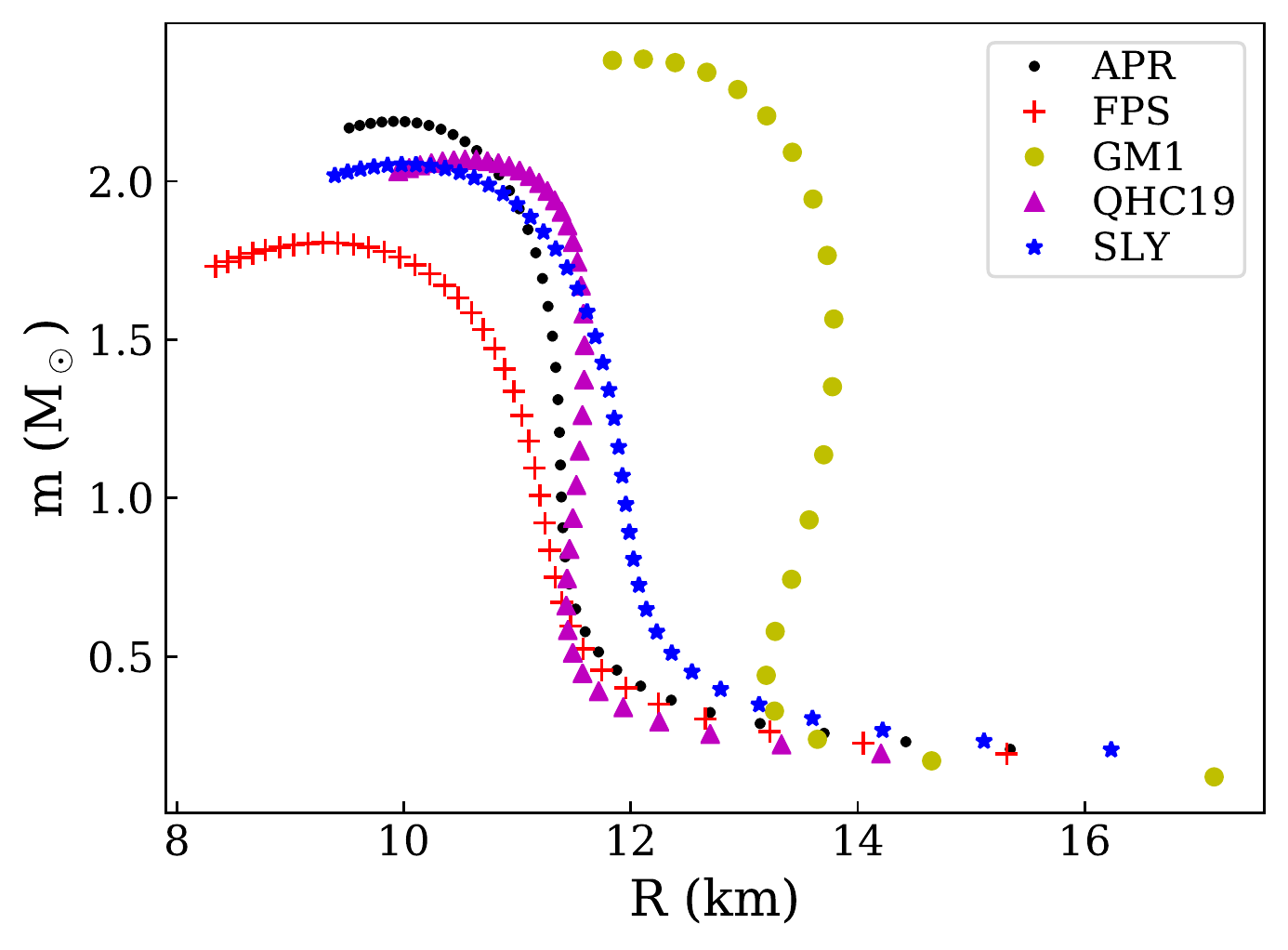}
  \caption{NS mass-radius relation with different EoS.}
 \label{fig:M-R}
\end{figure}

In the last section, we have seen that the effect of $r$-mode enters into the EOM through the normalized overlap $\bar{I}^r$. Here we shall identify a new universal relation between $\bar{I}^r$ and the tidal Love number $\lambda_f$. We consider five realistic EoS for cold NSs: APR \cite{Akmal:1998cf}, FPS \cite{pandharipande1989hot}, GM1 \cite{Glendenning:1991es,Douchin:2001sv}, QHC19 \cite{Togashi:2017mjp,Baym:2017whm,Baym:2019iky}, SLY \cite{Douchin:2001sv}, which are shown in Fig.\ \ref{fig:EOS}. Among them, the data of FPS are from Ref.\ \cite{Haensel:2004nu} and the rest of them are obtained from a EoS database CompOSE \cite{compose}. For comparison, we also include a polytropic EoS with $P\propto \rho^{\Gamma}$ and $\Gamma=2$. The mass-radius relations from those EoS are shown in Fig.\ \ref{fig:M-R}.

In Sec.\ \ref{sec:uni-Ir}, we first calculate $\bar{I}^r$ by solving Tolman–Oppenheimer–Volkoff (TOV) equations. Then in Sec.\ \ref{sec:uni-rel}, we explore the universal relation between $I^r$ and $\lambda_f$.

\subsection{The calculation of $I^r$}
\label{sec:uni-Ir}
We calculate $I^r$ at the zeroth order, i.e., using an unperturbed NS with spherical symmetry. Its metric is given by
\begin{align}
ds^2=-e^\nu dt^2+e^\lambda dr^2+r^2(d\theta^2+\sin^2\theta d\phi^2), \label{metric-equ}
\end{align}
we can obtain the density profile $\rho(r)$ by solving the TOV equations (see Appendix \ref{app:TOV} for more details). We then perform a numerical integration to get $I^r$ based on its definition in Eq.\ (\ref{Ir-def}). $\bar{I}^r$ is related to $I^r$ by
\begin{align}
I^r=\bar{I}^r\mathcal{C}^2,
\end{align}
where $\mathcal{C}$ is the compactness of the NS
\begin{align}
\mathcal{C}=\frac{m_{\rm NS}}{R_{\rm NS}}.
\end{align}

%
%
%

%

\subsection{Universal relations}
\label{sec:uni-rel}
\begin{figure*}[htb]
        \includegraphics[width=\columnwidth,height=7.2cm,clip=true]{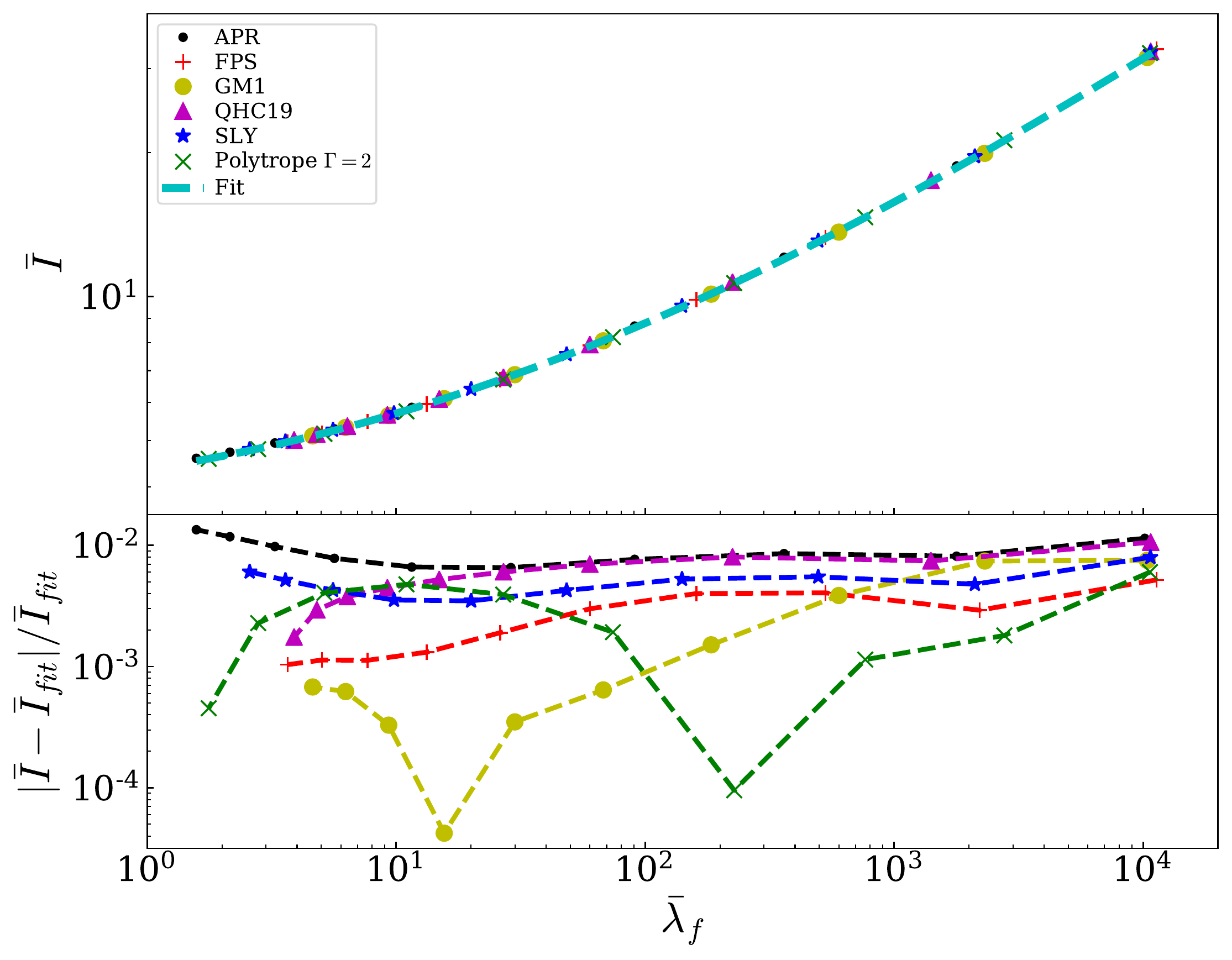}
         \includegraphics[width=\columnwidth,height=7.2cm,clip=true]{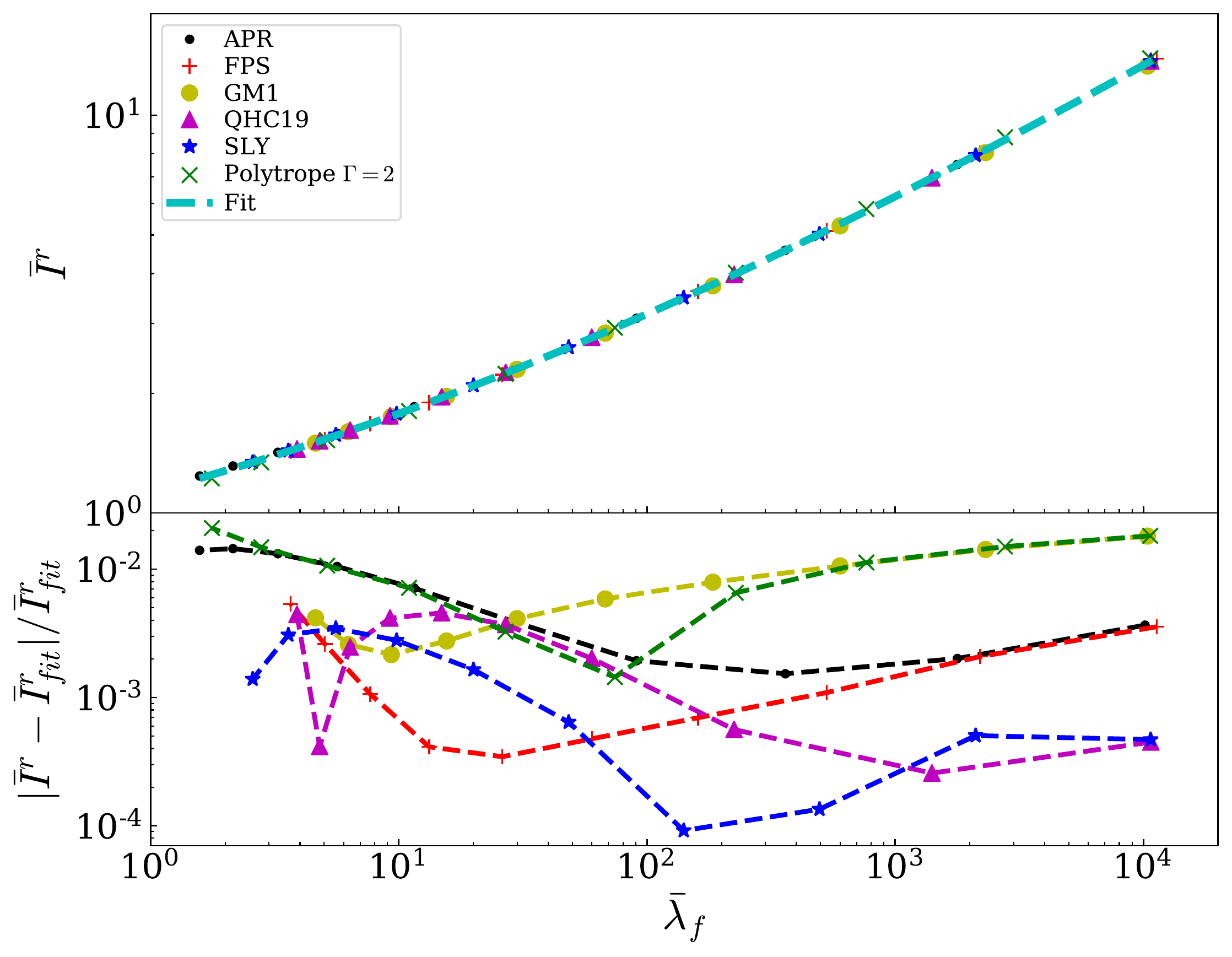}
  \caption{The I-Love and $\bar{I}^r$-Love universal relations for several EoS, as well as the fitting formulae in Eq.\ (\ref{uni-fit}). The bottom two plots are fractional errors between true values and fitted results; errors of both relations are within $10^{-2}$ for $\bar{\lambda}_f$ ranging from $\mathcal{O}(1)$ to $\mathcal{O}(10^4)$. }
 \label{fig:uni-I}
\end{figure*}

To establish the universal relation between the tidal Love number and the moment of inertia, Yagi \etal \cite{Yagi:2013awa,2013Sci...341..365Y} normalized the tidal Love number $\lambda_f$ and the moment of inertia $I$ as: $\bar{\lambda}_f=\lambda_f/m_{\rm NS}^5$ and $\bar{I}=I/m_{\rm NS}^3$ (The calculation of $\bar{\lambda}_f$ is summarized in Appendix \ref{app:uni-love}.). Universal relation between $\lambda_f$ and $\bar I$ are shown in the left panel of Fig.~\ref{fig:uni-I}. 


Similarly, we plot $\bar{I}^r$ as functions of $\bar{\lambda}_f$ for various EoS in the right panel of Fig.\ \ref{fig:uni-I}. We can see that their relation is also insensitive to EoS. Same as Yagi \textit{et al.}, we fit the relation with a polynomial on a log-log scale
\begin{align}
\log y=a+b\log\bar{\lambda}_f+c(\log\bar{\lambda}_f)^2+d(\log\bar{\lambda}_f)^3+e(\log\bar{\lambda}_f)^4, \label{uni-fit}
\end{align}
where $y=\bar{I}^r$ or $\bar{I}$. Results are shown in Table \ref{table:uni}, where we also list the I-Love relation for comparison. In Fig.\ \ref{fig:uni-I} we compare fitted results with true values. We can see that relative errors for both relations are similar and within $10^{-2}$.

Note that although the profile of $\rho$ comes from the solution of Einstein's equation (i.e., TOV equations), our definition of $\bar{I}^r$ in Eq.\ (\ref{def-Irbar}) still used a Newtonian model of the NS, and post-Newtonian equations for the gravitomagnetic coupling.  We conjecture that a universal relation will still exist after relativistic corrections are made, but we anticipate systematic corrections to the form of the relation.

\begin{table}
    \centering
    \caption{Coefficients for the fitting formulae of the NS I-Love and $\bar{I}^r$-Love relations.}
    \begin{tabular}{c c c c c c} \hline\hline
        $y$ &   $a$ & $b$ &  $c$ & $d$ & $e$   \\ \hline
$\bar{I}^r$ & $0.121$ & 0.169 &$1.25\times10^{-2}$ & $-9.38\times10^{-5}$ & $-1.92\times10^{-5}$\\ \hline
$\bar{I}$  & 1.47 & $8.17\times10^{-2}$ &$1.49\times10^{-2}$ & $2.87\times10^{-4}$ & $-3.64\times10^{-5}$ \\ \hline\hline
     \end{tabular}
     \label{table:uni}
\end{table}

\section{Gravitational waves} 
\label{sec:GW}
This section focuses on gravitational waves emitted by binaries that contain at least one spinning NS with $r$-mode resonance.  In Sec.\ \ref{sec:GW-hybrid}, we construct a hybrid waveform model that incorporates both $r$-mode and PN effects. In Sec.\ \ref{sec:toy-model}, we compare Flanagan and Racine's analytical formula for $r$-mode-resonance-induced GW phase with numerical results. 
%

\subsection{A hybrid $PN$-$r$-mode waveform model}
\label{sec:GW-hybrid}
Once we obtain the orbital evolution from the EOM, we can extract numerical GW waveform through the (mass) quadrupole formula \cite{Poisson+Will+14}
\begin{align}
\label{eq:hij}
h(t)=\frac{1}{D_L}(\ddot{Q}_{xx}-\ddot{Q}_{yy}),
\end{align}
where $D_L$ is the distance between the detector and the source. Here we assume the BNS is optimally oriented for the plus polarization. The quadrupole moment of the system is given by
\begin{align}
Q_{ij}=\mu(x_ix_j-r^2\delta_{ij}/3),
\end{align}
where only the orbital part is included, because $r$-mode does not induce additional mass quadrupole moment\footnote{R-mode does contribute to gravitational radiation through the current quadrupole moment [Eq.~(\ref{current-quad-moment})]. However, it is $10^{-7}$ smaller than the orbital mass quadrupole moment, which is negligible in our case.}.  


It is usually convenient to analyze data in the frequency domain (FD). Following Ref.\ \cite{Ma:2020rak}, we first sample the numerical $h(t)$ in the time domain with a rate  of $8192$\,Hz, and then use the fast Fourier Transform algorithm to transform the data to FD.   
Finally, we select data in the frequency band $[2F_0,2F_\text{contact}]$. (Note that at the mass quadrupole order, GW freuqency is twice the orbital frequency.)


Our numerical FD waveform  $\tilde{h}^{N+r}(f)$ contains the leading-order PP contribution as well as the effect of $r$-mode resonance; its phase $\Psi_{N+r}$ can be written as
\begin{align}
\Psi_{N+r}=\Psi_{N}+\Psi_{r}.
\end{align}
where $\Psi_{r}$ is the phase induced by the $r$-mode resonance; and $\Psi_N$ corresponds to the leading order of PP waveform. With the stationary phase approximation (SPA), $\Psi_N$ can be written as
\begin{align}
\Psi_{N}=2\pi ft_c-\phi_c+\frac{3}{128}(\pi \mathcal{M}f)^{-5/3},
\end{align}
where $t_c$ and $\phi_c$ are the time and GW phase of coalescence.

To incorporate other PN and ($f$-mode) tidal effects into the waveform, we introduce a phase correction within the SPA framework, writing 
\begin{align}
\tilde{h}(f)=\tilde{h}^{N+r}(f)e^{i\Psi_\text{SPA}}.
\end{align}
The total phase of $\tilde{h}(f)$, $\Psi_{\rm tot}$, can be written as
\begin{align}
\Psi_{\rm tot}=\Psi_{N}+\Psi_{r}+\Psi_\text{SPA}\,,
\end{align}
with
\begin{align}
\Psi_\text{SPA}=\Psi_\text{PP}+\Psi_{SO}+\Psi_{\bar{\lambda}^{(1)}_f}+\Psi_{\bar{\lambda}^{(2)}_f}\,. \label{PN-phase-main}
\end{align}
Here $\Psi_\text{PP}$ is the PN correction to the phase of non-spinning PP, up to 3.5PN\footnote{It excludes the leading-order contribution which  is already contained in $\Psi_N$.} \cite{Arun:2004hn}; $\Psi_{SO}$ is the spin-orbit coupling term \cite{Kidder:1992fr,Arun:2008kb,Blanchet:2011zv,2005PhRvD..71l4043M}; and $\Psi_{\bar{\lambda}^{(1)}_f}$ and $\Psi_{\bar{\lambda}^{(2)}_f}$ are $f$-mode tidal effects from $m_1$ and $m_2$, respectively \cite{2008ApJ...677.1216H,2012PhRvD..85l3007D,Henry:2020ski}. We have ignored the spin-spin coupling and the spin precession, because their effects are negligible (cf. Appendix \ref{app:precession}). Expressions of phase terms are shown in Appendix \ref{app:pn-phase}. 

Let us briefly review the parameter dependence of our waveforms. The Newtonian-plus-$r$-mode part of the waveform, $\tilde{h}^{N+r}$, depends on 8 independent parameters: chirp mass $\mc$; two individual spin frequencies $\Omega_{si}$; two individual $r$-mode coupling coefficients $\mathcal{I}_i$; luminosity distance, $D_L$; the coalescing time and phase, $t_c$ and $\phi_c$. The phase correction, $\Psi_{\rm SPA}$, depends on 5 additional parameters: mass ratio $\Xi=m_1/M$; the (anti-)symmetric tidal Love numbers $\bar{\lambda}_f^{s(a)}=(\bar{\lambda}_f^{1}\pm\bar{\lambda}_f^{2})/2$; the (anti-)symmetric dimensionless spin along the orbital angular momentum $\chi^{(z)}_{s(a)}=(\chi^{(z)}_1\pm\chi^{(z)}_2)/2$. Here we use the normalized Love number (Sec.\ \ref{sec:uni-rel}) because in this way it is more convenient to incorporate universal relations into the calculation. 


\subsection{Analytic model for $\Psi_r$}
\label{sec:toy-model}

\begin{figure}[htb]
        \includegraphics[width=\columnwidth,height=6.5cm,clip=true]{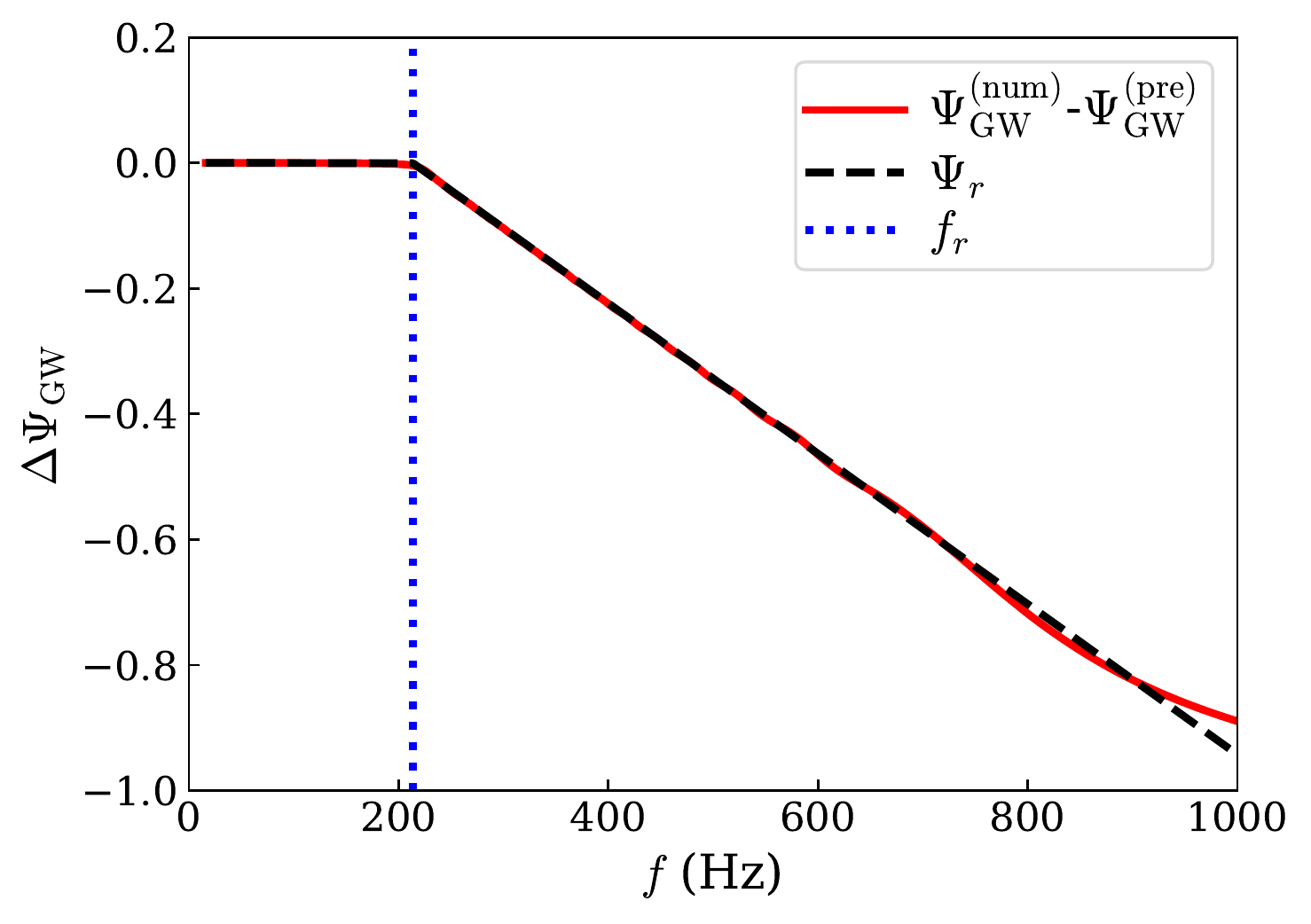}
  \caption{The GW phase difference induced by the $r$-mode DT versus GW frequency. It is compared with the expression of $\Psi_r$ in Eq.\ (\ref{toy_psi_r}). The BNS system is the same as the one we used in Fig.\ \ref{fig:orbit}, except that $\Omega_{s1}=80$ Hz.}
 \label{fig:GW-phase-compare}
\end{figure}

As discussed in Sec.\ \ref{sec:EOM-orbit}, DTs only affect the orbital evolution significantly near the resonance. In the post-resonance regime, the orbit is well described by a PP orbit. Similarly, in the FD,  resonance only leads to a phase shift $\delta\Phi_r$ and a time shift $\delta t_r$ to the waveform, i.e., \cite{Flanagan:2006sb}
%
\begin{align}
\Psi_r=(\delta\Phi_r-2\pi f \delta t_r)\Theta(f-f_r),
\end{align}
where the resonant GW frequency is given by [Eq.\ (\ref{reson-condi})]
\begin{align}
f_r=\frac{\dot{\phi}_r}{\pi}=\frac{4}{3\pi}\Omega_{s1}. \label{reson-gw-fre}
\end{align}
The Heaviside step function $\Theta(f-f_r)$ is introduced here because there is no DT when $f<f_r$\footnote{Here we assume that the pre-resonance PP orbit is aligned with the true orbit, so the phase and time shifts appear after the resonance. It is also equivalent to align the post-resonance PP orbit with the real orbit. Then the Heaviside step function should be changed to $\Theta(f_r-f)$. }. Using Eqs.\ (\ref{deltaphi-flanagan}), (\ref{deltat-flanagan}) and the relation 
\begin{align}
\delta\Phi_r=2\delta\phi_r,
\end{align}
we obtain
\begin{align}
\delta t_r=\frac{\delta\Phi_r}{2\pi f_r},
\end{align}
we arrive at
\begin{align}
\Psi_r=\left(1-\frac{f}{f_r}\right)\delta\Phi_r\Theta(f-f_r). \label{toy_psi_r}
\end{align}
We refer the interested reader to Appendix B in Ref.\ \cite{Yu:2017cxe} for a strict derivation. In Fig.\ \ref{fig:GW-phase-compare}, we compare Eq.\ (\ref{toy_psi_r}) with the numerical result, using the same BNS system as Fig.\ \ref{fig:orbit}, except that $\Omega_{s1}=80$ Hz. We can see that the phase difference induced by the $r$-mode, $\Psi_{\rm GW}^{\rm (num)}-\Psi_{\rm GW}^{\rm (pre)}$, agrees well with the expression of $\Psi_r$ in Eq.\ (\ref{toy_psi_r}). Here $\Psi_{\rm GW}^{\rm (num)}$ stands for the real GW phase, and $\Psi_{\rm GW}^{\rm (pre)}$ is the GW phase of the pre-resonance PP orbit (extending to the post-resonance regime).

The expression of $\Psi_r$ results in
\begin{subequations}
\begin{align}
&\frac{\partial \tilde{h}^{N+r}}{\partial\mathcal{I}}\sim \frac{\partial \tilde{h}^{N+r}}{\partial \delta\Phi_r}=i\left(1-\frac{f}{f_r}\right)\Theta(f-f_r)\tilde{h}^{N+r}, \label{phplove}\\
&\frac{\partial \tilde{h}^{N+r}}{\partial\Omega_{s}}\sim \frac{\partial \tilde{h}^{N+r}}{\partial f_r}+\frac{\partial \tilde{h}^{N+r}}{\partial\delta\Phi_r}\frac{\partial \delta\Phi_r}{\partial \Omega_{s1}}\sim\frac{f}{f_r^2}\delta\Phi_r\Theta(f-f_r)\tilde{h}^{N+r}, \label{phpspin}
\end{align}
\label{phppars}%
\end{subequations}
which are important to understand the result Fisher analysis, as we will discuss in the next section. In Eq.\ (\ref{phpspin}), we have ignored the term $\partial\tilde{h}^{N+r}/\partial\delta\Phi_r$, because it is suppressed by the factor $(1-f/f_r)$ when $f\sim f_r$.

With Eqs.\ (\ref{phppars}), we can learn two things. First, $\partial\tilde{h}^{N+r}/\partial\delta\mathcal{I}$ is proportional to $(1-f/f_r)$, which is close to 0 when $f\sim f_r$, so the constraint for $\mathcal{I}$ from GW detection is weaker than the one for $\Omega_s$. Second, the constraint for $\mathcal{I}$ is independent from the magnitude of $\delta\Phi_r$ (or $\mathcal{I}$); while the constraint for $\Omega_{s}$ is proportional to $\delta\Phi_r^{-1}\sim\mathcal{I}^{-1}$ (see also Ref.~\cite{Yu:2017cxe}).

\begin{figure}[htb]
        \includegraphics[width=\columnwidth,height=6.5cm,clip=true]{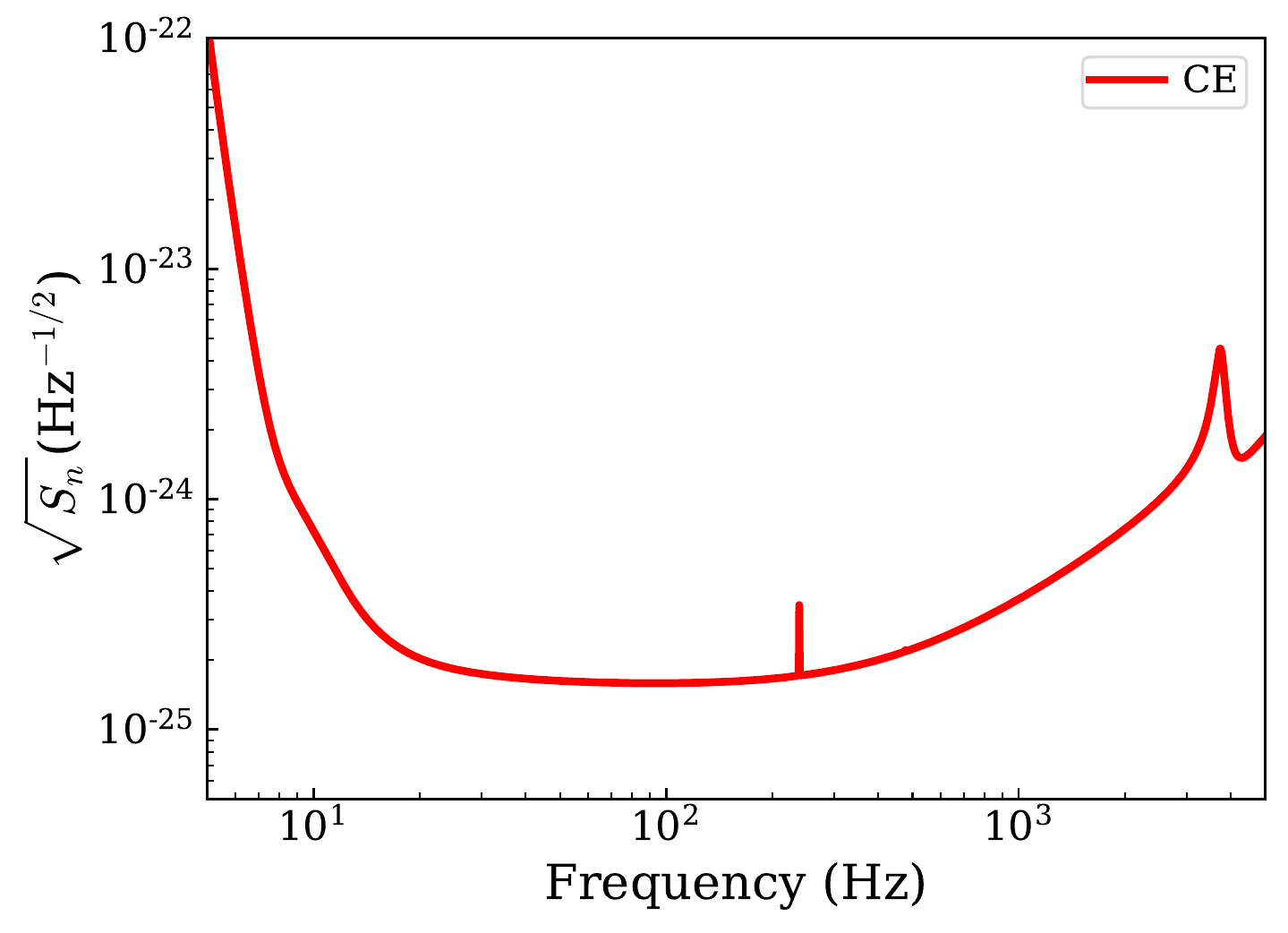}
  \caption{The noise spectral density of CE.}
 \label{fig:psd}
\end{figure}
\section{Constraints of parameters using $r$-mode dynamical tide} 
\label{sec:res}
In this section, we discuss parameter estimation based on the hybrid PN-$r$-mode waveform obtained in Sec.~\ref{sec:GW-hybrid}.  

\subsection{Fisher-matrix formalism and signal strength}

We shall do this by computing the  Fisher information matrix, defined by
\begin{align}
\Gamma_{ij}=\left(\left.\frac{\partial h}{\partial \theta^i}\right|\frac{\partial h}{\partial \theta^j}\right), \label{fisher-def}
\end{align}
where $\theta^i$ are parameters of the waveform, and derivative will be computed numerically.
The inner product between two waveforms $(h|g)$ is defined as
\begin{align}
(h|g)=4\Re \int\frac{\tilde{h}^*(f)\tilde{g}(f)}{S_n(f)}df,
\end{align}
with the superscript $*$ standing for complex conjugation, and $S_n(f)$ the noise spectral density of the detector. By inverting the Fisher information matrix, we obtain projected constraints on $\theta^i$ as
\begin{align}
\Delta\theta^i=\sqrt{(\Gamma^{-1})_{ii}},
\end{align}
and the covariance between $\theta^i$ and $\theta^j$ is characterized by the off-diagonal term $(\Gamma^{-1})_{ij}$. In this paper, we will mainly focus on the Cosmic Explorer (CE), whose $S_n(f)$ is shown in Fig.\ \ref{fig:psd} \cite{CE}.

\begin{figure}[htb]
        \includegraphics[width=\columnwidth,height=6.5cm,clip=true]{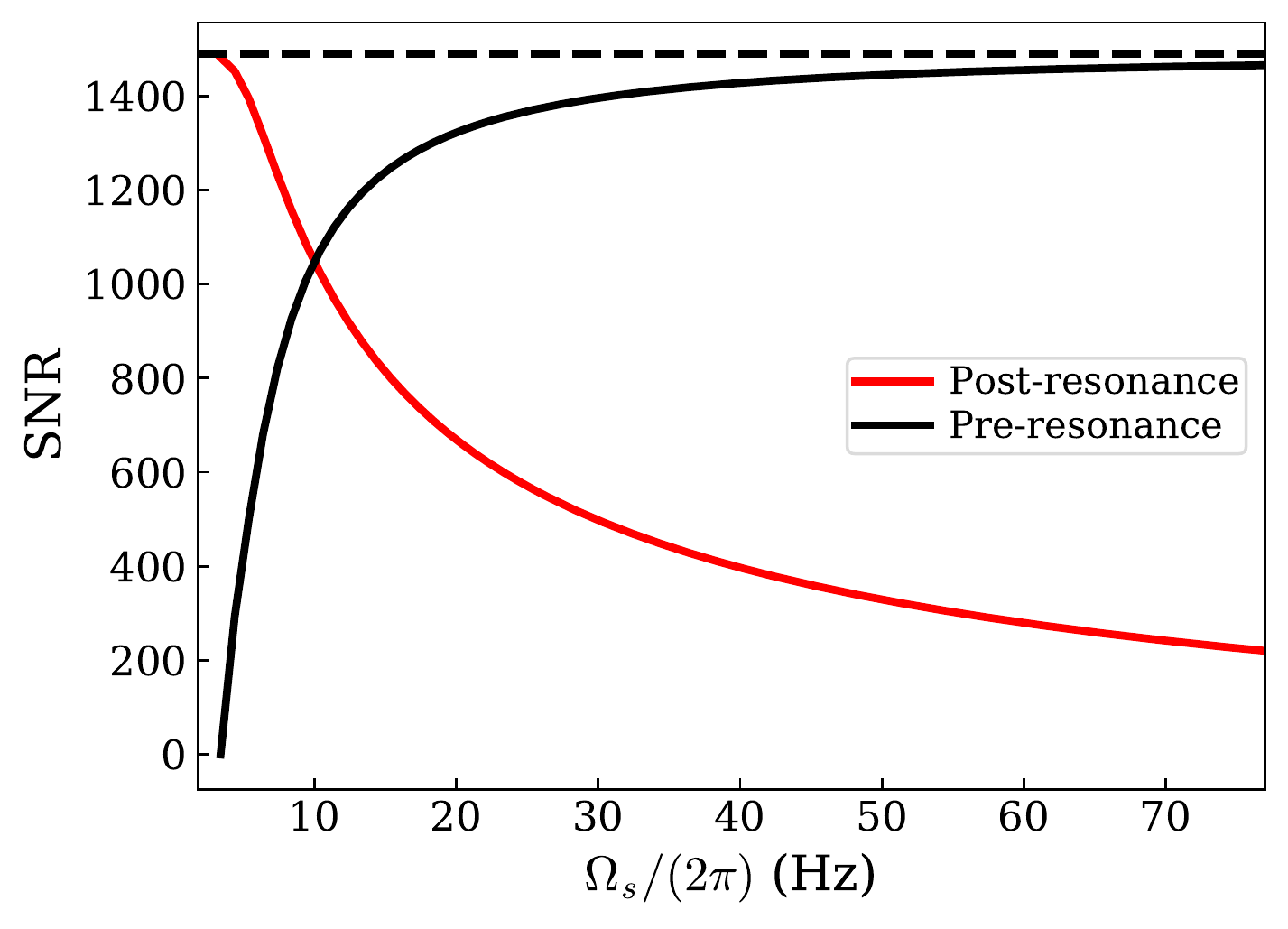}
  \caption{SNR of pre- and post-resonance GW signals as functions of spin frequency. The BNS system is $(1.4,1.35) M_\odot$, optimally oriented at 100Mpc. EoS is GM1. As a comparison, the horizontal dash line is the SNR of the entire in-band signals. There is no pre-resonance signals when $3/4F_0=3.38$ Hz, because our frequency band starts from there. Recalling that the resonant orbital frequency is proportional to the spin frequency, then the SNR of pre-resonance signal increases with the spin frequency.}
 \label{fig:snr-post-reson}
\end{figure}

To see the loudness of BNS signals, we choose a $(1.4,1.35)M_\odot$ BNS system as an example, and plot the SNR of pre- and post-resonance signals as functions of spin frequency in Fig.\ \ref{fig:snr-post-reson}. The system is assumed to be  100Mpc away and optimally oriented. Unless stated otherwise, our lower limit of GW frequency band is $2F_0=9$ Hz. We note that CE is also sensitive to the frequency below 9 Hz, yet it is computationally expensive to simulate the low-frequency evolution.  We have checked that neglecting those signals does not lead to a significant change on parameter estimation if $\Omega_s\gtrsim 2\pi\times10$ Hz (cf.~Fig.~\ref{fig:F0}).  On the other hand, low-frequency signals are important if $\Omega_s<3/4F_0=3.38$ Hz. Nonetheless, as we will show shortly, $r$-mode sector does not provide as strong constraints in that regime, hence that part of parameter space is not of our interest.
%
Because resonant orbital frequency is proportional to spin frequency, SNR of the pre-resonance signals increases with the spin frequency, and there is no pre-resonance signals when the spin frequency is below $3/4F_0=3.38$ Hz. For $10$ Hz $\lesssim\Omega_s\lesssim80$ Hz, both the pre- and post-resonance signals are loud enough to be detected. Thus phase and time shift induced by the $r$-mode resonance (Sec.\ \ref{sec:toy-model}) can be extracted from the waveform, and can be used for parameter estimation. This is the foundation for our later discussions.
 
\begin{table}
    \centering
    \caption{A summary of properties of NS for GM1 and FPS.}
    \begin{tabular}{c c c c c c} \hline\hline
        EOS &   $m/M_\odot$ & $R /{\rm km}$ &  $k_2$ & $\bar{I}^r$ & $\bar{I}$  \\ \hline
GM1 & 1.4 & 13.79 & 0.116 & 6.151 & 15.85 \\ \hline
GM1 & 1.35 & 13.78 & 0.121 & 6.578 &16.87  \\ \hline
FPS  & 1.4 & 10.90 &0.0668 &3.709 &10.09\\ \hline
FPS & 1.35 & 10.95 &0.0711&4.000 &10.77 \\ \hline
     \end{tabular}
     \label{table:eos-summary}
\end{table}

\begin{figure*}[htb]
        \includegraphics[width=\columnwidth,height=7.3cm,clip=true]{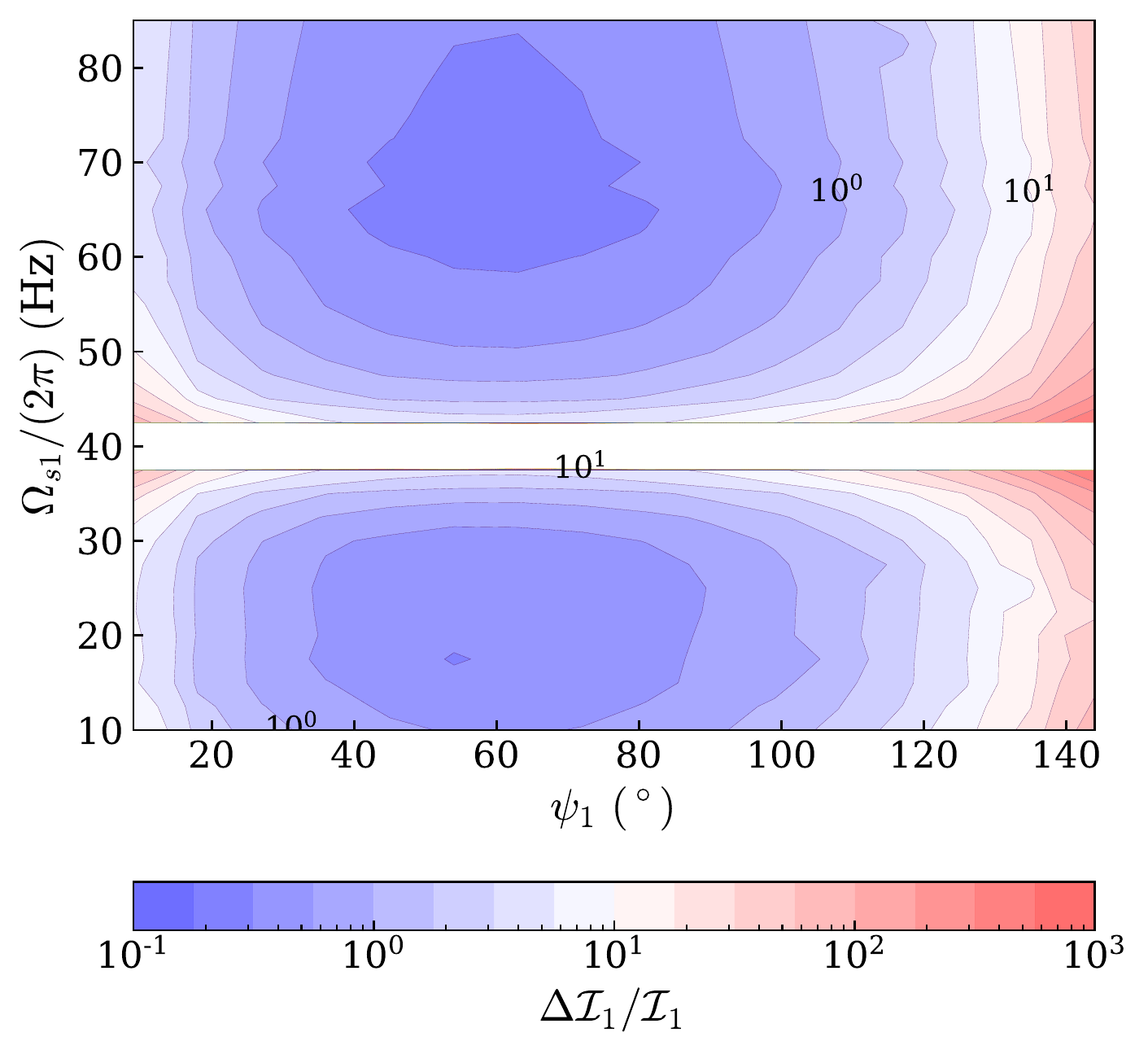}
         \includegraphics[width=\columnwidth,height=7.3cm,clip=true]{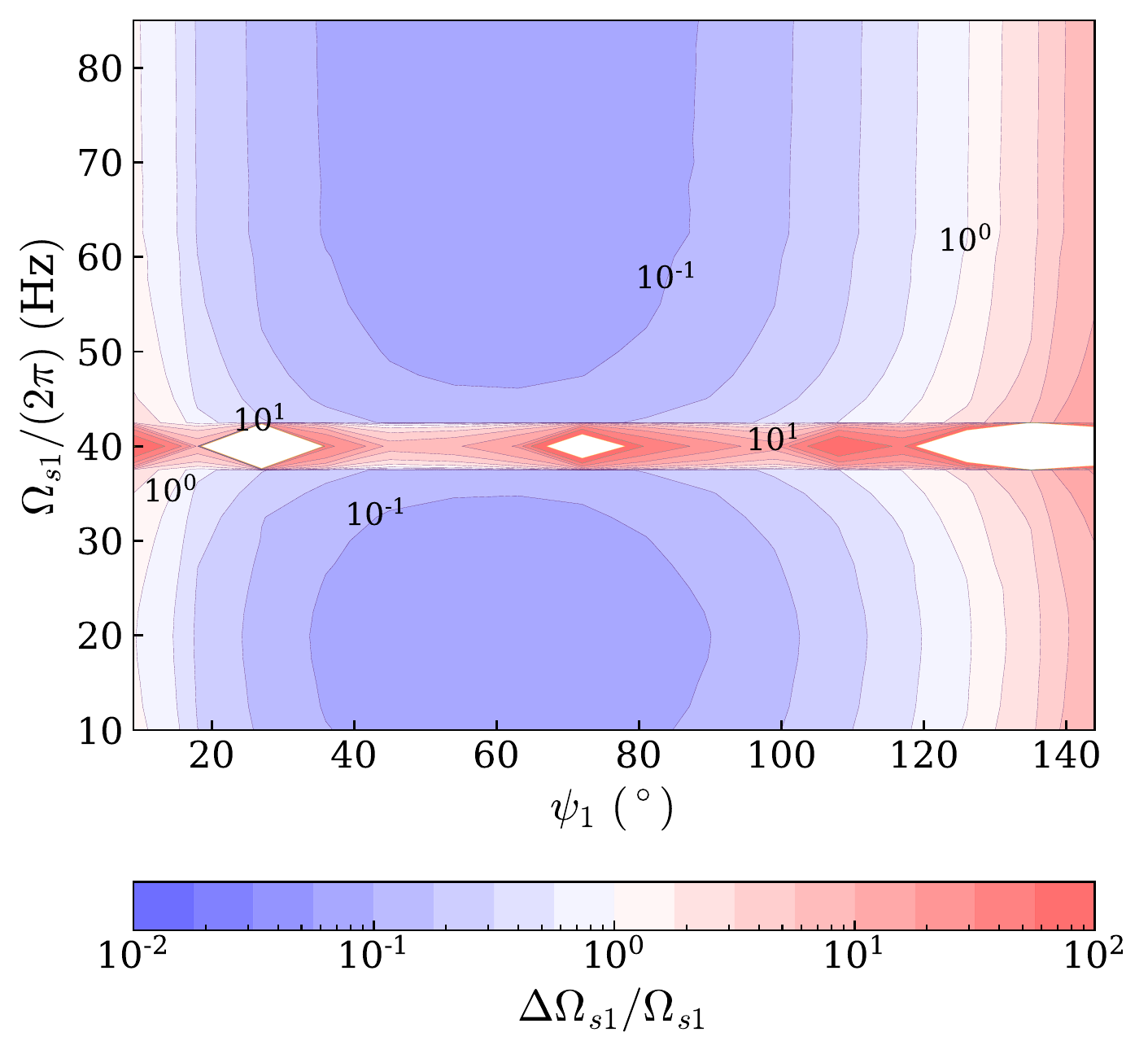} 
  \caption{Case I: relative errors of $\mathcal{I}_1$ (left) and $\Omega_{s1}$ (right) as functions of $\psi_1$ and $\Omega_{s1}$, i.e., the spin configuration of $m_1$. The errorbar is in log scale. For $m_2$, we fix its spin configuration as $\Omega_{s2}=40$ Hz,  $\psi_2=7\pi/18$. The EoS is GM1. Eq.\ (\ref{deltaI-deltaspin-dep}) shows that both $\Delta\mathcal{I}_1/\mathcal{I}_1$ annd $\Delta\Omega_{s1}/\Omega_{s1}$ have $\psi_1$ dependence: $\sin^{-2}\psi_1\cos^{-4}\psi_1/2$. Therefore they diverge to infinite as $\psi_1\to0,\pi$, and become the best when $\psi=\pi/3$. In the best scenario, the constraint on $\Omega_{s1}$ is around 6\%, and the one on $\mathcal{I}_1$ is 21.7\%. $\mathcal{I}_1$ is less constrained because $\partial h/\partial\mathcal{I}_1$ is suppressed by the factor $(1-f/f_r)$ as $f\sim f_r$ [Eq.\ (\ref{phplove})]. Constraints get bad when $\Omega_{s1}\sim\Omega_{s2}=40$ Hz, because two resonances are degenerated in GW.}
 \label{fig:nouni}
\end{figure*}

In the rest of this section, we use the same BNS system to explore the effect of $r$-mode resonance on parameter estimation. Since the sky location, the inclination between the orbital angular momentum and the line of sight, and the polarization angle are not of our interest, below we simply fix their values and consider a Fisher matrices involving only the intrinsic parameters. We mainly consider three situations. In Sec.\ \ref{sec:fim-I}, we first investigate the case where resonances take place in both NSs. The $r$-mode DTs are treated as an independent degree of freedom (i.e., the universal relations between NS properties are not used). Then in Sec.\ \ref{sec:fim-II}, we discuss the improvements on parameter estimation by incorporating the universal relations. Finally in Sec.\ \ref{sec:fim-one-reson}, we study BHNS systems. For comparison, we consider two EoSs, GM1 and FPS, as they respectively give the largest and smallest radii for the same mass (see Fig.\ \ref{fig:M-R}). The NS properties for these two EoSs are summarized in Table \ref{table:eos-summary}. Because the features for both EoSs are similar, we mainly discuss GM1 in the main text, and put the results of FPS in Appendix \ref{app:Case I} and \ref{app:Case II}.

\subsection{Case I: two resonant NSs without the universal relations}
\label{sec:fim-I}

As we discussed in Sec.\ \ref{sec:intro-dt}, for a BNS system where $r$-mode excitation takes place in each NS, $\Omega_{si}$ and $\mathcal{I}_i$ can be constrained by GW. Since $r$-mode DT adds independent contributions to the GW phase evolution,
simply adding this effect does not improve the existing measurement accuracy provided by other PN effects. 
For example, individual tidal Love numbers, as well as individual dimensionless spins, are still degenerate. 
Therefore, in this subsection, we mainly focus on the measurement of $r$-mode sector itself, and study estimation accuracy for $\Omega_{si}$ and $\mathcal{I}_i$.
The full waveform depends on  13 parameters, including 9 PP parameters, $\mc$, $\Xi$, $\chi_{s(a)}^{(z)}$, $\bar{\lambda}_f^{s(a)}$, $D_L$, $t_c$, $\phi_c$, and 4 $r$-mode parameters $\Omega_{si}$, and $\mathcal{I}_i$.  The Fisher matrix is 13 dimensional. 
We note again that here we use the normalized Love number [Eq.~(\ref{def-Irbar})] instead of the Love number listed in Table \ref{table:rmode-uni-intro}.



To be concrete, let us 
fix the spin vector of $m_2$,  $\Omega_{s2}=40$ Hz and $\psi_{2}=7\pi/18$, and explore the following parameter space: $\Omega_{s1}\in[10,85]$ Hz, $\psi_{1}\in[\frac{1}{20}\pi,\frac{17}{20}\pi]$. In Fig.\ \ref{fig:nouni}, we show the relative errors of $\mathcal{I}_{1}$ and $\Omega_{s1}$ as functions of $\Omega_{s1}$ and $\psi_1$. We first observe that the relative errors of $\mathcal{I}_1$ and $\Omega_{s1}$ depend on $\psi_1$ in a similar way; both become worse as $\psi_1\to0$ and $\pi$. However, their behaviors are caused by different reasons. Recall the discussion in Sec.\ \ref{sec:toy-model}, $\Delta\mathcal{I}_1$ is independent of $\mathcal{I}_1$, while $\Delta\Omega_{s1}\propto\mathcal{I}_1^{-1}$, hence we can obtain
\begin{align}
\frac{\Delta\mathcal{I}_1}{\mathcal{I}_1}\sim\frac{\Delta\Omega_{s1}}{\Omega_{s1}}\sim \mathcal{I}_1^{-1} \sim (\bar{I}_{i}^{r})^{-2}\sin^{-2}\psi_1\cos^{-4}\psi_1/2, \label{deltaI-deltaspin-dep}
\end{align}
where Eq.\ (\ref{mathI}) is used. We have checked that the dependence above is consistent with our numerical calculation. With this knowledge, we can know that constraints become the best when $\psi_1=\pi/3$, where $\mathcal{I}_1$ is maximal. In the best case, the constraint on $\Omega_{s1}$ is around 6\%, and on $\mathcal{I}_1$ around $22\%$. Meanwhile, the measurement error $\Delta\mathcal{I}_1/\mathcal{I}_1$ is less than 1 when $\psi_1\in\left[18^\circ,110^\circ\right]$, therefore there is a large range of parameter space where one can extract meaningful constraints from GW.

Secondly, constraints on both $\Omega_{si}$ and $\mathcal{I}_i$ become worse when $\Omega_{s1}\sim\Omega_{s2}=40$ Hz, because two resonances take place at similar locations, and are therefore indistinguishable from each other. In this regime, Fisher-matrix based analysis becomes invalid, and a more detailed Bayesian analysis will be required. Out of this regime, two resonances do not interfere anymore. We have checked that constraints on $\Omega_{s2}$ and $\mathcal{I}_2$ are insensitive to the values of $\psi_1$ and $\Omega_{s1}$ (except when $\Omega_{s1}\sim\Omega_{s2}$). In the best-case scenario, $\Delta\Omega_{s2}/\Omega_{s2}$ is around 5\%, and $\Delta\mathcal{I}_2/\mathcal{I}_2$ around 17\%.

\begin{table}
    \centering
    \caption{ The comparison between  constraints for $\Omega_{si}$ and $\mathcal{I}_i$ with two EoS: FPS and GM1. We explore the parameter space:  $\Omega_{s1}\in[10,85]$ Hz, $\psi_{1}\in[\frac{\pi}{20},\frac{17}{20}\pi]$, while choose $\Omega_{s2}=40$ Hz and $\psi_{2}=7\pi/18$. The median values of the ratio between two EoS for several parameters are shown in the first four columns. Eq.\ (\ref{deltaI-deltaspin-dep}) shows that the constraint is proportional to $(\bar{I}_i^r)^{-2}$, so we also show their ratios in the last two columns, for comparison. All numbers are close.}
    \begin{tabular}{c c c c c c c} \hline\hline
    Parameters & $\Delta\Omega_{s1}$ & $\Delta\Omega_{s2}$ & $\Delta\mathcal{I}_{1}$ & $\Delta\mathcal{I}_{2}$ & $(\bar{I}^r_1)^{-2}$ & $(\bar{I}^r_2)^{-2}$ \\ \hline
    FPS/GM1 & 2.67 & 2.65 & 3.23 & 2.78 & 2.77 & 2.72 \\ \hline\hline
     \end{tabular}
     \label{table:EOS-ratio}
\end{table}


\begin{figure}[htb]
        \includegraphics[width=\columnwidth,height=6.1cm,clip=true]{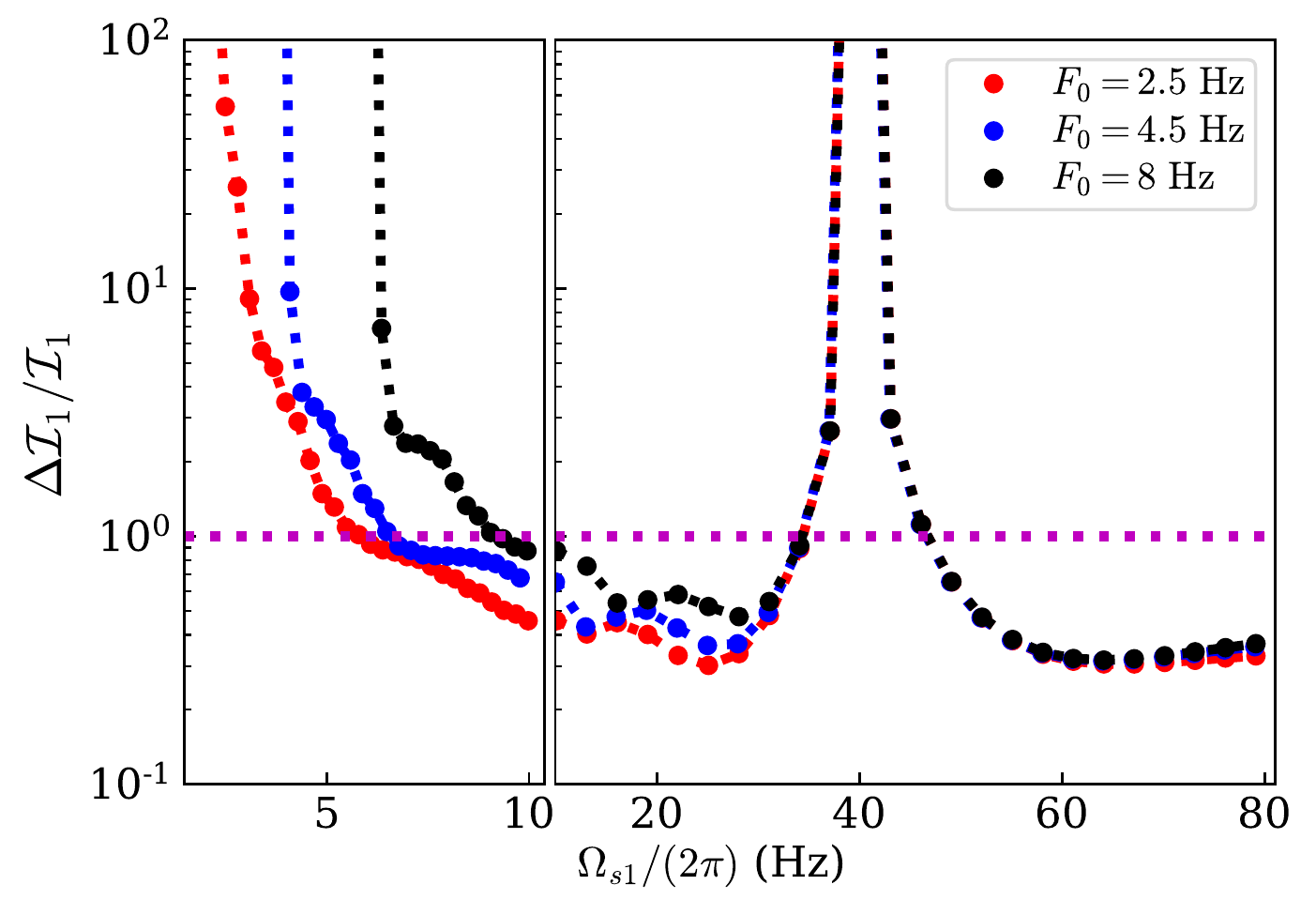}
        \includegraphics[width=\columnwidth,height=6.1cm,clip=true]{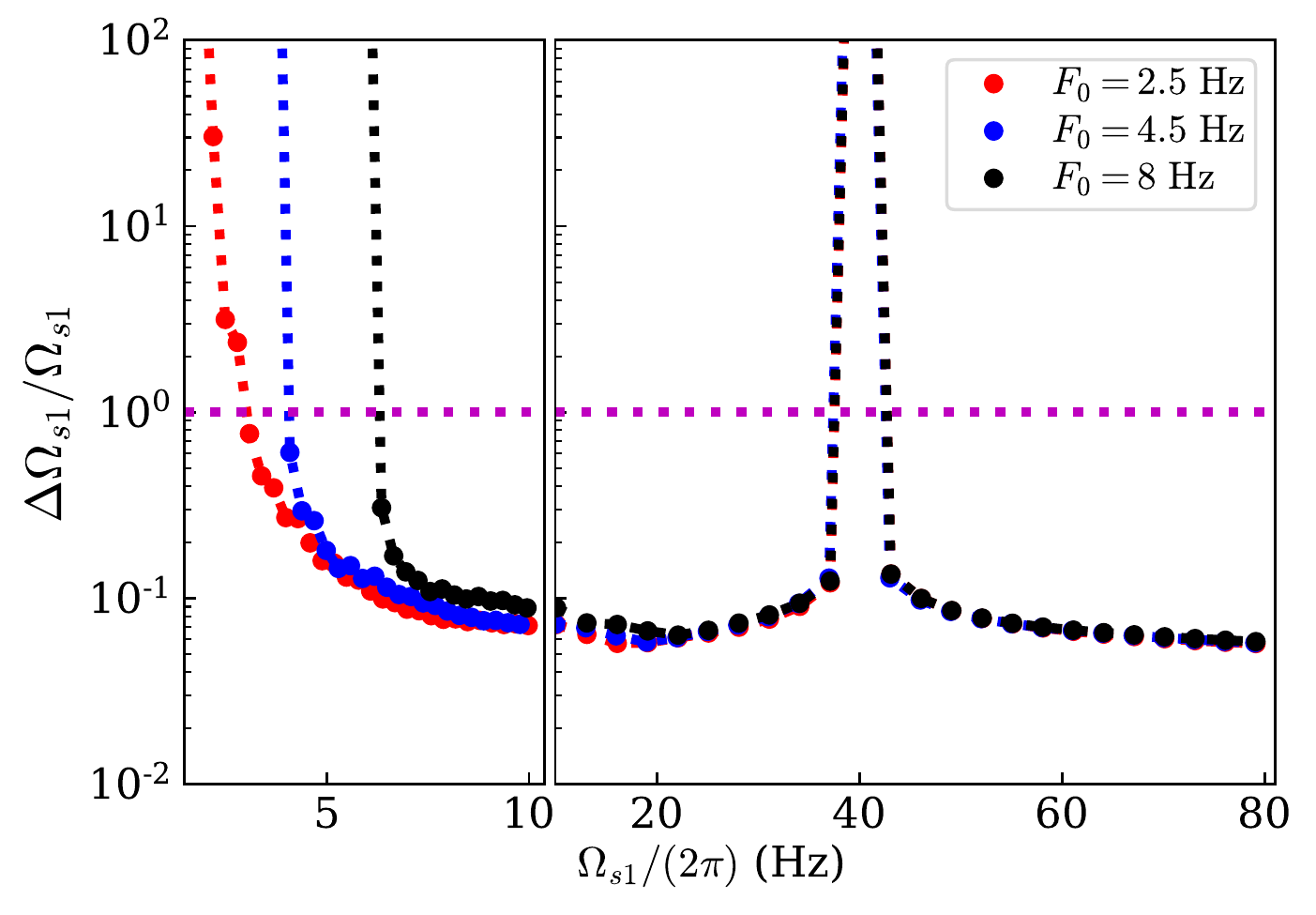}
  \caption{Fractional errors of $\Omega_{s1}$ and $\mathcal{I}_1$ as functions of $\Omega_{s1}$, with different values of $F_0$. We set the spin configuration for $m_2$ to be the same as Fig.~\ref{fig:nouni}, and $\psi_1=\pi/3$. Fractional errors first decrease with $\Omega_{s1}$, because there are more in-band pre-resonance signals. Then it becomes bad as $\Omega_{s1}\sim\Omega_{s2}$, since two resonances are not distinguishable.
   The lower limit of $\Omega_{s1}/(2\pi)$ is taken to be $3/4F_0$, i.e., resonance takes place initially (at orbital frequency $F_0$). We cannot get constraints on $\mathcal{I}_1$ and $\Omega_{s1}$ if we further decrease the spin frequency. Those curves show that the value of $F_0$ only affects these constraints mildly.}
 \label{fig:F0}
\end{figure}

Thirdly, constraints on $\Omega_s$ are better than those on $\mathcal{I}$, because $\partial \tilde{h}^{N+r}/\partial\mathcal{I}$ is suppressed by the factor $(1-f/f_r)$ as $f\sim f_r$ [Eq.\ (\ref{phplove})]. Recalling that the spin frequency determines where the resonance takes place (in time or frequency domain), while  $\mathcal{I}$ characterizes the phase shift (strength of the DT), therefore we can conclude that GW signals are more sensitive to the location of resonance than the strength of the resonance. 
 
Finally, we investigate the effect of EoS. We compute the ratio of relative errors with FPS to those with GM1: $\left[\Delta \Omega_{si}^{\rm (FPS)}/\Omega_{si}^{\rm (FPS)}\right]\Big{/}\left[\Delta\Omega_{si}^{\rm (GM1)}/\Omega_{si}^{\rm (GM1)}\right]$ and $\left[\Delta \mathcal{I}_{i}^{\rm (FPS)}/\mathcal{I}_{i}^{\rm (FPS)}\right]\Big{/}\left[\Delta \mathcal{I}_{i}^{\rm (GM1)}/\mathcal{I}_{i}^{\rm (GM1)}\right]$. In Table \ref{table:EOS-ratio}, we provide the median values of these ratios over the parameter space we have explored. The results for different parameters are similar:  those for GM1 are better than FPS by a factor of $2.6\sim3.2$. The numbers are also close to the ratio of $(\bar{I}^r)^{-2}$, recalling that the relative errors are proportional to $\mathcal{I}^{-1}\sim(\bar{I}^{r})^{-2}$. We can then conclude that, constraints are more stringent for less compact NSs, i.e., those with harder EoS, with error inversely proportional to $r$-mode overlap. Detailed results for FPS are shown in Appendix \ref{app:Case I}. Since the two EoSs are representative for hard and soft EoSs, the fractional error on the spin for other EoSs can be between 6\% to 16\%. As a result, the $r$-mode resonance provides an important channel to put constraints on the spin frequency.

If $\Omega_{s1}$ is further decreased to below 3.38 Hz, we need to lower the value of $F_0$ to include enough pre-resonance signals\footnote{Constraints on $\Omega_{s2}$ and $\mathcal{I}_2$ are unaffected, because the two resonances are independent from each other. }. In Fig.~\ref{fig:F0}, we compare $\Delta\mathcal{I}_1/\mathcal{I}_1-\Omega_{s1}$ and $\Delta\Omega_{s1}/\Omega_{s1}-\Omega_{s1}$ relations with different values of $F_0$. We can see that the value of $F_0$ affects the constraint mildly if $\Omega_{s}>10$ Hz. Furthermore, the fractional error of $\mathcal{I}_1$ exceeds $100\%$ when $\Omega_{s1}\lesssim6$ Hz. In this parameter regime, the $r$-mode cannot be unambiguously detected. Therefore, we here only focus on $\Omega_{s}>10$ Hz, and it is enough to choose $F_0=4.5$ Hz for a general calculation.

\begin{figure}[htb]
         \includegraphics[width=\columnwidth,height=9cm,clip=true]{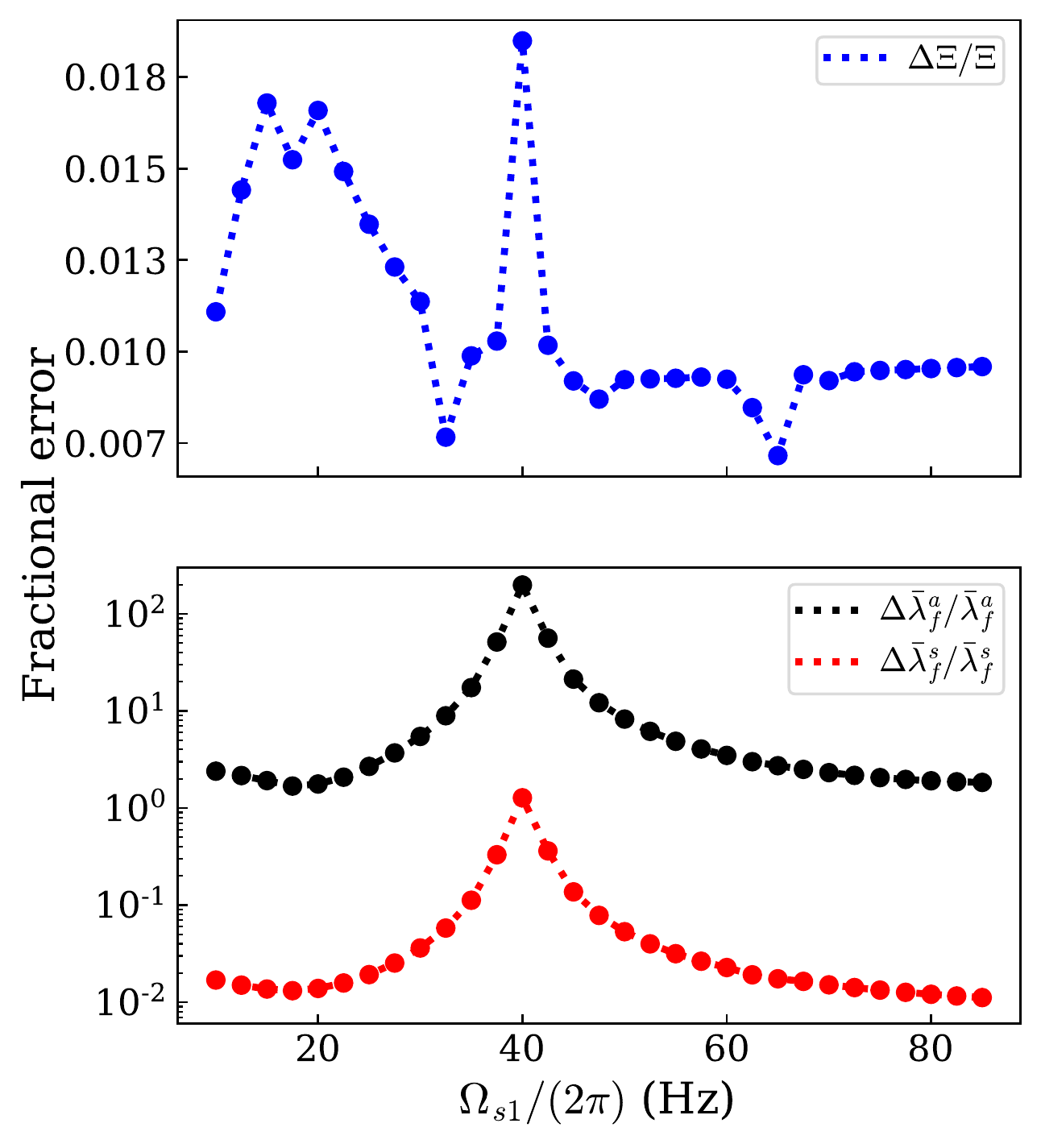}
  \caption{Case II: fractional errors as functions of $\Omega_{s1}$ after incorporating universal relations. The spin configuration for $m_2$ is same as Fig.\ \ref{fig:nouni}, and $\psi_1=3\pi/10$. EoS is still GM1. The top panel is for $\Xi$ while the bottom one corresponds to $\bar{\lambda}_f^{s(a)}$. }
 \label{fig:uni-GM1}
\end{figure}

\begin{figure}[htb]
        \includegraphics[width=\columnwidth,clip=true]{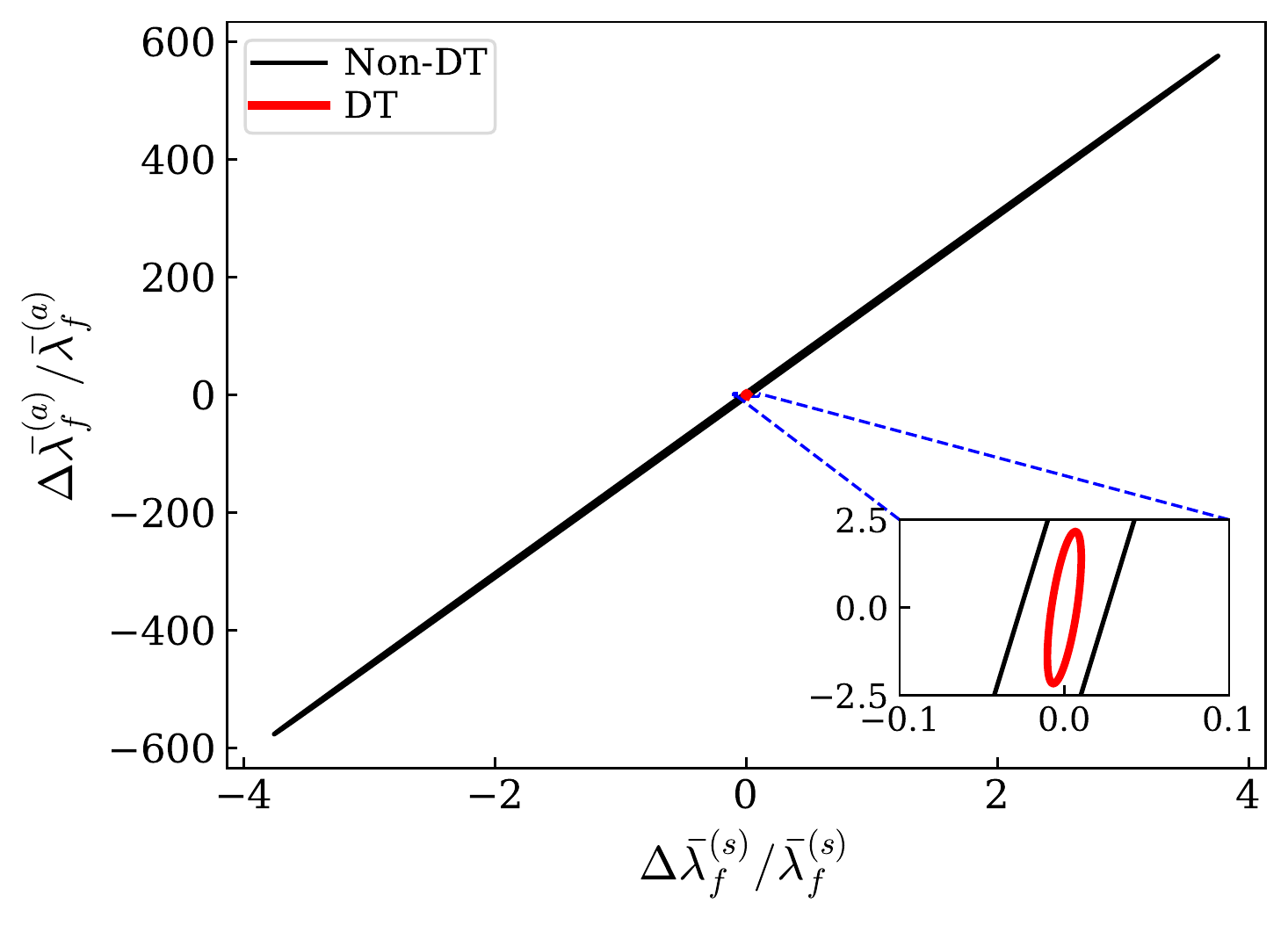} \\
        \includegraphics[width=\columnwidth,clip=true]{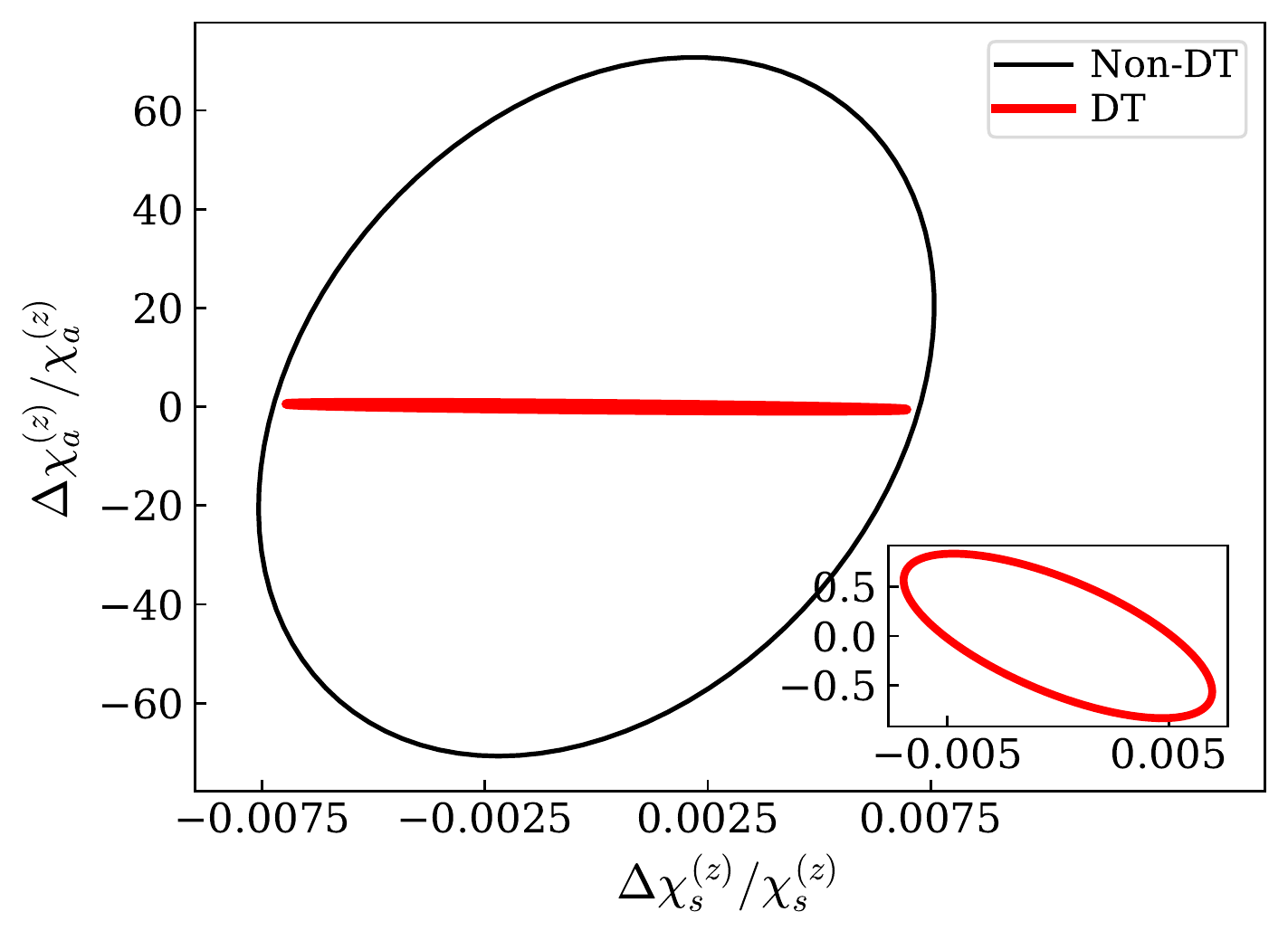}
  \caption{The error ellipses of $(\Delta\bar{\lambda}_f^{a},\Delta\bar{\lambda}_f^{s})$ and $(\Delta\chi_a^{(z)},\Delta\chi_s^{(z)})$, with $\Omega_{s1}=80$ Hz, $\Omega_{s2}=40$ Hz, $\psi_1=\pi/3$ and $\psi_2=7\pi/18$. The black curve is the result of PN effects (including adiabatic tidal effect and spin-orbit coupling). The red curve is the result after including $r$-mode resonances (with universal relations). For those ellipses, both directions are improved by resonances.}
 \label{fig:contours}
\end{figure}

\begin{figure}[htb]
        \includegraphics[width=\columnwidth,height=8.cm,clip=true]{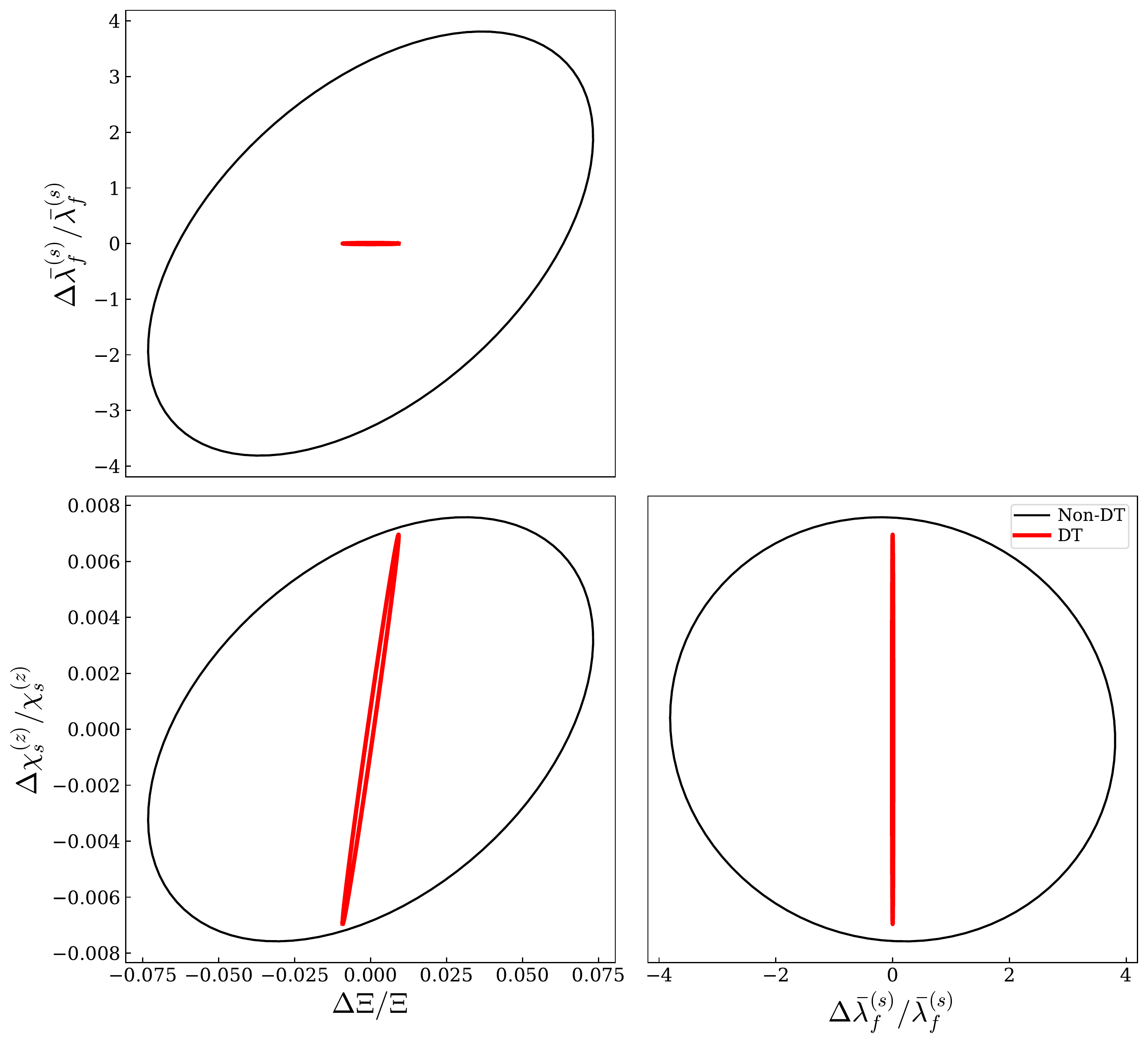} 
  \caption{Same as Fig.\ \ref{fig:contours}. The error ellipses between $\Xi-\bar{\lambda}_f^{s}-\chi_s^{(z)}$.}
 \label{fig:cor}
\end{figure}

\subsection{Case II: two resonant NSs + universal relations}
\label{sec:fim-II}

In the previous subsection, we only studied the estimation of parameters in the $r$-mode sector because this sector is independent of other PN effects. However, if we take into account the universal relations between NS properties,   the $r$-mode sector will behave like a bridge that connects tidal and spin parameters.  We will then have enough number of degrees of freedom in the waveform to constrain all parameters, as summarized in Table\ \ref{table:rmode-uni-intro}. 

\subsubsection{Neutron-Star Parameters}

To begin with, we need to express $\chi_{s(a)}^{(z)}$ in terms of spin frequency and (normalized) moment of inertia [Sec.\ \ref{sec:uni-rel}]:
\begin{align}
\chi_{s(a)}^{(z)}=\frac{1}{2}\left(\bar{I}_1\Omega_{s1}m_1\cos\psi_1\pm\bar{I}_2\Omega_{s2}m_2\cos\psi_2\right),
\end{align}
We will then have 12 intrinsic parameters, including: chirp mass $\mc$; mass ratio $\Xi$; (anti-)symmetric normalized Love numbers $\bar{\lambda}_f^{s(a)}$; two inclination angles $\psi_i$; two spin frequencies $\Omega_{si}$; two $r$-mode coupling coefficients $\mathcal{I}_i$; and two normalized momentum of inertia $\bar{I}_i$. Meanwhile, we have 8 constraints from GW
\begin{align}
\mc,\quad \Xi,\quad \chi^{\rm eff},\quad \lambda_f^{\rm eff}, \quad \Delta\Omega_{si}, \quad \Delta\mathcal{I}_{i}, \label{uni-deltaspin}
\end{align}
and 4 constraints from the two universal relations
\begin{align}
&\mathcal{I}_i=\mathcal{I}_i(\bar{\lambda}_f^{s(a)},\Xi,\psi_i,\mc), 
&\bar{I}_i=\bar{I}_i(\bar{\lambda}_f^{s(a)}),  \label{love-uni}
\end{align}
where we have used the definitions of $\mathcal{I}$ in Eq.\ (\ref{mathI}). 
In principle, these 12 constraints are enough to decode the BNS system. However, as we have seen in the last subsection, constraints in Eq.\ (\ref{uni-deltaspin}) in practice may not be measured with extremely high statistical accuracy; the relative error on $\mathcal{I}$ can be  as bad as 100\%. Therefore, we should be prepared that  degeneracy may not be completely broken in practice.

After using the universal relations, the Fisher matrix is 11D, corresponds to $\mc$, $\Xi$, $\Omega_{si}$, $\bar{\lambda}_f^{s(a)}$, $\psi_i$, $t_c$, $\phi_c$, $D_L$.

\subsubsection{Parameter Constraints: tidal Love numbers $\bar{\lambda}_f^{s,a}$ }

In the lower panel of Fig.\ \ref{fig:uni-GM1}, we plot the dependence of $\Delta\bar{\lambda}_f^{s(a)}/\bar{\lambda}_f^{s(a)}$ on $\Omega_{s1}$, with  $\psi_1=3\pi/10$, $\psi_2=7\pi/18$ and $\Omega_{s2}=2\pi\times 40$ Hz. (we use the GM1 EoS; data for the FPS EoS are shown in Appendix \ref{app:Case II}.) 
Constraints become worse when $\Omega_{s1}\sim\Omega_{s2}=2\pi\times 40$ Hz because two resonances take place simultaneously, making it impossible to resolve their individual features. 
For favorable values of $\Omega_{s1}$, $\Delta\bar{\lambda}_f^{s}$ can be  constrained as well as $\sim1.0\%$, while $\Delta\bar{\lambda}_f^{a}/\bar{\lambda}_f^{a}$ is $\sim1.7$. The degeneracy between two individual tidal Love numbers is still not broken, but substantially reduced. To see this more clearly, we plot the error ellipses between $\Delta\bar{\lambda}_f^{a}$ and $\Delta\bar{\lambda}_f^{s}$ in Fig.\ \ref{fig:contours}, for $\Omega_{s1}=80$ Hz and $\psi_1=\pi/3$, when $r$ mode is either included or not included.   Constraints on both directions are substantially improved by $r$-mode, but the fractional error of $\bar{\lambda}_f^{a}$ is still greater than 1. 

Let us note that breaking of $\bar{\lambda}_f^{a}$ degeneracy is difficult because the predicted values of $\bar{\lambda}_f^{a}$ is intrinsically very small, if we make the tacit assumption that neutron stars all have the {\it same}, albeit unknown,  EoS.  For example, in our case, the two neutron stars are $(1.4,1.35)\,M_\odot$, and with the GM1 EoS, $\bar{\lambda}_f^{a}/\bar{\lambda}_f^{s}\approx 0.1$.  Nevertheless, it is theoretically possible that the two neutron stars do not follow the same EoS --- and for this reason, it is still very meaningful to dramatically reduce $\Delta\bar{\lambda}_f^{a}$, even if our measurement error is somewhat larger than the predicted value of $\bar{\lambda}_f^{a}$. 

As we turn to the individual Love numbers of the two neutron stars,  $\bar{\lambda}_f^{(i)}$, we find that their errors are still correlated in general. For some special cases, the individual Love numbers can be sufficiently well constrained. In Table \ref{table:fim-uni}, we show one example with $\Omega_{s1}=80$ Hz, $\psi_1=\pi/3$, $\Omega_{s2}=40$ Hz, $\psi_1=7\pi/18$, and GM1 EoS. Relative errors of two individual tidal Love numbers are around 20\%.

\begin{table}
    \centering
    \caption{A special case where the individual normalized Love numbers and inclination angles are well constrained. Two NSs have spin $\Omega_{s2}=40$Hz, $\psi_2=7\pi/18$, $\Omega_{s1}=80$Hz and $\psi_1=\pi/3$. EoS is GM1.}
    \begin{tabular}{c c c c c} \hline\hline
    Parameters & $\Delta \bar{\lambda}^{(1)}_f/ \bar{\lambda}^{(1)}_f$ & $\Delta \bar{\lambda}^{(2)}_f/ \bar{\lambda}^{(2)}_f$ & $\Delta\psi^{(1)}$ (rad) & $\Delta\psi^{(2)}$ (rad) \\ \hline
    Constraints & 0.26 & 0.22 & 0.18 & 0.28 \\ \hline\hline
     \end{tabular}
     \label{table:fim-uni}
\end{table}

We then study the improvement of constraints by comparing results above with those of pure PN effects (i.e., adiabatic tidal effect). The improvement factor, Imp. $\Delta\bar{\lambda}_f^{s(a)}/\bar{\lambda}_f^{s(a)}$, is defined to be the ratio between constraints from two sides (the factor is larger than 1 when the $r$-mode enhances the constraint). Results are shown in Table \ref{table:fim-impr-fac}. We can see constraints on $\Delta\bar{\lambda}_f^{s(a)}/\bar{\lambda}_f^{s(a)}$ are improved. In the best case, the factor is around 300--400.

\begin{table}
    \centering
    \caption{A summary of the effect of including $r$-mode resonance on parameter constraints. The second column gives the best fractional errors for $\Xi$ and $\bar{\lambda}_f^{s(a)}$ achievable when we vary $\psi_{1}$ and $\Omega_{s1}$.  These fractional errors are generally improved, when compared with those achievable only  including PN effects. In the third and forth columns, we list the best and worst improvement factors for each parameter, as we vary $\psi_{1}$ and $\Omega_{s1}$.}
    \begin{tabular}{c c c c} \hline\hline
        \multirow{2}{*}{Parameters} & Best & Best  & Worst \\ 
    & constraints &  improvement &   improvement\\ \hline 
     $\Delta \bar{\lambda}^{s}_f/ \bar{\lambda}^{s}_f$ & $10^{-2}$ & 389 & 1 \\ \hline 
     $\Delta \bar{\lambda}^{a}_f/ \bar{\lambda}^{a}_f$ & 1.84& 336&1 \\ \hline
     $\Delta\Xi/\Xi$ & $5\times10^{-3}$ & 11.6 &1\\ \hline\hline
     \end{tabular}
     \label{table:fim-impr-fac}
\end{table}

\subsubsection{Parameter Constraints: mass ratio $\Xi$ and spins $\chi_{a,s}^{(z)}$ }

The measurement of $\Delta\chi_a^{(z)}$ and $\Delta\chi_s^{(z)}$ can also benefit from universal relations\footnote{These are not independent variables in this subsection, their constraints are obtained by error propagation.}.  As shown in the lower panel of Fig.\ \ref{fig:contours}, while the improvement in $\chi_{s}^{(z)}$ is mild, the constraint on $\chi_{a}^{(z)}$ is improved by a factor of $\sim120$. Therefore, the degeneracy between individual dimensionless spin is also reduced.

The case for $\Xi$ is similar. Its correlation with other parameters is reduced by the $r$-mode DT and the universal relations. Compared with constraints from pure PN effects, the error in $\Xi$ can be improved by a factor of $1-11.6$, as summarized in Table \ref{table:fim-impr-fac}. In the first row of Fig.\ \ref{fig:uni-GM1}, we also present its fractional error as a function of $\Omega_{s1}$. We can see $\Delta\Xi/\Xi$ is insensitive to $\Omega_{s}$. Generally, it can be constrained to $\sim 1.3\%$.

 It is also well-known that the estimation errors of $\Xi$, $\bar{\lambda}_f^{s}$ and $\chi_s^{(z)}$ are correlated in absence of $r$-mode resonance. The effects of incorporating $r$-mode resonances are shown in Fig.\ \ref{fig:cor}. Whereas in Refs.~\cite{TheLIGOScientific:2017qsa,Abbott:2018exr} the error reduction relies on imposing an observational-based prior on $\chi_{z}$, once the $r$-mode resonances are taken into account, the mutual correlations between parameters are significantly reduced. As suggested in Ref.~\cite{Yu:2017zgi}, the $r$-mode DT indeed  improves dramatically the inference accuracy on both the tidal deformability  and the NS component masses. The later can be further converted to an indication of the maximum mass of NSs with a population of events. We thus conclude that the $r$-mode DT plays a crucial role in constraining the nature of NS EoS. 

\subsubsection{Parameter Constraints: inclination angles}

\begin{figure}[htb]
        \includegraphics[width=\columnwidth,height=6.4cm,clip=true]{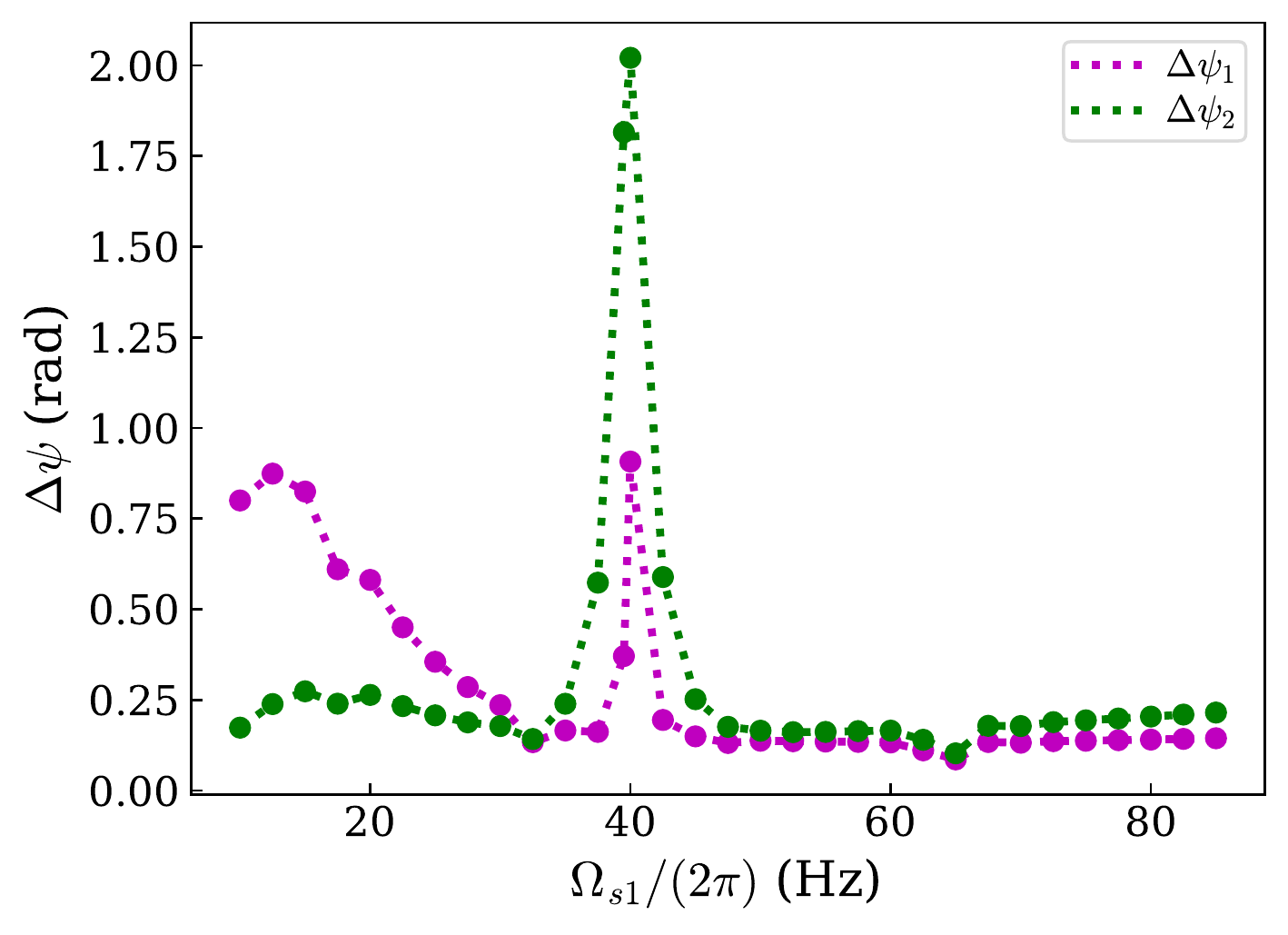}
  \caption{Case II: constraints on inclination angles $\psi_i$ as functions of $\Omega_{s1}$. Spin configurations are same as Fig.~\ref{fig:nouni} and $\psi_1=3\pi/10$. Generally speaking, $\psi_1$ and $\psi_2$ are correlated. In the best case, $\Delta\psi_1\sim\Delta\psi_2\sim0.09$ rad.}
 \label{fig:uni-psi-GM1}
\end{figure}

Besides improving constraints, universal relations also provide the information of inclination angles $\psi_i$, which is hard to be accessible by other PN effects.  In Figs.\ \ref{fig:uni-psi-GM1}, we show $\Delta\psi_{i}$ as functions of $\Omega_{s1}$. Generally speaking, $\psi_1$ and $\psi_2$ are correlated. In the best case, $\Delta\psi_1\sim\Delta\psi_2\sim0.09$ rad.
We note that determining this angle may play significant roles in astrophysics, as it allows the constraining of the NS natal kicks, i.e., kicks received by NSs at their formation due to asymmetric supernova explosions (see, e.g., Refs.~\cite{Shklovskii:70, Brandt:95, Martin:09}). This may further shed light on models and theories of core-collapse supernova~\cite{Janka:07, Burrows:13}.

\subsubsection{If only one of NS is resonant}

\begin{figure}[htb]
        \includegraphics[width=\columnwidth,height=6cm,clip=true]{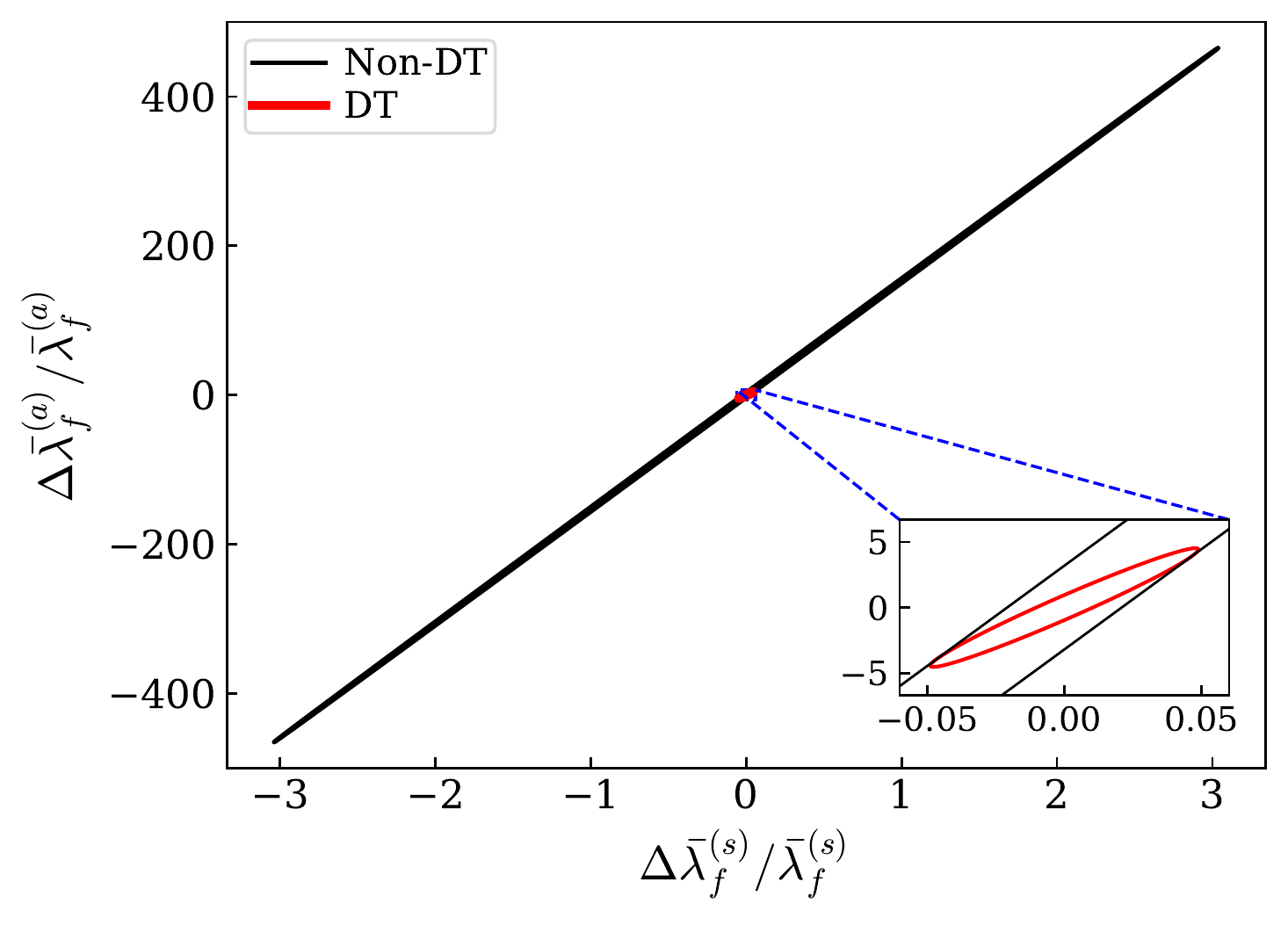} 
  \caption{Similar to the top panel of Fig.~\ref{fig:contours}. But with only one in-band $r$-mode resonance. For the binary system, we choose $\Omega_{s1}=40$ Hz, $\psi_1=\pi/3$. The value of $\Omega_{s2}$ is taken to be small enough such that the $r$-mode of $m_2$ is not excited in-band. The degeneracy between $\bar{\lambda}_f^{s(a)}$ can still be reduced a lot.}
 \label{fig:love-contour-one-reson}
\end{figure}

If one of NS (e.g. $m_2$) in the binary system rotates at very low frequency, its $r$-mode resonance is not in-band anymore, and  $r$-mode resonance does not provide enough constraints to decode the whole system. In fact, we find that the two inclination angles $\psi_{i}$ are highly correlated and are totally unmeasurable. Nevertheless, the measurement of Love number can still benefit from the single, in-band resonance. As an example, we still adopt a BNS system with $(1.4,1.35)M_\odot$ and $\psi_{1}=\pi/3,~\Omega_{s1}=40$ Hz. The value of $\Omega_{s2}$ is taken to be small enough such that the $r$-mode of $m_2$ is not excited in-band. In Fig.~\ref{fig:love-contour-one-reson}, we show contours between $\bar{\lambda}_f^{s}$ and $\bar{\lambda}_f^{a}$ with or without DT. We can see they are similar to the case of two resonances in Fig.~\ref{fig:contours}: even with only one resonance, the degeneracy between $\bar{\lambda}_f^{s(a)}$ can still be substantially reduced.

\subsection{Case III: BHNS}
\label{sec:fim-one-reson}

\begin{figure*}[htb]
        \includegraphics[width=\columnwidth,height=6.1cm,clip=true]{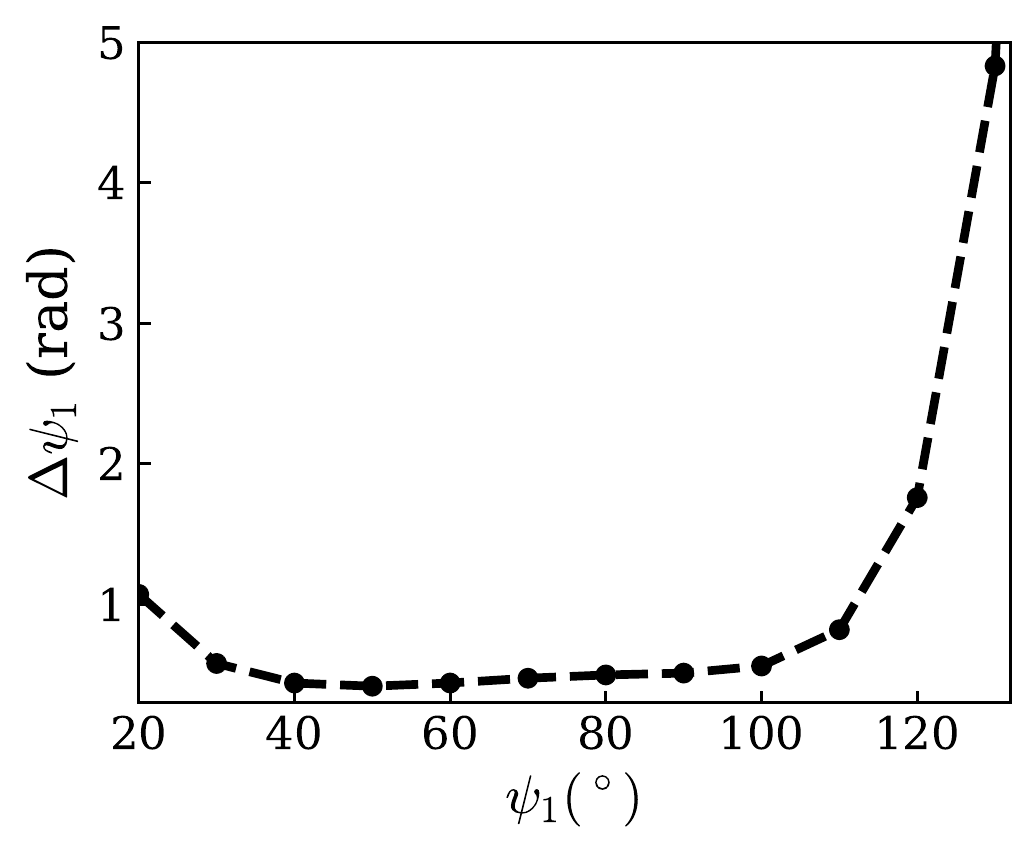} 
        \includegraphics[width=\columnwidth,height=6.1cm,clip=true]{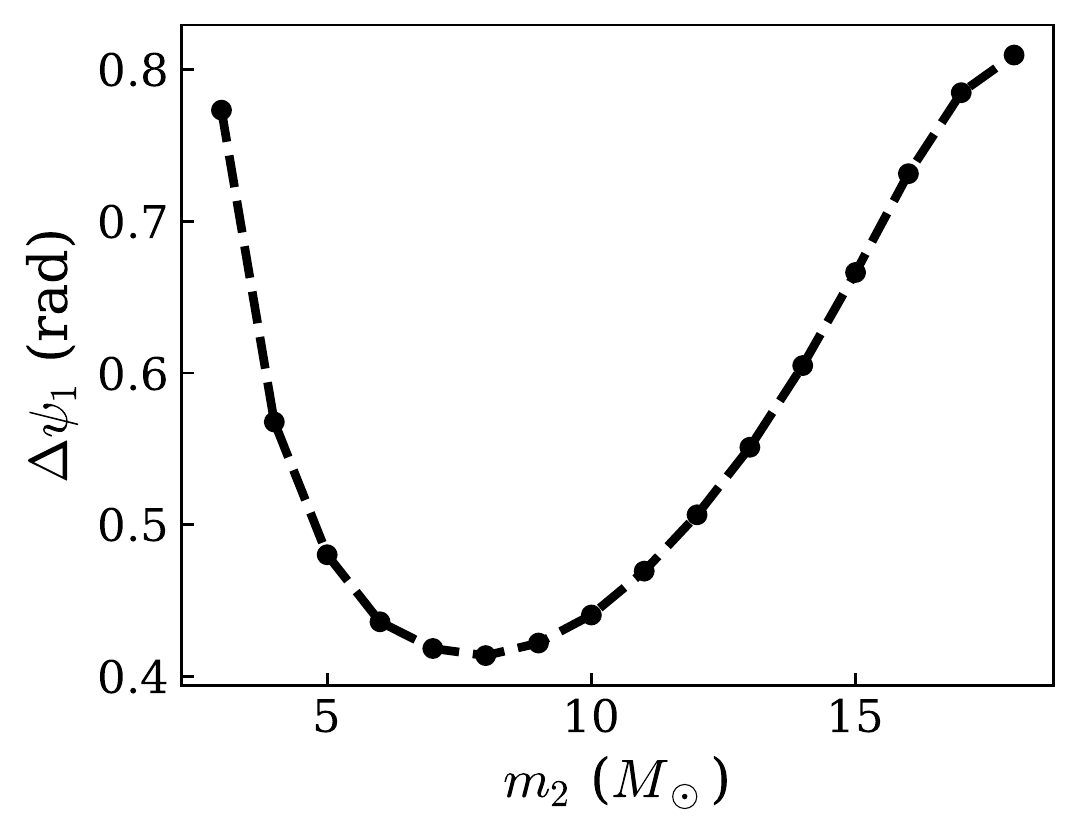} \\
        \includegraphics[width=\columnwidth,height=6.1cm,clip=true]{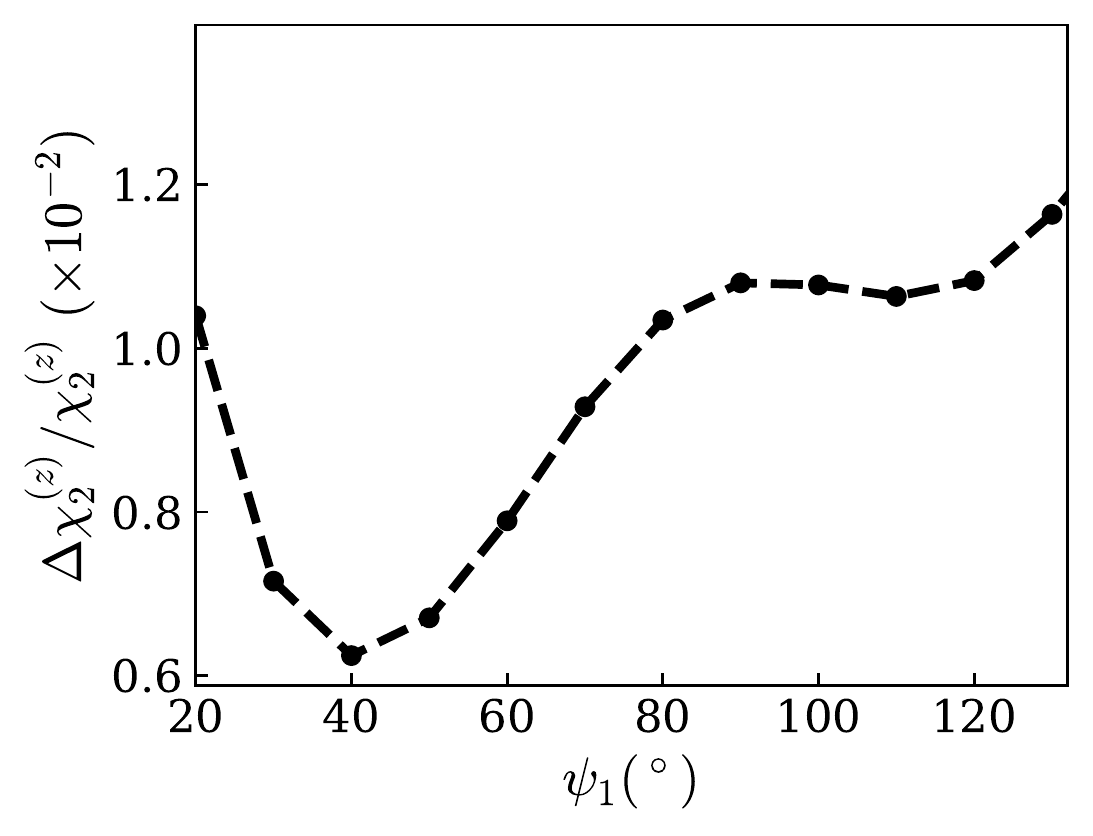}
        \includegraphics[width=\columnwidth,height=6.1cm,clip=true]{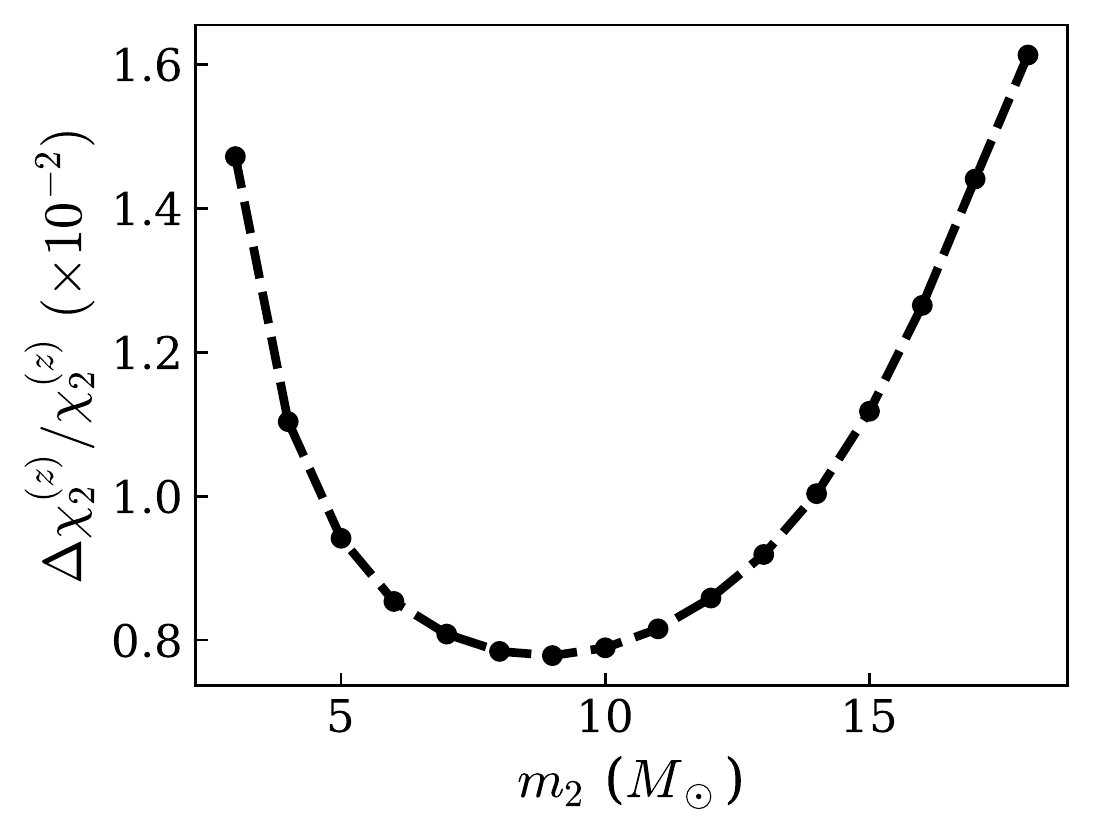}
  \caption{Case III: constraints on several parameters in the case of BHNS system. We choose $\Omega_{s1}=30$ Hz and $\chi_{2}^{(z)}=0.1$. The EoS is GM1, and the binary system is 100Mpc away from the detector. In the left panel, we show constraints as functions of $\psi_1$, where we fix the mass of BH to be $10M_\odot$; whereas in the right panel, we study their dependence on the mass of BH $m_2$, with $\psi_1=\pi/3$. Using universal relations, the degeneracy of parameters is totally broken, where $\Delta\psi$ is $\sim 1$ rad, and the relative errors of $\chi_{2}^{(z)}$ are $\sim1$\%.}
 \label{fig:fim-BHNS}
\end{figure*}


For a BHNS system, the $r$-mode resonance only takes place once before merger. As shown in Table \ref{table:rmode-uni-bhns-intro}, there are 8 parameters for the system: $\mc$, $\Xi$, $\bar{\lambda}_f^{(1)}$, $\bar{I}_1$, $\Omega_{s1}$, $\mathcal{I}_1$, $\psi_1$ and $\chi_2^{(z)}$. And there are 8 constraints from GW and universal relations. Hence the degeneracies can be reduced, even broken.

To study this case, we choose a BHNS system with $(1.4,10)M_\odot$. The NS is assumed to spin at 30 Hz, and EoS is GM1. As for BH, we assume $\chi_{2}^{(z)}=0.1$. The distance of the system is 100 Mpc. Results are shown in the left column of Fig.\ \ref{fig:fim-BHNS}. In this case, the degeneracy between two individual spins is completely broken. For $\chi_{2}^{(z)}$, it is constrained to $\sim1\%$, and $\psi_1$ is constrained to $\sim1$ rad.

We further investigate the effect of BH mass on constraints by varing $m_2$, while fixing $\Omega_{s1}$=30 Hz and $\psi_1=\pi/3$. The distance of the system is still 100 Mpc. In the right column of Fig.\ \ref{fig:fim-BHNS}, we show constraints as functions of BH mass. We can see both constrains first decrease with $m_2$, because of the increase of SNR. If we further increase $m_2$, post-resonance signals then will be reduced, and the constraints will become worse accordingly.

\section{Conclusion} 
\label{sec:conclusion}

In this paper, we studied the tidal excitation of $r$-mode by 
the 
gravitomagnetic force in coalescencing NS binaries. We began by a brief review on the dynamics of 
these 
systems: the $r$-mode is excited by 
the
gravitomagnetic field from the companion, while the induced current quadrupole moment gives rise to an acceleration back to the orbit. By assuming the orbit to be quasicircular, we obtained a coupled EOM. Next, we numerically integrated the coupled set of EOM and discussed features of solutions. We confirmed that the pre- and post-resonance orbital evolution can be well described by two PP orbits, as proposed in FR07 \cite{Flanagan:2006sb}. The post-resonance PP orbit is related to the pre-resonance one through a ``jump'' in orbital frequency at the resonance. We subsequently investigated the tidal evolution, by extending Ref.\ \cite{Ma:2020rak} to the $r$-mode, and providing analytic formulae for tidal evolution that are accurate in all regimes: adiabatic, resonance and post-resonance. Separately, using the TOV equation, we found a universal relation between the normalized $r$-mode overlap $\bar{I}^r$ of a neutron star and its normalized tidal Love number $\bar{\lambda}_f$.

We then moved on to the wave zone and studied gravitational waves emitted by such binaries. We constructed a hybrid GW waveform that combines several SPA models with results from numerical integrations of the coupled EOM. This waveform contains information from $r$-mode resonance, adiabatic tidal effect, and spin-orbit coupling. To understand the effect of $r$-mode DT  on  GW waveforms, we adopted the model in FR07: the $r$-mode induces phase and time shifts in GW. We found  this model to be quantitatively accurate. Finally, with the hybrid waveform, we calculated Fisher information matrix to investigate how $r$-mode resonances improves parameter estimation accuracy. We mainly studied three cases: binary NS systems with $r$-mode resonances in NSs,  binary NS systems including $r$-mode resonances together with universal relations between NS properties, and BHNS binary systems.

We found a variety of interesting results. First, the excitation of $r$-mode is mainly described by two parameters: spin frequency $\Omega_s$ and $r$-mode coupling coefficient $\mathcal{I}$. The spin frequency $\Omega_s$ determines when the resonance takes place during the inspiral, and $\mathcal{I}$ determines the phase and time shifts induced by the $r$-mode resonance. Choosing a $(1.4,1.35)M_\odot$ BNS system at 100Mpc away as example, we found that when the inclination angle is within the range of $\left[18^\circ,110^\circ\right]$ (and if $\Omega_{s1,2}$ are not too close), the measurements errors of $\mathcal{I}_i$ and $\Omega_{si}$ are less than 100\%, where Fisher analysis is valid and we can extract meaningful information from GW. The best constraint on $\Omega_s$ is around $6\%$ with 3G detectors; whereas for $\mathcal{I}$, the value is around 22\%. The constraint on $\mathcal{I}$ is worse than $\Omega_s$ because $\partial \tilde{h}^{N+r}/\partial\mathcal{I}$ is suppressed by the factor $(1-f/f_r)$ as $f\sim f_r$ [Eq.\ (\ref{phplove})]. In other words, the waveform is more sensitive to the location of the resonance than to the phase shift. With the analytic model for the $r$-mode, we found $\Delta\mathcal{I}/\mathcal{I}\sim\Delta\Omega_s/\Omega_s\sim \sin^{-2}\psi\cos^{-4}\psi/2$ [Eq.\ (\ref{deltaI-deltaspin-dep})]. This is consistent with our numerical calculations. The formula shows that the constraint is the best when $\psi=\pi/3$, while there is no constraint as $\psi\to0,\pi$. We also found that two resonances in each star do not get strongly correlated except for $\Omega_{s1}\sim\Omega_{s2}$, when effects from the two resonances become indistinguishable.

Secondly, with the help of the universal relations, $r$-mode resonance behaves like a bridge that connects adiabatic tidal effect and the spin-orbit coupling. In principle, for systems which have two $r$-mode resonances, we can obtain as many constraints as free parameters in GW. This situation is in contrary to the case without DT, where the universal relation requires additional parameters to be incorporated. This is because if one wants to use I-Love relation to connect adiabatic tidal effect and the spin-orbit coupling, four more free parameters: inclination angle $\psi_i$ and spin frequency $\Omega_{si}$ should be introduced. In the absence of $r$-mode resonance, they cannot be constrained at all. 

Although the $r$-mode resonance provides enough constraints to decode the BNS system, some constraints are not very accurate in practice. For example, errors on $\mathcal{I}_i$ sometimes are as large as 100\%, where the information is not meaningful. This will diminish the role of $r$-mode in degeneracy breaking. Nevertheless, our calculations show that two individual normalized Love numbers are still significantly correlated in the most cases. The  best relative errors of symmetric normalized Love number is $\sim1.3\%$, while is 1.84 for the anti-symmetric normalized Love number. Both of them are improved by factors of $\sim300-400$ in the best-case scenario, comparing with those that come solely from PN effects. In favorable cases, the normalized Love numbers of individual NS can be sufficiently well constrained. As shown in Table \ref{table:fim-uni}, each normalized Love number is constrained to $20\%$, significantly improve our understanding on the NS EoS. Meanwhile, the $r$-mode DT allows us to put constraints on the inclination angle between the spin and orbital angular momentum, which is hard to be accessible by other PN effects. In the best-case scenario, each inclination angle is constrained to 0.09 rad. This could potentially constrain the NS natal kick and hence the supernovae explosion mechanism. The other benefit from the universal relations is constraints on mass ratio $\Xi$, which is known to have correlated errors with $\bar{\lambda}_f^{s}$ and $\chi_s^{(z)}$ in absence of DT. After including $r$-mode DT and universal relations, $\Xi$ measurement can be improved by actors of $1-11.6$. For most cases, its fractional error is around $\sim1\%$. An improved estimation accuracy on $\Xi$ means better accuracy on the component masses. This could constrain the maximum mass of NSs with a large number of detection and shed light on the NS EoS in a way complementary to the information derived from tidal deformability.



Thirdly, for BHNS systems, we can also obtain as many constraints as parameters. As a result, degeneracies are totally broken. Choosing a BHNS system as example, we found $\Delta\psi_1\sim 1$ rad, and $\Delta\chi_2^{(z)}/\chi_2^{(z)}\sim1\%-2\%$.


Our results show that $r$-mode resonance will be an important channel for 3G detectors to extract information of NSs. Since the excitation only requires NSs to spin at tens of Hertz, these events are quite generic in coalescing binaries that have NSs. Therefore, to develop an accurate GW waveform from these systems seems necessary in the future. Our numerical calculations of $r$-mode are only on Newtonian order, and PN effects are incorporated through a naive way. Also, the corrections of DT onto PN terms are not considered here.  Therefore, one possible avenue for future work is to perform a systematic study on the scenario with relativity. This includes to couple the gravitomagnetic force with rotational modes of relativistic stars\footnote{There is no pure $r$-mode in relativistic barotropic stars.} by the formalism in Ref.\ \cite{Lockitch:2000aa}, and to study the orbital evolution with either PN approach \cite{Racine:2004hs} or EOB formalism \cite{Hinderer:2016eia,Steinhoff:2016rfi}. As pointed out by Idrisy \etal \cite{Idrisy:2014qca}, there is more than 10\% variance for the mode frequency of relativistic stars, depending on the compactness. This is on the same order of statistical accuracy of $\Omega_s$ in our paper. Therefore, the relativistic corrections would be important in this case. It is also interesting to see how relativistic corrections to the $\bar{I}^r-\bar{\lambda}_f$ universal relation changes the parameter estimation. 

The other direction would be numerical relativity. Although the excitation of $f$-mode has been observed by recent study \cite{Foucart:2018lhe}, the simulation of $r$-mode is more difficult to achieve with the current version of numerical relativity code, such as SpEC \cite{spec}, since the mode amplitude of $r$-mode is much smaller. A typical Lagrangian displacement of $r$-mode is only $10^{-4}$ of the radius of a NS. This requires much more resolutions to resolve $r$-mode. However, with the upgraded version of SpEC, SpECTRE \cite{Kidder:2016hev}, this may be doable in the near future.

Furthermore, our paper mainly focus on the $(2,2)$ $r$-mode in barotropic NSs. As pointed out by Poisson \etal recently \cite{Poisson:2020eki}, four inertial modes can be excited by the gravitomagnetic force before merger. Meanwhile, for NSs with buoyancy, there is also the $(2,1)$ $r$-mode, which plays a role as important as the $(2,2)$ mode \cite{Flanagan:2006sb}. These modes have different $\psi$-dependence, and contain different information of NSs. Therefore, they can further reduce the degeneracy of parameters, and put more constraints on the inclination angle $\psi$. On the other hand, the detection of $(2,1)$ mode can confirm the existence of buoyancy in cold NSs, so it is worth to incorporate these modes in the future. 

A caveat, however, is that our analysis as well as the studies mentioned above, all assumed that the matter inside the NS behaves as a normal fluid. In reality, superfluidity may be expected in cold NSs~\cite{Yakovlev:99} and may lead to richer dynamics than what we considered here thanks to its two-fluid nature~(see e.g., Ref.~\cite{Andersson:01}). Ref.~\cite{Lindblom:00} showed that the superfluid $r$-modes characterized by a common flow of neutrons and protons reduce to their normal-fluid counterparts (i.e., the $r$-mode studied in our work) in the slow-rotation limit. On the other hand, Ref.~\cite{Andersson:01} argued the existence of a new class of $r$-modes whose fluid motion is such that neutrons and protons are counter moving. This new class of $r$-modes are not accounted for in our current study and are deferred to future studies.

Meanwhile, we ignored damping on the $r$-mode due to microphysical processes in the NS. While viscous damping and nonlinear saturation may play a critical role for the $r$-mode instability in newly-born NSs and/or X-ray binaries~\cite{Andersson:2000mf}, its effect might be subdominant in coalescing NS binaries given the very short duration of tidal excitation [$<1\,{\rm S}$; Eq.~(\ref{eq:tau_r})] compared to the typical viscous damping timescale of $10^4\,{\rm s}$~\cite{Lindblom:00}. Nevertheless, sufficiently large uncertainty remains in our current understanding of the NS microphysics, and future explorations on the dissipation of saturation of $r$-modes under various astrophysical contexts will be of great value.

Finally, we want to emphasize that we treated the two universal relations as exact relations. However, as pointed out in Ref.~\cite{Carson:2019rjx}, even the most insensitive relations still have residual variability with respect to EoSs and could lead to systematic bias in parameter estimations for 3G detectors pontentially comparable to the statistical uncertainties. Therefore, studying the impact of such EoS variability would be another interesting direction to go.

\begin{acknowledgments}
We thank the LIGO Extreme Matter working group for the useful comments during the preparation of this work. We also thank Chris van den Broeck for useful comments. Research of S.M.\ and Y.C.\ are supported by the Simons Foundation (Award Number 568762), the Brinson Foundation, and the National Science Foundation (Grants 
PHY-2011968
PHY-2011961 and PHY-1836809). H.Y. is supported by the Sherman Fairchild Foundation. The computations presented here were conducted on the Caltech High Performance Cluster, partially supported by a grant from the Gordon and Betty Moore Foundation.
\end{acknowledgments}  

\appendix

\section{Justification of ignoring the PP precession}
\label{app:precession}

When the spins $\bm{S}_{1,2}$ are misaligned with respect to the orbital angular momentum $\bm{L}$, various precession effects will happen and, in principle, modify the dynamics when the orbit is both close to and far away from a mode's resonance. We will show quantitatively that all the precession-induced corrections are small and therefore can be safely ignored. 

Specifically, we have~\cite{Barker:75, Schmidt:2014iyl}
\begin{align}
    & \frac{d\uvect{S}_1}{dt} = \vect{\Omega}_{\rm dS}^{(1)}\times \uvect{S}_{1},
    \label{eq:dS1_v_dt} \\ 
    & \frac{d\uvect{S}_2}{dt} = \vect{\Omega}_{\rm dS}^{(2)}\times \uvect{S}_{2},
    \label{eq:dS1_v_dt} \\
    & \frac{d\uvect{L}}{dt} = \left[\vect{\Omega}_{\rm LT}^{(1)} + \vect{\Omega}_{\rm LT}^{(2)}\right] \times \uvect{L},
\end{align}
where
\begin{equation}
    \vect{\Omega}^{(1, 2)}_{\rm dS} = \frac{3\left(m_{2, 1} + \mu/3\right)}{2D}\Omega_{\rm orb} \uvect{L}, \label{eq:omega_dS} 
\end{equation}
is the rate of the leading-order de Sitter precession of the spins induced by the orbital angular momentum $\vect{L}$ and 
\begin{equation}
    \vect{\Omega}_{\rm LT}^{(1,2)} = \frac{S_{1,2}\left(4+3m_{2,1}/m_{1,2}\right)}{2 D^3}\uvect{S}_2,
\end{equation}
is the rate of the Lense-Thirring precession of $\vect{L}$ due to $\vect{S}_{1,2}$. The hat stands for the unit vector along the direction of the corresponding quantity.

As a result of the de Sitter precession of $\vect{S}_1$, the resonance frequency of the $r$-mode in $m_1$ should be shifted by $\Omega_{\rm dS}^{(1)}$ with
\begin{equation}
    \frac{\Omega_{\rm dS}^{(1)}}{2\pi} \simeq 0.8 \times \left(\frac{f_r}{80\,{\rm Hz}}\right)^{5/3},
\end{equation}
assuming a binary with $m_1\simeq m_2\simeq 1.4\,M_\odot$ and evaluating at $f=f_{r}$ with $f_r$ the resonant frequency [Eq.\ (\ref{reson-gw-fre})]. The above shift is $\sim1\%$ of the spin frequency, about 10 times smaller than the typical statistical error of $\sim 10\%$ (e.g.,  Fig.~\ref{fig:nouni}), it can thus be neglected. 

Meanwhile, the Lense-Thirring precession of the orbital angular momentum together with the spin-spin interaction will cause the inclination angle of the spin of NS, $\psi_{1(2)}$, evolves slowly as~\cite{Yu:20b}
\begin{equation}
    \frac{d}{dt}\cos\psi_1 = \Omega_{\Delta \psi_1} \uvect{S}_1 \cdot \left(\uvect{S}_2\times \uvect{L}\right),
\end{equation}
where $\Omega_{\Delta \psi_1} = 3 (1+q)S_2/2qD^3 \simeq (6/7)\Omega_{\rm LT}^{(2)}$ for nearly equal-mass binaries with $q\equiv m_2/m_1\simeq 1$. We can thus define a timescale $\tau_{\Delta \psi_1}=1/\Omega_{\Delta \psi_1}$, which is given by  
\begin{equation}
    \tau_{\Delta \psi_1} \simeq 70\,{\rm s}\left(\frac{f_{r}}{80\,{\rm Hz}}\right)^{-2}\left(\frac{\chi_2\sin\psi_2}{0.02}\right)^{-1}. 
\end{equation}
On the other hand, the duration of resonance is given by [Eq.\ (\ref{ttilde})]
\begin{align}
    \tau_{\rm r} \simeq \left(\frac{1}{\omd}\right)^{1/2}= 0.52\,{\rm s}\left(\frac{f_r}{80\,{\rm Hz}}\right)^{-11/6}. 
    \label{eq:tau_r}
\end{align}
We thus see that $\psi_{1}$ is a well-defined quantity at resonance. Moreover, it is well defined throughout the entire post-resonance evolution which lasts about 4\,s (Fig.~\ref{fig:AB}), much shorter than $\tau_{\Delta \psi_1}$. 

The above two points allow us to conclude that the modifications due to precession is indeed negligible during mode resonances. We now consider their effects away from resonance. To do so, we drop the $r$-mode effects and use the \texttt{LALSuite}~\cite{lalsuite} to generate PP waveforms with \texttt{IMRPhenomPv2} approximation~\cite{Schmidt:2014iyl, Hannam:14}. 

\begin{figure}[htb]
        \includegraphics[width=\columnwidth,height=6.1cm,clip=true]{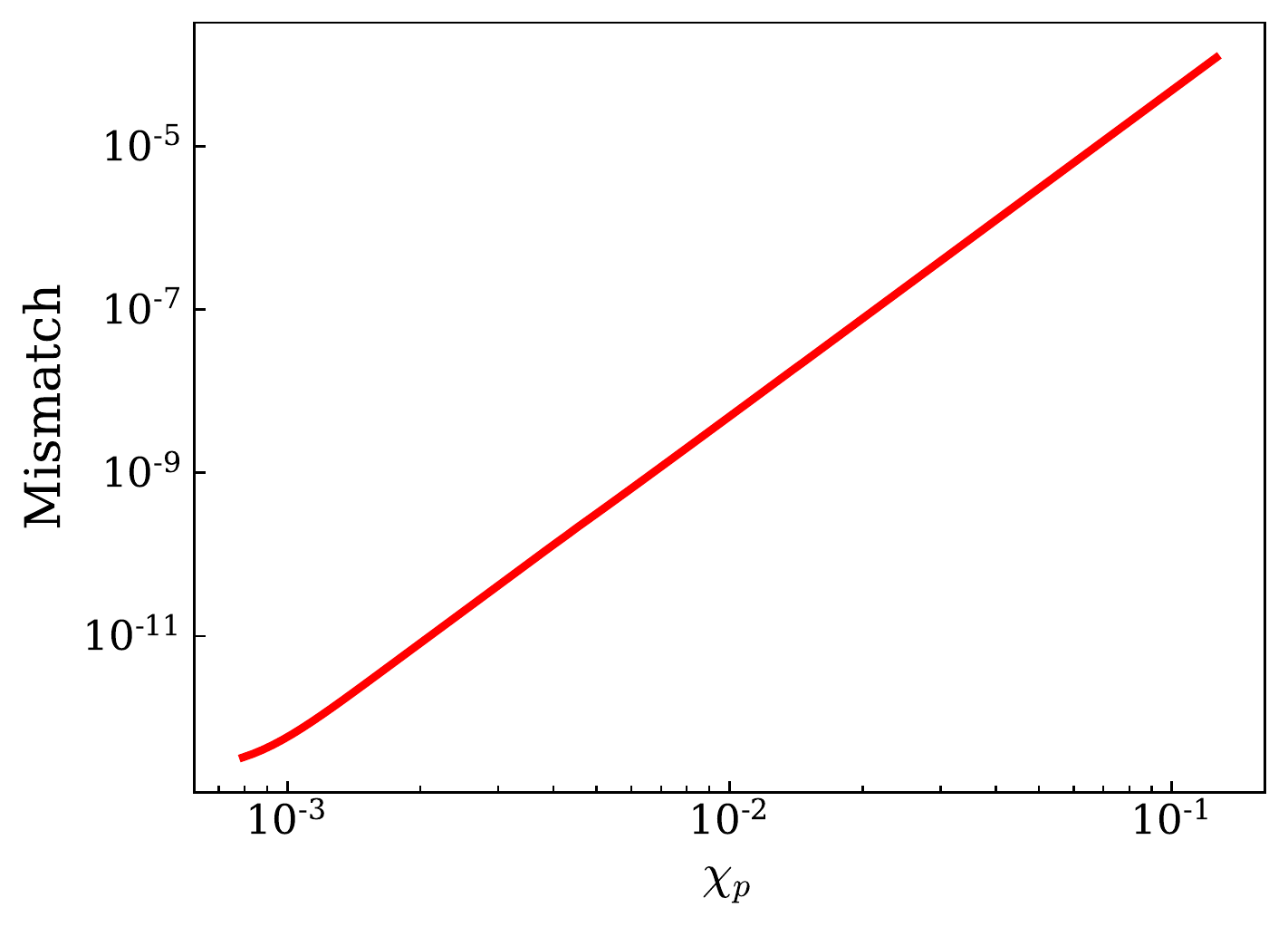}
  \caption{Mismatch between a precessing waveform $\chi_{\rm p}\neq 0$ and an a non-precessing but otherwise identical one.}
 \label{fig:mismatch}
\end{figure}

In Fig.~\ref{fig:mismatch} we compute the mismatch of a precessing waveform (parameterized in terms of $\chi_{\rm p}$) with a  non-precessing but otherwise identical one. Here the mismatch is defined as
\begin{equation}
\text{Mismatch}(h_1, h_2) = 1-\max_{t_c, \phi_c}\frac{(h_1|h_2)}{\sqrt{(h_1|h_1)(h_2|h_2)}}. 
\end{equation}
For a typical NS with spin $< 100 $ Hz, it corresponds to a precession parameter $\chi_{\rm p} \lesssim 0.1$ (even assuming $\psi=\pi/2$; $\chi_{\rm p}$ is smaller for harder EOSs). Consequently, neglecting precessions will only cause small errors on the PP waveforms. 

\begin{figure}[htb]
        \includegraphics[width=\columnwidth,height=7.3cm,clip=true]{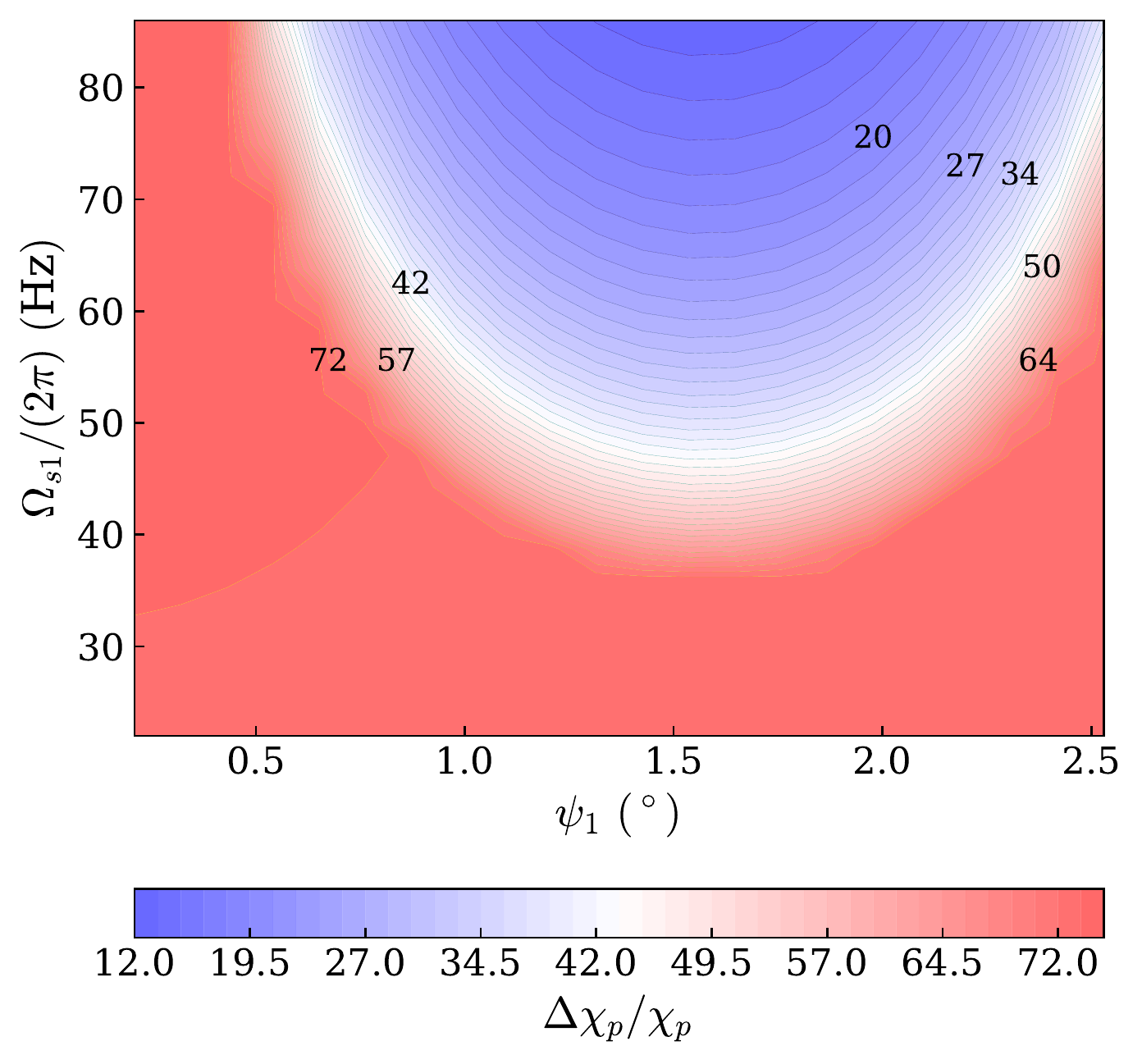}
  \caption{The fractional error in inferring the PP precession parameter $\chi_p$. }
 \label{fig:err_chi_p}
\end{figure}

Lastly, we show in Fig.~\ref{fig:err_chi_p} the parameter estimation uncertainties of $\chi_{\rm p}$ using the PP waveform alone. Here we have assumed a fiducial relation of $\chi_{\rm p} =0.06 \Omega_{s_1}\sin\psi_1/(2\pi \times100\,{\rm Hz})$ so that we can show the y-axis in physical spin units. As the fractional error $\Delta \chi_{\rm p}/\chi_{\rm p} > 10$ for the parameter space of interest, we thus do not expect we would be able to further improve the parameter estimation accuracy by incorporating the precession effects.





\section{The Tolman–Oppenheimer–Volkoff equations}
\label{app:TOV}
The stress-energy tensor $T_{\mu\nu}$ for a perfect fluid is given by
\begin{align}
T_{\mu\nu}=(\rho+p)u_\mu u_{\nu}+pg_{\mu\nu},
\end{align}
where $p$ and $\rho$ stand for the pressure and energy density of the star, and $u_\mu$ is four-velocity. The metric $g_{\mu\nu}$ is given by Eq.\ (\ref{metric-equ}). With Einstein field equation, quantities shown above can be solved by the TOV equations
\begin{align}
&\frac{dm}{dr}=4\pi r^2\rho, \label{TOV-M}\\
&\frac{dp}{dr}=-\frac{(4\pi r^3p+m)(\rho+p)}{r(r-2m)}, \\
&\frac{d\nu}{dr}=2\frac{4\pi r^3p+m}{r(r-2m)}, \label{TOV-nu}
\end{align}
where 
\begin{align}
m\equiv\frac{1-e^{-\lambda}}{2}r. \label{m-lambda}
\end{align}
In practice, it is preferable to cast them into a new form for numerical integration. Following the procedure of Ref.\ \cite{Lindblom:1998dp}, we use the specific enthalpy $h$, defined by 
\begin{align}
dh=\frac{dp}{\rho+p},
\end{align}
to replace $r$, where the integration constant is set by the condition $h\rightarrow0$ as $\rho\rightarrow0$ and $p\rightarrow0$. Defining two new dependent variables, $u=r^2$ and $v=m/r$, then we have
\begin{align}
&\frac{du}{dh}=-\frac{2u(1-2v)}{4\pi up+v}, \label{udot}\\
&\frac{dv}{dh}=-(1-2v)\frac{4\pi u\rho-v}{4\pi up+v},\label{vdot}\\
&\frac{d\nu}{dh}=-2. \label{nudot}
\end{align}
At the center of the star, we have
\begin{align}
h=h_c, \quad u = 0, \quad v=0,  \label{ini}
\end{align}
with $h_c$ a free parameter. The surface of star locates at $h=0$. 

Therefore, we can find the structure of the star by integrating Eqs.\ (\ref{udot})--(\ref{nudot}) from $h=h_c$ to $h=0$, with the initial conditions in Eq.\ (\ref{ini}). The total mass of the star can be obtained from the formula $m_{\rm NS}=\sqrt{u}v|_{h=0}$, and the radius is given by $R_{\rm NS}=\sqrt{u}|_{h=0}$. The quantity $\nu$ is linear in $h$, where the integration constant is set by the condition $\nu|_{h=0}=\log(1-2m_{\rm NS}/R_{\rm NS})$, to connect the value outside of star.
\section{The calculation of tidal Love number $\lambda_f$}
\label{app:uni-love}
Let us consider linearized even-parity perturbations to the equilibrium metric in Eq.\ (\ref{metric-equ}). Following Refs.\ \cite{1967ApJ...149..591T,PhysRevD.43.1768,PhysRevD.56.2118}, the perturbed metric in the Regge-Wheeler gauge can be written as
\begin{align}
g_{\mu\nu}=g_{\mu\nu}^{(0)}+h_{\mu\nu},
\end{align}
with 
\begin{align}
h_{\mu\nu}=\text{diag}[e^{-\nu}H_0,e^{\lambda} H_2,r^2K,r^2\sin^2\theta K]Y_{lm}(\theta,\phi),
\end{align}
where $H_0$, $H_2$, $K$ are functions of $r$. The perturbation on the stress-energy tensor is given by \cite{1967ApJ...149..591T}
\begin{align}
&\delta T^0_0=-\delta\rho_lY_{lm}(\theta,\phi)=-\frac{d\rho}{dp}\delta p_lY_{lm}(\theta,\phi), \\
&\delta T^i_i=\delta p_lY_{lm}(\theta,\phi).
\end{align}
With Einstein field equation, we obtain \cite{PhysRevD.56.2118}
\begin{align}
&H_2=H_0=H,\\
&H''+H'\left\{\frac{2}{r}+e^{\lambda}\left[\frac{2m}{r^2}+4\pi r(p-\rho)\right]\right\} \notag \\
&+H\left\{e^\lambda\left[-\frac{6}{r^2}+4\pi(\rho+p)\frac{d\rho}{dp}+4\pi(5\rho+9p)\right] \right. \notag \\
&\left.-\left(\frac{d\nu}{dr}\right)^2\right\}=0, \label{Hddot}
\end{align}
where we only focus on the $l=2$ component. The mass function $m$ is related to the metric function $\lambda$ by Eq.\ (\ref{m-lambda}).

Imposing regularity condition at $r=0$ yields the initial condition $H\propto r^2$. The proportionality constant does not matter here, so we simply choose it as 1. Functions $\lambda$, $m$, $p$, $\rho$ in the above equation can be obtained from the solutions of TOV equations. 

Integrating Eq.\ (\ref{Hddot}) from $r=0$ to $r=R_{\rm NS}$ leads to the dimensionless tidal Love number $k_2$ as \cite{Hinderer:2007mb}
\begin{align} 
&k_2=\frac{8}{5}\mathcal{C}^5(1-2\mathcal{C})^2[2+2\mathcal{C}(y-1)-y]\left\{2\mathcal{C}[6-3y\right. \notag \\
&+3\mathcal{C}(5y-8)]+4\mathcal{C}^3[13-11y+\mathcal{C}(3y-2)+2\mathcal{C}^2(1+y)] \notag \\
&+3(1-2\mathcal{C})^2[2-y+2\mathcal{C}(y-1)]\log(1-2\mathcal{C})\}^{-1},
\end{align}
where $y=R_{\rm NS}H'(R_{\rm NS})/H(R_{\rm NS})$, and the tidal Love number $\lambda_f$ is given by
\begin{align}
\lambda_f=\frac{2}{3}k_2R_{\rm NS}^5.
\end{align}

\section{The GW phase with SPA}
\label{app:pn-phase}

From Ref.~\cite{Arun:2004hn,Kidder:1992fr,Arun:2008kb,Blanchet:2011zv,2005PhRvD..71l4043M,2008ApJ...677.1216H,2012PhRvD..85l3007D,Henry:2020ski}, we obtain the frequency-domain gravitational-wave phasing, up to 3.5\,PN order for point-particle contributions, up to 3PN for spin terms, and up to 2.5\,PN for adiabatic, $f$-mode tide.  Here are terms {\it in addition to} the leading Newtonian phasing $\Psi_N$: 
\begin{widetext}
\begin{subequations}
\begin{align}
&\Psi_\text{PP}=\frac{3}{128}(\pi\mc f)^{-5/3}\left\{\left(\frac{3715}{756}+\frac{55}{9}\eta\right)x-16\pi x^{3/2} +\left(\frac{15293365}{508032}+\frac{27145}{504}\eta+\frac{3085}{72}\eta^2\right)x^2 +\left(\frac{38645}{756}\pi-\frac{65}{9}\pi\eta\right)     \right. \notag \\
&\times (1+3\log v)x^{5/2}+\left[\frac{11583231236531}{4694215680}-\frac{640\pi^2}{3}-\frac{6848}{21}\gamma_E-\left(\frac{15737765635}{3048192}-\frac{2255}{12}\pi^2\right)\eta+\frac{76055}{1728}\eta^2-\frac{127825}{1296}\eta^3\right. \notag \\
&\left.\left.-\frac{6848}{21}\log(4v)\right]x^3+\left(\frac{77096675}{254016}+\frac{1014115}{3024}\eta-\frac{36865}{378}\eta^2\right)\pi x^{7/2}\right\}, \label{psi-pp}\\
&\Psi_{SO}=\frac{3}{128}(\pi\mc f)^{-5/3}\left\{4\left(\frac{113}{12}-\frac{19}{3}\eta\right)(\bm{\hat{L}}\cdot\bm{\chi}_s)x^{3/2} -10\left[\frac{719}{48}\delta_m(\bm{\hat{L}}\cdot\bm{\chi}_s)(\bm{\hat{L}}\cdot\bm{\chi}_a)+\left(\frac{719}{96}+\frac{\eta}{24}\right)(\bm{\hat{L}}\cdot\bm{\chi}_s)^2\right.  \right. \notag \\
&\left.+\left(\frac{719}{96}-30\eta\right)(\bm{\hat{L}}\cdot\bm{\chi}_a)^2\right]x^2-(1+3\log v)\left[\left(\frac{732985}{2268}-\frac{24260}{81}\eta-\frac{340}{9}\eta^2\right)(\bm{\hat{L}}\cdot\bm{\chi}_s)+\left(\frac{732985}{2268}+\frac{140}{9}\eta\right)\delta_m(\bm{\hat{L}}\cdot\bm{\chi}_a)\right] x^{5/2}\notag \\
&\left.+\frac{2270\pi}{3}\left[\left(1-\frac{227}{156}\eta\right)(\bm{\hat{L}}\cdot\bm{\chi}_s)+\delta_m(\bm{\hat{L}}\cdot\bm{\chi}_a) \right]x^3\right\}, \notag \\
&\Psi_{\bar{\lambda}^{(1)}_f}=-\frac{3}{16\eta}x^{5/2}(12-11\Xi)\bar{\lambda}^{(1)}_f\Xi^4\left\{1+\frac{5(3179-919\Xi-2286\Xi^2+260\Xi^3)}{672(12-11\Xi)}x-\pi x^{3/2}+\frac{1}{12-11\Xi}\left[\frac{39927845}{508032}-\frac{480043345}{9144576}\Xi\right.\right. \notag \\
&\left.\left.+\frac{9860575}{127008}\Xi^2-\frac{421821905}{2286144}\Xi^3+\frac{4359700}{35721}\Xi^4-\frac{10578445}{285768}\Xi^5\right]x^2-\frac{\pi(27719-22127\Xi+7022\Xi^2-10232\Xi^3)}{672(12-11\Xi)}x^{5/2}\right\}, \label{app-lambda1} \\
&\Psi_{\bar{\lambda}^{(2)}_f}=(1\rightarrow2 ~\text{and} ~\Xi\rightarrow 1-\Xi),\label{app-lambda2}
\end{align}
\end{subequations}
\end{widetext}
Here we have defined $x=v^2=(\pi Mf)^{2/3}$, $\Xi=m_1/M$, $\delta_m=(m_1-m_2)/M$. $\hat{L}$ is the unit vector along the orbital angular momentum. Symmetric and anti-symmetric dimensionless spins are defined as $\bm{\chi}_s=(\bm{\chi}_1+\bm{\chi}_2)/2$ and $\bm{\chi}_a=(\bm{\chi}_1-\bm{\chi}_2)/2$ with $\bm{\chi}_i=\bar{I}_i\bm{\Omega}_{si}m_i$. Here we use the normalized momentum of inertia, as well as the normalized tidal Love number in Eqs.\ (\ref{app-lambda1}) and (\ref{app-lambda2}), in order to use the I-Love universal relation.


\section{Case I for FPS EoS (without universal relations)}
\label{app:Case I}
\begin{figure}[htb]
        \includegraphics[width=\columnwidth,height=7.3cm,clip=true]{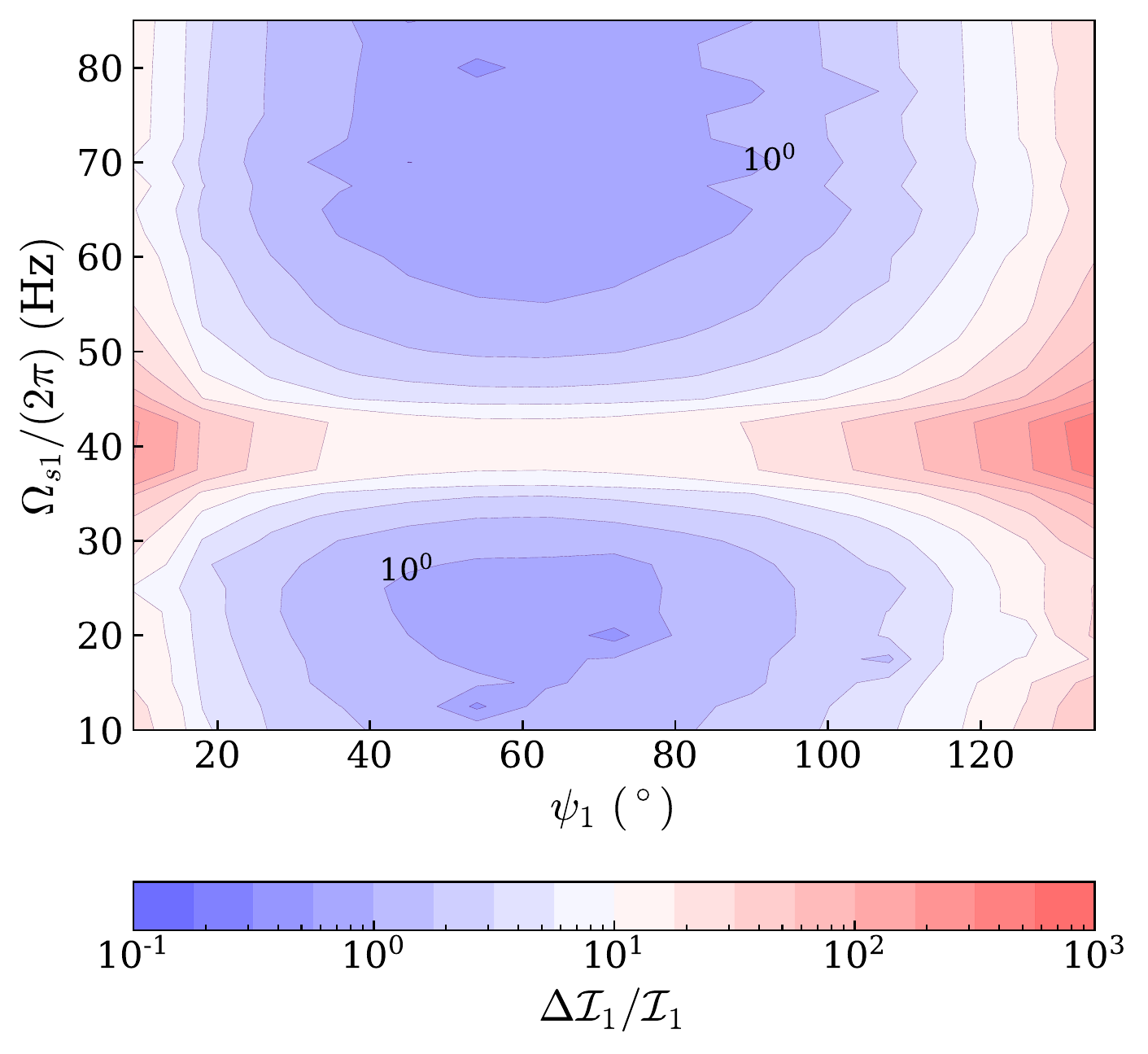}
         \includegraphics[width=\columnwidth,height=7.3cm,clip=true]{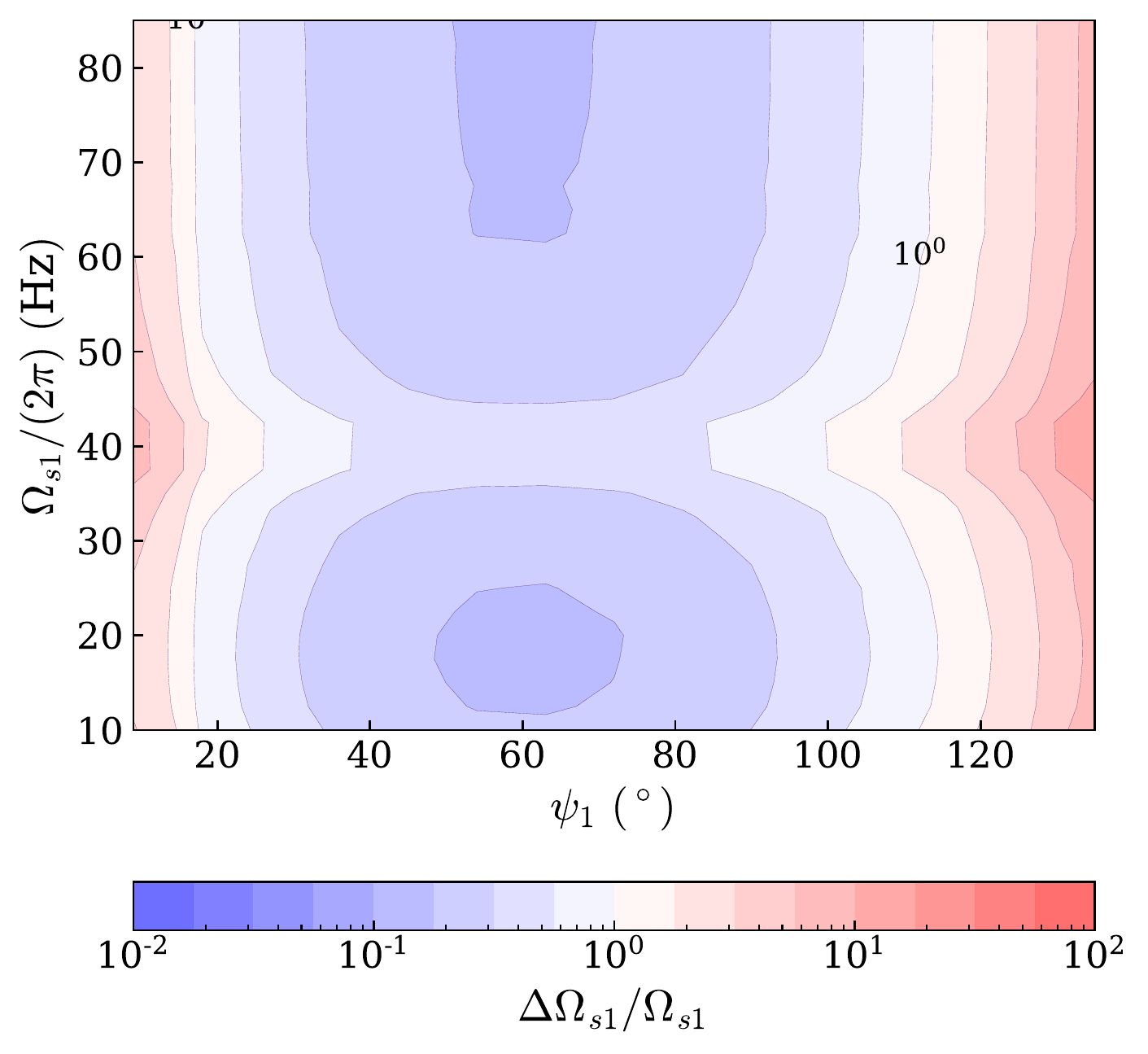} 
  \caption{Same as Fig.\ \ref{fig:nouni}, but with FPS EoS. The constraints are worse than those for GM1 by factors of $\sim2.6-2.7$}
 \label{fig:nouni-fps}
\end{figure}

In Fig.\ \ref{fig:nouni-fps}, we present the results for a BNS system with FPS EoS. Following Sec.\ \ref{sec:fim-I}, we still assume that both NSs are excited before merger. The spin vector of $m_2$ is fixed at $\Omega_{s2}=2\pi\times40$ Hz and $\psi_{2}=7\pi/18$. And we vary the spin of $m_1$: $\Omega_{s1}\in2\pi\times[10,85]$ Hz, $\psi_{1}\in[\frac{1}{20}\pi,\frac{17}{20}\pi]$. Without universal relations, we can put constraints on $\Omega_{si}$ and $\mathcal{I}_i$. From Table \ref{table:EOS-ratio}, we know that the constraints for FPS are worse than those for GM1 by factors of $\sim2.6-3.2$. But their dependence on $\Omega_{s1}$ and $\psi_{s1}$ is the same as the case of GM1.

\section{Case II for FPS EoS (with universal relations)}
\label{app:Case II}
\begin{figure}[htb]
         \includegraphics[width=\columnwidth,height=9cm,clip=true]{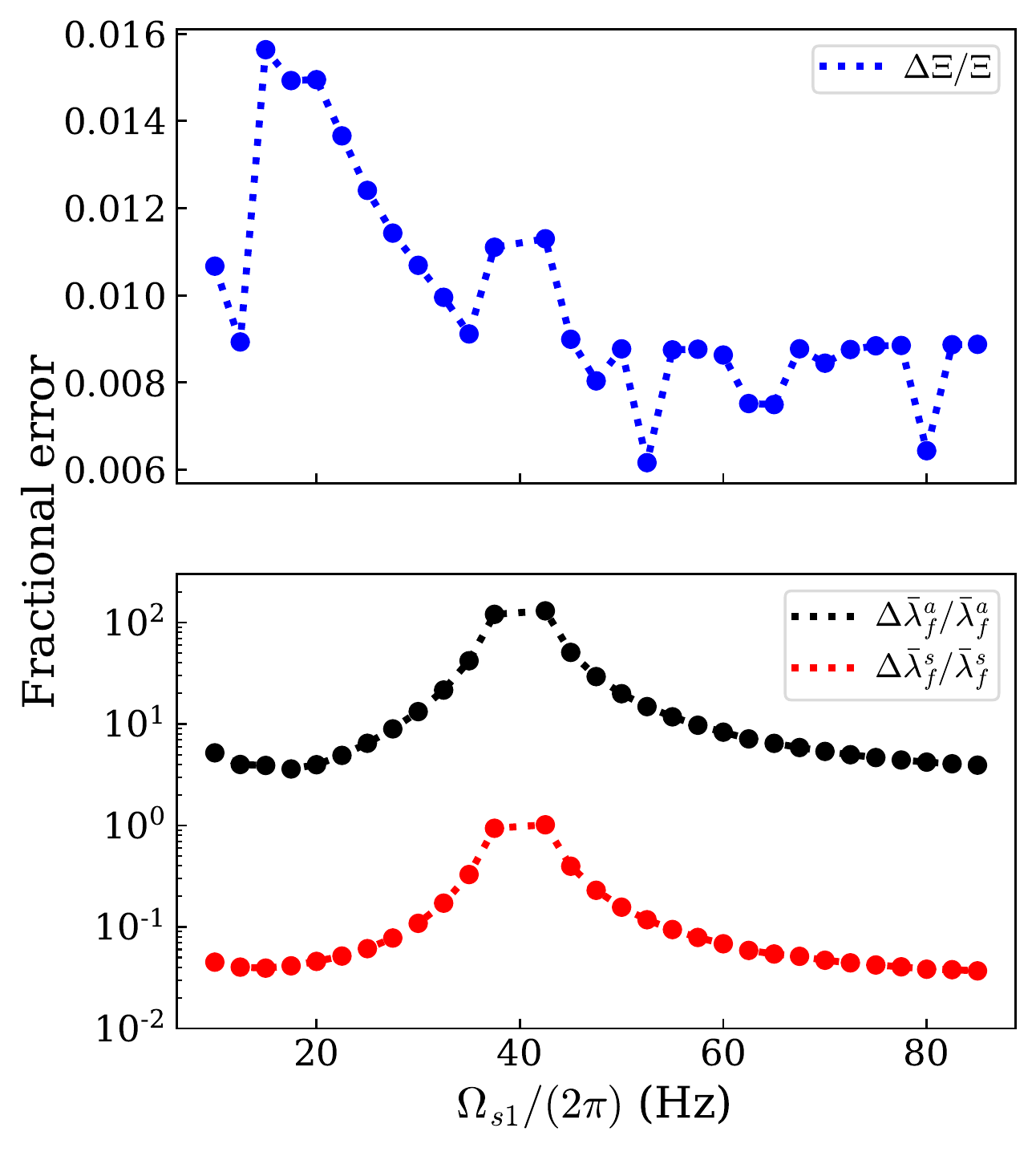}
  \caption{Same as Fig.\ \ref{fig:uni-GM1}, but with FPS EoS.}
 \label{fig:uni-fps}
\end{figure}

\begin{figure}[htb]
        \includegraphics[width=\columnwidth,height=6.4cm,clip=true]{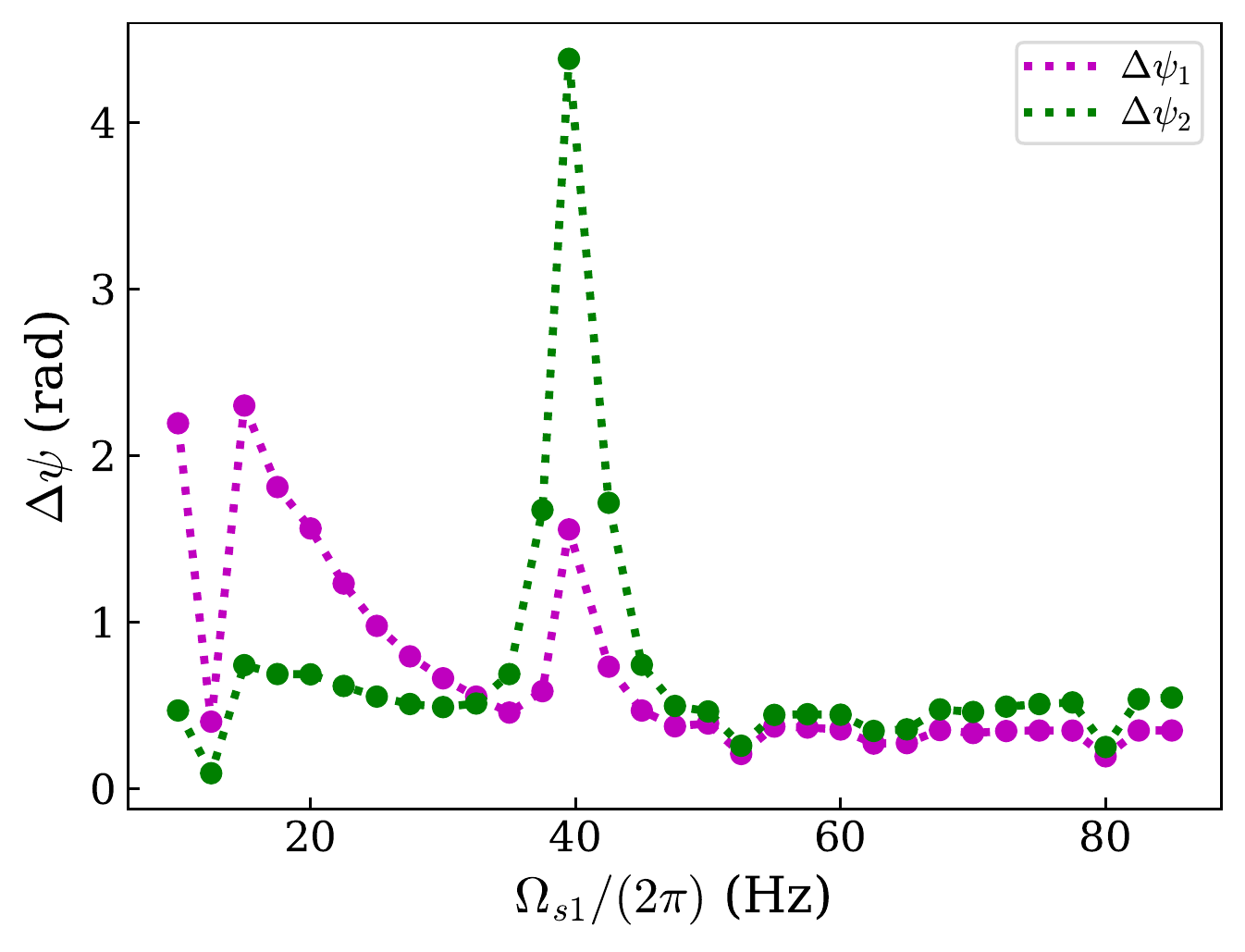}
  \caption{Same as Fig.\ \ref{fig:uni-psi-GM1}, but with FPS EoS.}
 \label{fig:uni-fps-psi}
\end{figure}

After incorporating universal relations into the calculations of Appendix \ref{app:Case I}, we can obtain the constraints on $\bar{\lambda}_f^{s(a)}$, $\Xi$, and $\psi_{i}$, as shown in Fig.\ \ref{fig:uni-fps} and \ref{fig:uni-fps-psi}.


\clearpage
\bibliography{References}

\end{document}